دانشکده مهندسی برق و کامپیوتر

پایان نامه‌ی کارشناسی ارشد در رشته‌ی مهندسی کامپیوتر- هوش مصنوعی

# مدل سازی انتشار اطلاعات در شبکه های اجتماعی

به وسیله‌ی
نیما حیدری

مهر ماه 1394

بسم الله الرحمن الرحیم

# Abstract

## Information Diffusion in Social Networks and Representing of a Paradigm for that

By
**Nima Heidari**

One major feature of social networks (e.g., massive online social networks) is the dissemination of information, such as news, rumors and opinions. Information can be propagated via natural connections in written, oral or electronic forms. The physics of information diffusion has been changed with the mainstream adoption of the Internet and Web. Until a few years ago, the major barrier for someone who wanted a piece of information to spread through a community was the cost of the technical infrastructure required to reach a large number of people. Today, with widespread access to the Internet, this bottleneck has largely been removed. Information diffusion has been one of the focuses in social network research area, due to its importance in social interactions and everyday life. More recently, during the last twenty to thirty years, there has been interest and attention not just in observing information and innovation flow, but also in influencing and creating them. Modeling information diffusion in networks enables us to reason about its spread. In this paper, we present a new method for modeling information diffusion process in social networks namely PSODM. This model takes advantage of using 'chromosomes' to represent individuals and schemas generated by HDF functions as pieces of information or knowledge on a certain topic. We use the particle swarm optimization as an interaction tool between network individuals to overcome the 'information lost' problem. Each node of the underlying social network is considered as a particle in a swarm. A state vector represents a particle position which demonstrates its information. The position (information level) update of these particles takes place whenever a particle contacts its neighbors. In addition, in order to compare different proposed models in the area, we use a community detection based approach that has been introduced in our previous work. With modeling information flow in a large online social network derived from Enron Email dataset, we experimentally demonstrate the effectiveness of PSODM. On the other hand, we use the concept of community detection in social networks for the purpose of comparison between the proposed method and other state-of-the art models.

# چکیده

## انتشار اطلاعات در شبکه های اجتماعی و ارائه ی مدلی برای آن

بسیاری از سیستم های پیچیده طبیعی و اجتماعی را می توان در قالب یک شبکه یا گراف نشان داد، مجموعه ای از رأس ها و یال هایی که این رأس ها را به هم متصل می کنند. به عنوان نمونه هایی از شبکه ها، می توان به شبکه های اجتماعی، شبکه های فنی (مانند اینترنت) و شبکه های زیستی (مانند شبکه های عصبی) اشاره کرد. رأس ها در این شبکه ها، موجودیت ها و یال ها، ارتباط بین آن ها را نشان می دهند. مثلاً در شبکه اجتماعی، مردم را با رأس ها، و ارتباط های داده ای بین کامپیوترها و یا روابط دوستی بین مردم را با یال ها نمایش می دهیم.
شبکه ها دارای ویژگی های آماری مشترکی هستند. یکی از این ویژگی ها، ویژگی پدیده دنیای کوچک بوده که به 6 درجه جدایی نیز معروف است و بیان می کند که در یک شبکه، فاصله متوسط بین رأس ها، کوتاه و معمولا تابعی لگاریتمی از تعداد آن هاست. ویژگی بعدی، ویژگی خوشه‌بندی یا انتقال پذیری شبکه است و بیان می‌کند که دو رأس مجاور با رأس سوم، با احتمال زیادی با یکدیگر مجاور بوده و با ضریب خوشه بندی مقداردهی می شود. جدا از ویژگی های آماری مشترک میان شبکه های اجتماعی ، ویژگی دیگری را نیز می توان در نظر گرفت که به تازگی کانون توجهات را به خود جلب کرده و از آن به عنوان راه حلی برای بسیاری از مشکلات موجود در شبکه های اجتماعی استفاده می شود. این ویژگی فرآیند انتشار در شبکه های اجتماعی نام دارد. انتشار اطلاعات را می توان به عنوان یکی از نمونه های فرآیند انتشار نام برد. انتشار اطلاعات یک تعریف عمومی است که شامل هر چیزی که در یک شبکه گسترش می یابد می شود. بیشینه کردن گسترش اعتبار مانند یک ایده ی برتر و یا حتی تشخیص سریع یک فاجعه مانند دزدی. همه ی اینها نمونه هایی از انتشار اطلاعات هستند.
در نهایت از فرایند انتشاری که در شبکه های اجتماعی رخ می دهد، از جمله انتشار اطلاعات، جهت ارائه ی مدلی برای دسته ای از پدیده ها در این شبکه ها استفاده می شود. پدیده هایی مانند گسترش ویروس ها و بد افزارها در کامپیوتر ها، گسترش اطلاعات مربوط به کالاها در میان مردم و غیره. ارائه این مدل های آماری می تواند راهنمایی باشد جهت بررسی بیشتر ساختار شبکه های اجتماعی ، نحوه گسترش و انتشار اطلاعات در آنها و همچنین تشخیص تاثیرگذارترین رئوس در این شبکه ها ؛ مبحثی که به تازگی مورد توجه بسیاری از محققین قرار گرفته است.

## فهرست مطالب

عنوان
صفحه

**فصل اول: مقدمه**

**فصل دوم: پیشینه ی تحقیق**

**فصل سوم: ارائه راه حل و روش های پیشنهادی**

عنوان
صفحه

**فصل چهارم**

**عنوان**

**صفحه**

**فصل پنجم: بحث و نتیجه گیری**

# فهرست جدول ها

عنـــــــــــــــــــــــــــوان صفحه

# فهرست شکل‌ها

عنــــــــــــــــــــــــــــــــــــــــــــــــــــــــــــوان
صفحه

# فصل اول

تعامل انسان با کامپیوتر[1] از زمان ایجاد اولین کامپیوترها همواره مورد توجه بوده است و شامل مطالعه، برنامه ریزی و طراحی رابطه بین مردم (کاربران) و کامپیوترها است. معمولا از HCI به عنوان نقطه تقاطع علوم کامپیوتر، علوم رفتاری[2] ، علم طراحی[3] و چند زمینه ی دیگر یاد می شود. این اصطلاح برای اولین بار توسط کارد، موران و نیوول در کتاب "روانشناسی تعامل انسان با کامپیوتر" مطرح شده است و دلالت ضمنی بر این مطلب دارد که بر خلاف ابزارهایی که کاربرد محدود دارند، کامپیوتر دارای مزایا و کاربردهای بیشماری بوده که در یک دیالوگ بدون انتها بین آن و کاربر انجام می شوند [1].

متخصصان این حوزه در ابتدا به دنبال تولید سخت افزارهایی با ارگونومی مناسب برای راحتی انسان بودند. طی دهه ی 1980، این حوزه با تحولی پارادایمیک مواجه شد و تمرکز اصلی آن به جنبه های ادراکی رفتار کاربر و ایجاد نرم افزارهای کاربر پسند سوق یافت. اما طولی نکشید که در دهه 1990، موج جدیدی گفتمان غالب متخصصان این حوزه را متحول ساخت. در این گفتمان جدید، افزایش کیفیت ارتباط میان انسان ها با کامپیوترها هدف نبود، بلکه در این دیدگاه کامپیوتر به عنوان ابزاری برای ایجاد تعاملات انسانی نگاه می شد. با توجه به این رویکرد، شبکه های اجتماعی اینترنتی به عنوان عامل ایجاد تعامل میان انسان ها در فضای مجازی از اهمیت خاصی برخوردار گشت. یکی از این تاثیرات ظهور مفاهیم جدیدی چون وب اجتماعی 2 بود [2].

با گسترش وب اجتماعی، نیاز به تحلیل ساختارها و رفتارهای شبکه های اجتماعی، به عنوان یکی از نیازمندی های اساسی شرکت های تجاری مبدل گشت. تحلیل شبکه های اجتماعی در بسیاری از کاربردها از جمله مدیریت شبکه اجتماعی، تحلیل گرایش بازار، شناسایی افراد تاثیرگذار و حامیان، ارتقاء کارایی سامانه های توصیفگر و.... قابل استفاده است. نیازمندی های تجاری باعث شده است در سال های اخیر در بعد آکادمیک توجه زیادی به تحلیل شبکه های اجتماعی گردد. امروزه این ابزار قدرتمند نه تنها

[1] Human-Computer Interaction (HCI)

[2] Behavioral Science

[3] Design Science

مورد توجه متخصصان فناوری اطلاعات می باشد، بلکه پژوهشگران سایر رشته هایی چون علوم تربیتی، زیست شناسی، علوم ارتباطات، اقتصاد، ....، به عنوان یک تکنیک کلیدی از تحلیل شبکه اجتماعی بهره می برند. برای تحلیل شبکه ها، از معیارها و نرم افزارهای متفاوتی استفاده می شود. نرم‌افزارهای تجزیه و تحلیل شبکه اجتماعی جهت شناسایی، تجزیه و تحلیل، تجسم و شبیه سازی رأس ها و یال‌ها از انواع مختلف داده‌های ورودی (رابطه ای و غیر رابطه ای)، از جمله مدل‌های ریاضی شبکه‌های اجتماعی است. ابزار تجزیه و تحلیل شبکه به محققان اجازه می‌دهد تا شبکه‌هایی با اندازه‌های مختلف (شبکه‌های کوچک مانند خانواده و شبکه‌های بزرگ مانند اینترنت) را بررسی کنند این نرم‌افزارها با فراهم آوردن ابزارهای مختلف اجازه اعمال رویه‌های ریاضی و آماری را روی مدل شبکه می‌دهند. این نرم‌افزارها با نمایش‌های بصری شبکه‌های اجتماعی به درک و تحلیل نتایج کمک زیادی می‌کنند. معیارهای زیر در تحلیل شبکه های اجتماعی کاربرد وسیعی دارند:

Betweenness: تعداد افرادی در شبکه که یک شخص بطور غیر مستقیم از طریق خطوط مستقیم آنها متصل شده است.

Closeness: تنوع مجموعه کوتاهترین مسیرها بین هر فرد و دیگر افراد در شبکه.

Centrality degree: محاسبه میزان پیوندهایی که فرد با دیگر افراد در شبکه دارد.

Centralization: تفاوت بین تعداد پیوندها برای هر نود تقسیم شده توسط بیشترین مجموع تفاوتها. یعنی در یک شبکه همیشه رأس هایی وجود دارند که نسبت به دیگر رأس ها تعداد پیوندهای بیشتری دارند. در شبکه‌ای که دچار عدم تمرکز است تفاوت کمی بین پیوندهای هر نود وجود دارد.

Cohesion: اشاره به درجه‌ای دارد که افراد بطور مستقیم با همدیگر ارتباط دارند.

Path length: مسافت بین هر دو نود در یک شبکه را می‌گویند، میانگین Path length در واقع میانگین مسافتهای بین تمامی جفت رأس ها است.

Structural hole: تعداد کمی از افراد که اگر از گروه خارج شوند گروه از همدیگر جدا می‌شوند و اتصالات قطع می‌شود.

شبکه ی اجتماعی ساختاری اجتماعی است که از گره هایی(که عموما فردی یا سازمانی هستند) تشکیل شده است که توسط یک یا چند نوع خاص از وابستگی به هم متصل اند، برای مثال: قیمت‌ها، الهامات، ایده‌ها و تبادلات مالی، دوست‌ها، خویشاوندی، تجارت، لینک‌های وب، سرایت بیماری‌ها (اپیدمولوژی) یا مسیرهای هواپیمایی. ساختارهای حاصل اغلب بسیار پیچیده هستند. تحلیل شبکه‌های اجتماعی روابط اجتماعی را با اصطلاحات رأس و یال می‌نگرد. رأس‌ها بازیگران فردی درون شبکه‌ها هستند و یال‌ها روابط میان این بازیگران هستند. انواع زیادی از یال‌ها می‌توانند میان رأس‌ها وجود داشته باشد. تحقیق در تعدادی از زمینه‌های آکادمیک نشان داده است که شبکه‌های اجتماعی در بسیاری از سطوح به کار گرفته می‌شوند از خانواده‌ها گرفته تا ملت‌ها و نقش مهمی در تعیین راه حل مسائل، اداره کردن تشکیلات و میزان موفقیت افراد در رسیدن به اهدافشان ایفا می‌کند. در ساده‌ترین شکل یک شبکه ی اجتماعی نگاشتی از تمام یال‌های مربوط، میان رأس‌های مورد مطالعه است. شبکه ی اجتماعی هم چنین می‌تواند برای تشخیص موقعیت اجتماعی هر یک از بازیگران مورد استفاده قرار گیرد. این مفاهیم غالبا در یک نمودار شبکه ی اجتماعی نشان داده می‌شوند که درآن، نقطه‌ها رأس‌ها هستند و خطها نشانگر یال‌ها [3]. اما آنچه که در این تحلیل، پایه و اساس محسوب می شود، نظریه ی گراف ها است که در کنار کاربرد های بیشمار آن، در تحلیل شبکه های اجتماعی نیز نقشی مهم ایفا می کنند. در ادامه به شرح خلاصه ای از تاریخچه گراف و مواردی از کاربرد آن خواهیم پرداخت.
نظریه گراف شاخه‌ای از ریاضیات بوده که درباره ی گراف‌ها بحث می‌کند و در واقع شاخه‌ای از توپولوژی است که با جبر و نظریه ماتریس‌ها پیوند مستحکم و تنگاتنگی دارد. نظریه ی گراف برخلاف شاخه‌های دیگر

ریاضیات خاستگاه مشخصی دارد و آن، حل معمای پل های کونیگزبرگ[1] در سال 1736 و توسط اویلر ریاضیدان سوئیسی است[4]. از آن زمان تاکنون تحقیقات زیادی بر روی گراف ها انجام شده که این امر منجر به جمع آوری مطالب بسیاری در مورد آن ها شده است. از جمله ی این تحقیقات می توان به موارد زیر اشاره کرد.

در سال ۱۸۴۷، گوستاو کیرشهف نوع خاصی از گرافها به نام درخت را مورد بررسی قرار داد. کیرشهف این مفهوم را هنگام تعمیم قوانین اهم برای جریان الکتریکی در کاربردهایی که حاوی شبکه‌های الکتریکی بودند به کار گرفت. ده سال بعد، آرتور کیلی همین نوع گراف را برای شمارش ایزومرهای متمایز هیدروکربنهای اشباع شده ی $C_n H_{2n+2}$ به کار برد. در همین دوران شاهد حضور دو ایده ی مهم دیگر در صحنه هستیم. ایده ی اول حدس چهار رنگ بود که نخستین بار توسط فرانسیس گوثری در حدود سال ۱۸۵۰ مورد تحقیق قرار گرفت. این مسئله سرانجام در سال ۱۹۷۶، توسط کنث اپل و ولفگانگ هیکن و با استفاده از یک تحلیل رایانه‌ای پیچیده حل شد. ایده ی مهم دوم، دور همیلتونی بود. این دور به افتخار سر ویلیام روآن همیلتون نامگذاری شده است. او این ایده را در سال ۱۸۵۹ برای حل معمای جالبی حاوی یال‌های یک دوازده وجهی منتظم به کار گرفت. یافتن جوابی برای این معما چندان دشوار نیست ، ولی ریاضیدانان هنوز در پی یافتن شرایطی لازم و کافی هستند که گراف‌های بیسوی حاوی مسیر یا دورهای همیلتونی را مشخص کنند. پس از این کارها تا بعد از سال ۱۹۲۰ فعالیت اندکی در این زمینه صورت گرفت. مسئله ی مشخص کردن گراف‌های مسطح را کازیمیر کوراتوفسکی، ریاضیدان لهستانی، در سال ۱۹۳۰ حل کرد. نخستین کتاب درباره ی نظریه ی گراف در سال ۱۹۳۶ منتشر شد. این کتاب را ریاضیدان مجارستانی، دنش کونیگ، که خود محقق برجسته‌ای در این زمینه بود، نوشت. از آن پس فعالیت‌های بسیاری در این زمینه صورت گرفته و کامپیوتر نیز در چهار دهه ی اخیر به یاری این فعالیت‌ها آمده است [5].

پیشرفت‌های اخیر در ریاضیات، به ویژه در کاربردهای آن موجب گسترش چشمگیر نظریه ی گراف شده است به گونه‌ای که هم‌اکنون نظریه ی گراف ابزار بسیار

[1] Puzzle of Königsberg’s bridges

مناسبی برای تحقیق در زمینه‌های گوناگون مانند نظریه کدگذاری، تحقیق در عملیات، آمار، شبکه‌های الکتریکی، علوم رایانه، شیمی، زیست‌شناسی، علوم اجتماعی و سایر زمینه‌ها گردیده است.
از گرافها برای حل مسایل زیادی در ریاضیات و علوم کامپیوتر استفاده می‌شود. ساختارهای زیادی را می‌توان به کمک گرافها به نمایش در آورد. برای مثال برای نمایش چگونگی رابطه وب سایت‌ها به یکدیگر می‌توان از گراف جهت دار استفاده کرد. به این صورت که هر وب سایت را به یک رأس در گراف تبدیل می‌کنیم و در صورتیکه در این وب سایت لینکی به وب سایت دیگری بود، یک یال جهت دار از این رأس به رأسی که وب سایت دیگر را نمایش می‌دهد وصل می‌کنیم. از گرافها همچنین در شبکه‌ها، طراحی مدارهای الکتریکی، اصلاح هندسی خیابان‌ها برای حل مشکل ترافیک، و.... استفاده می‌شود. مهم‌ترین کاربرد گراف مدلسازی پدیده‌های گوناگون و بررسی بر روی آنهاست. با گراف می‌توان به راحتی یک نقشه بسیار بزرگ یا شبکه‌ای عظیم را در درون یک ماتریس به نام ماتریس وقوع گراف ذخیره کرد و یا الگوریتم های مناسب مانند الگوریتم دایکسترا یا الگوریتم کروسکال و... را بر روی آن اعمال نمود. در این جا به بررسی گرافهایی می‌پردازد که می‌توان آن‌ها را به نحوی روی صفحه کشید که یال‌ها جز در محل رأس‌ها یکدیگر را قطع نکنند. این نوع گراف در ساخت جاده‌ها و حل مساله کلاسیک و قدیمی سه خانه و سه چاه آب به کار می‌رود. کاربرد گراف بازه‌ها از گرافها برای حل مسایل زیادی در ریاضیات و علوم کامپیوتر استفاده می‌شود. ساختارهای زیادی را می‌توان به کمک گرافها به نمایش در آورد. درخت و ماتریس درخت در رشته‌های مختلفی مانند شیمی مهندسی برق و علم محاسبه کاربرد دارد. کیرشهف در سال ۱۸۴۷ میلادی هنگام حل دستگاههای معادلات خطی مربوط به شبکه‌های الکتریکی درختها را کشف و نظریه درختها را بارور کرد. کیلی در سال ۱۸۵۷ میلادی درختها را در ارتباط با شمارش ایزومرهای مختلف هیدروکربنها کشف کرد وقتی مثلا می‌گوییم در ایزومر مختلف $C_4H_{10}$ وجود دارد منظورمان این است که دو درخت متفاوت با ۱۴ رأس وجود دارند که درجه ۴ رأس از این ۱۴ رأس چهار و درجه هر یک از ۱۰ رأس باقیمانده یک است. اگر هزینه کشیدن مثلا راه آهن بین هر دو شهر ازp شهر مفروض مشخص باشد ارزان

ترین شبکه ای که این p شهر را به هم وصل می کند با مفهوم یک درخت از مرتبه p ارتباط نزدیک دارد. به جای مساله مربوط به راه آهن میتوان وضعیت مربوط به شبکه های برق رسانی و لوله کشی نفت و لوله کشی گاز و ایجاد کانالهای آب رسانی را در نظر گرفت. برای تعیین یک شبکه با نازلترین هزینه از قاعده ای به نام الگوریتم صرفه جویی استفاده می‌شود که کاربردهای فراوان دارد. از گراف ها می توان به عنوان کد های کمکی نام برد که به DVB Playerها در بالا بردن قابلیت های آنها کمک می کنند. گراف ها دارایی مزایای مختلفی هستند که شفاف تر کردن و واضحتر کردن تصویر و کاهش مصرف CPU به عنوان یکی از اصلی ترین مزایای آنها بشمار میرود [5].

یکی دیگر از کاربرد های مهم گراف ها در نمایش سیستم های پیچیده طبیعی و اجتماعی و در قالب یک شبکه[1] است؛ مجموعه ای از رأس ها و یال هایی که این رأس ها را به هم متصل می کنند[6]. به عنوان نمونه هایی از شبکه ها، می توان به شبکه های اجتماعی[2] [7]، شبکه های فنی[3](مانند اینترنت [8]) و شبکه های زیستی (مانند شبکه های عصبی [9]) اشاره کرد. رأس ها در این شبکه ها، موجودیت ها و یال ها، ارتباط بین آن ها را نشان می دهند. مثلاً در شبکه اینترنت، کامپیوترها یا مسیریاب ها و در شبکه اجتماعی، مردم را با رأس ها، و ارتباط های داده ای بین کامپیوترها و یا روابط دوستی بین مردم را با یال ها نمایش می دهیم.

---

[1] Network
[2] Social networks
[3] Technological networks

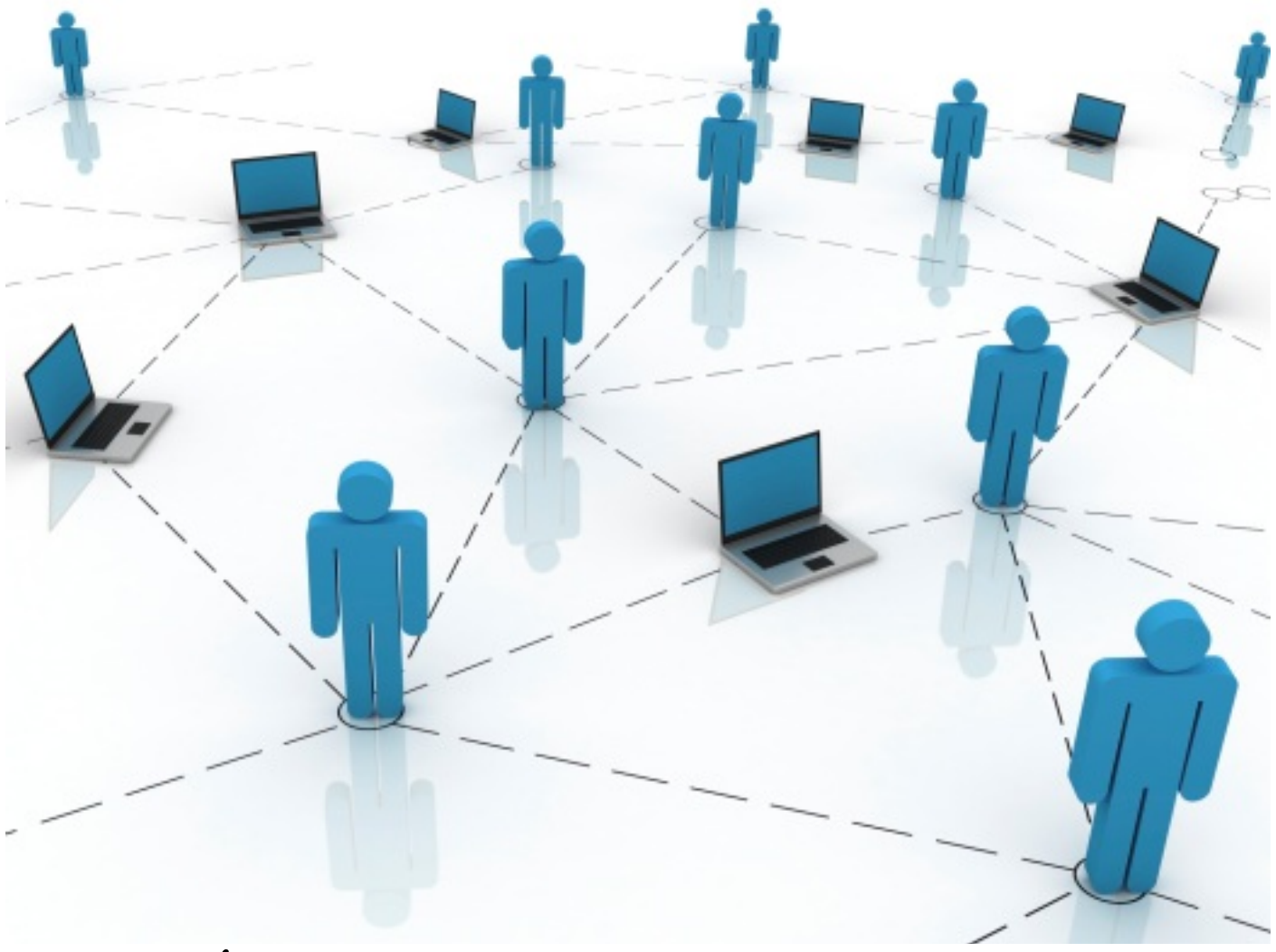

**شکل1-1- شبکه های اجتماعی مجموعه ای از رأس هایی هستند که با هم در ارتباط هستند.
این رأس ها می توانند افراد، رایانه ها و ... باشند**

شبکه ها دارای ویژگی های آماری مشترکی هستند. یکی از این ویژگی ها، ویژگی پدیده دنیای کوچک[1] [10] بوده که به 6 درجه جدایی[2] [11] نیز معروف است و بیان می کند که در یک شبکه، فاصله متوسط بین رأس ها، کوتاه و معمولا تابعی لگاریتمی از تعداد آن ها است. ویژگی دیگر، ویژگی توزیع های درجه اریب به رأس[3] [12] بوده و بیان می کند که درجه بیشتر رأس های یک شبکه، کم و تنها تعداد محدودی از آن ها درجه بالا دارند و توزیع درجات، غالبا به فرم نمایی یا قانون توانی[4] می باشد[6]. ویژگی بعدی، ویژگی خوشه بندی[5] یا انتقال پذیری شبکه[6]است و بیان می کند که دو رأس مجاور با رأس سوم، با احتمال زیادی با یکدیگر مجاور بوده و با ضریب خوشه بندی مقدار دهی می شود[13]. اما مهم ترین ویژگی شبکه که توجه بیشتری را به خود جلب کرده است، ویژگی ساختار تشکل[7]می باشد[14] [15] که به معنی وجود گروه های متراکم از رأس ها و ارتباطات تنک بین این گروه ها است (شکل 1). (به این ویژگی،

---

[1]Small world phenomenon
[2]Six degrees of separation
[3]Right-skewed degree distributions
[4]Power-law
[5]Clustering
[6]Network transitivity
[7]Community structure

خوشه بندی نیز گفته می شود، اما برای جلوگیری از تداخل با ویژگی قبل، از این نام استفاده نمی کنیم). این گروه ها، معمولاً خوشه[1]، تشکل [2]، گروه های به هم پیوسته[3] و یا مدول[4] نامیده می شوند. در ادامه، از عنوان "تشکل" برای ارجاع به چنین گروه هایی استفاده خواهیم کرد.

جدا از ویژگی های آماری مشترک میان شبکه های اجتماعی ، ویژگی دیگری را نیز می توان در نظر گرفت که به تازگی کانون توجهات را به خود جلب کرده و از آن به عنوان راه حلی برای بسیاری از مشکلات موجود در شبکه های اجتماعی استفاده می شود. این ویژگی فرآیند انتشار[5] در شبکه های اجتماعی نام دارد. انتشار اطلاعات[6] را می توان به عنوان یکی از نمونه های فرآیند انتشار نام برد. انتشار اطلاعات یک تعریف عمومی است که شامل هر چیزی که در یک شبکه گسترش می یابد می شود. بیشینه کردن گسترش اعتبار[7] مانند یک ایده ی برتر و یا حتی تشخیص سریع یک فاجعه مانند دزدی. همه ی اینها نمونه هایی از انتشار اطلاعات هستند.

در نهایت از فرایند انتشاری که در شبکه های اجتماعی رخ می دهد، از جمله انتشار اطلاعات، جهت ارائه ی مدلی برای دسته ای از پدیده ها در این شبکه ها استفاده می شود. پدیده هایی مانند گسترش ویروس ها و بد افزارها در کامپیوتر ها، گسترش اطلاعات مربوط به کالاها در میان مردم و غیره. ارائه این مدل های آماری می تواند راهنمایی باشد جهت بررسی بیشتر ساختار شبکه های اجتماعی ، نحوه گسترش و انتشار اطلاعات در آنها و همچنین تشخیص تأثیرگذارترین رئوس در این شبکه ها [17،19] ؛ مبحثی که به تازگی مورد توجه بسیاری از محققین قرار گرفته است.

---

1 Cluster

2 Community

3 Cohesive groups

4 Module

5 Diffusion Process

6 Information Diffusion

7 Maximizing Spread of Influence

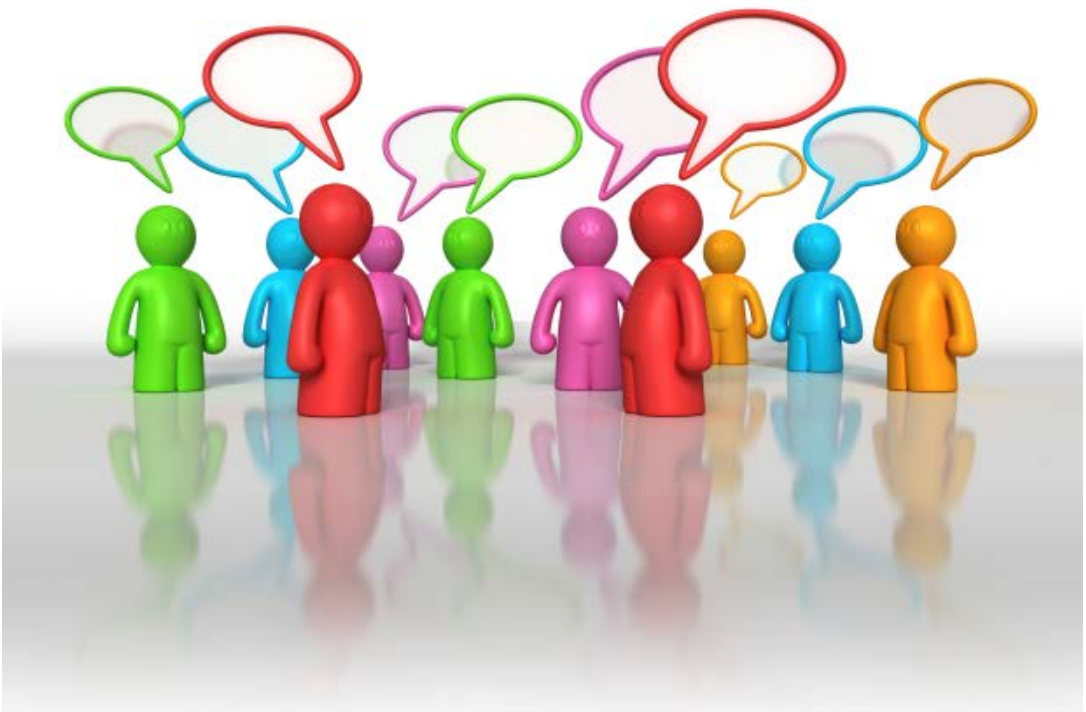

**شکل1-2- فرآیند گسترش در تمامی جنبه های شبکه های اجتماعی قابل مشاهده است.**
**افرادی که با یکدیگر در ارتباط هستند با تبادل اطلاعات باعث گسترش اطلاعات می شوند.**

## 1-1- انگیزه از انجام این پایان نامه

همانطور که پیشتر نیز بیان شد، تحلیل شبکه های اجتماعی به عنوان یک تکنیک کلیدی در جامعه شناسی، انسان شناسی، جغرافیا، روانشناسی اجتماعی، جامعه شناسی زبان، علوم ارتباطات، علوم اطلاعات، مطالعات سازمانی، اقتصاد و زیست شناسی مدرن همانند یک موضوع محبوب در زمینه ی تفکر و مطالعه پدیدار شده است.
امروزه یکی از مسائل اساس که در شبکه های اجتماعی مطرح می شود بحث در مورد انتشار اطلاعات در این شبکه ها است. اطلاعات می تواند باعث تشکیل یک ایده ی عمومی، ایجاد ترس و اضطراب در یک جامعه، پذیرش یک محصول توسط خریداران و غیره شود و به این طریق نقش بسیار اساسی را در سازمان های اجتماعی ایفا می کند. اطلاعات از طریق افراد مختلف در سطح شبکه منتشر می شود. افراد می توانند با استفاده از شیوه های گوناگون (شیوه های مرسومی نوشتاری، زبانی و الکترونیکی) باعث انتشار اطلاعات در شبکه ها شوند. ساختار اطلاعاتی با توجه به گسترش روز افزون استفاده از اینترنت و وب بسیار تغییر کرده و از شکل سنتی خود خارج شده است. تا چند سال پیش اگر فردی نیازمند دریافت اطلاعات خاصی می بود، می بایست هزینه ی ایجاد زیربنای ارتباطی با افراد مختلف را نیز می پرداخت. امروزه با توجه به دسترسی گستره و آسان به اینترنت این محدودیت ها از بین رفته است.
طی چند دهه ی اخیر، دانشمندان تنها به مشاهده و انتشار اطلاعات در شبکه های اجتماعی دقت نمی کنند

و هدف بزرگتری را دنبال می نمایند. امروزه هدف اصلی ایجاد، گسترش و بیشینه کردن اطلاعات در یک شبکه است. انتشار اطلاعات یک اصل کلی است و هر چیزی که بتواند در یک شبکه منتشر شود زیر چتر انتشار اطلاعات قرار می گیرد. اهداف بررسی انتشار اطلاعات را می توان به دو دسته ی اصلی تقسیم کرد:

1. بیشینه کردن گسترش تاثیر[1]
   - مثال: ایده ها[2]، نظرات[3]، پیشنهادات[4] و غیره.
2. تشخیص سریع حادثه[5]
   - بیماری ها، نفوذ ها و غیره.

برای مثال، دومینگوس و ریچاردسون [18] و کِمپه و همکاران [19] به طور مشخص مسئله ی بیشینه کردن گسترش تاثیر را مورد بررسی و مطالعه قرار داده اند. در این مسئله هدف اصلی یافتن یک مجموعه حاصل از k عضو است که ما با انتخاب مناسب این مجموعه و فعال کردن آن ها می خواهیم به بیشترین تعداد اعضای فعال برسیم. بالفرض اگر تعداد k تا کالای رایگان در اختیار داشته باشیم می خواهیم این کالا ها را بین افرادی تقسیم کنیم که بدانیم بیشترین تاثیر را بر روی سایرین خواهند داشت و آن ها را به خریدن کالای مورد نظر مشتاق خواهند کرد. جهت تحلیل و بررسی بیشتر این نوع مسائل و مشکلات، ما نیاز به درک عمیق تر از ساختارهای درگیر داریم که همین امر باعث می شود که بحث مدل سازی و پیش بینی انتشار اطلاعات بیشتر مورد توجه قرار گیرند. به طور دقیق تر می توان گفت که ما برای حل مشکلات در این زمینه نیازمند مدل های قوی جهت بررسی انتشار اطلاعات داریم.

مدل های انتشار اطلاعاتی که تاکنون مطرح شده اند پایه و اساسی مبتنی بر کارهای انجام شده در زمینه های گوناگونی از جمله جامعه شناسی، فیزیک، بیماری شناسی و بازاریابی دارند. تمامی این مدل ها فرض را بر این قرار می دهند که رأس های یک شبکه دو حالت دارند: فعال[6] و غیرفعال[1]. رأس های فعال

[1] Maximizing the Spread of Influence
[2] Idea
[3] Review
[4] Comment
[5] Early Detection of Outbreak
[6] Active

اطلاعات را پخش کرده و به سایر رأس ها که در حالت غیر فعال هستند انتقال می دهند. مدل های گوناگون در این زمینه فرضیات متفاوتی را در مورد نحوه ی انتقال اطلاعات میان رأس ها در شبکه ها را در نظر می گیرند و همین امر است که موجب تمایز بین آن ها می شود. برای مثال دو مدل آبشاری مستقل[2] و آستانه ی خطی[3] در این زمینه بسیار پر کاربرد هستند.
به تازگی نیز مدل های گوناگونی طی سال های اخیر در این زمینه ارائه شده است. برای مثال لاهیری و سربیان [20] مدلی را ارئه کردند که بر اساس الگوریتم های ژنتیکی[4] کار می کند. این مدل با ترکیب شدن با توابع چند بعدی هلند[5] ساختاری را ارئه می دهد که می توان انتقال اطلاعات میان رأس ها را با استفاده از عملگر متقاطع[6] نمایش داد. در این مدل برای هر رأس به جای یک بیت تنها که یک وضعیت ساده را نشان می دهد از یک بردار وضعیت[7] استفاده می کنیم. ارائه دهندگان این روش معتقدند که می توانند در این مدل بر خلاف سایر مدل های قبلی واحد های چندگانه ی اطلاعات[8] را نمایش دهند. هرچند که این ادعا به راحتی قابل نقض خواهد بود و این مدل نیز مانند سایر مدل های پیشین توانایی چندانی در نمایش اطلاعات چند گانه را نخواهد داشت. از بین رفتن خاصیت مدل سازی واحد های چندگانه ی اطلاعات در این روش به دلیل استفاده از عملگر متقاطع است که باعث از بین رفتن اطلاعات می شود.
پس مشکلاتی که روش های مطرح شده تاکنون داشته اند به طور خلاصه شامل موارد زیر می شود:

1. نیاز به بهینه سازی تعدادی پارامتر در روش های اولیه
2. عدم توانایی در نمایش مسائل پیچیده ی دنیای امروز
3. عدم توانایی در نمایش واحد های چندگانه ی اطلاعاتی

---

[1] Inactive
[2] Independent Cascade
[3] Linear Threshold
[4] Genetic Algorithms
[5] Holland Hyper-defined Functions
[6] Crossover
[7] State Vector
[8] Multiple Units of Information

4. عدم توانایی در تطابق با شبکه های بزرگ و پیچیده ی کنونی

اهمیت ارائه ی مدلی برای انتشار اطلاعات در شبکه های اجتماعی با توجه به تمامی مطالب ذکر شده در بالا و موارد و مشکلات موجود در اکثر مدل های متداول برای انتشار اطلاعات، ما را بر آن داشت تا در این پایان نامه، روشی بر مبنای الگوریتم بهینه سازی ازدحام ذرات[1] جهت مدل سازی انتشار اطلاعات در شبکه های اجتماعی ارائه دهیم به طوریکه بسیاری از مشکلات ذکر شده را برطرف سازد. بررسی ها نشان می دهند که روش پیشنهادی، روشی کارا بوده و نتایج بهتری در مقایسه با سایر روش های موجود، ارائه می دهد.

## 1-2- نگاه کلی به فصول رساله

این رساله مشتمل بر پنج فصل می باشد که در ادامه ی مقدمه، در فصل دوم به بررسی و تحلیل روش های ارائه شده در این حوزه خواهیم پرداخت. در فصل سوم، روش های پیشنهادی با تمامی جزئیات ارائه خواهد شد و در فصل چهارم نتایج حاصل از اجرای آن ها بر روی مجموعه داده[2] های معرفی شده به همراه تحلیل نتایج آورده شده است. در انتها، در فصل پنجم به شرح نتیجه گیری و کارهای آینده پرداخته ایم.

[1] Particle Swarm Optimization

[2] Dataset

# فصل دوم

# پیشینه ی تحقیق

## 2-1- مقدمه

همانطور که پیش از این نیز مطرح کردیم، بررسی فرآیند های انتشار در شبکه های اجتماعی یکی از مهمترین مسائل تحقیق در حوزه ی شبکه های اجتماعی است. از دیرباز دانشمندان در رشته های جامعه شناسی، فیزیک، علوم کامپیوتری و غیره مسائل مربوط به فرآیند های انتشار را مورد بررسی قرار داده اند. بررسی فرآیند انتشار بیماری های واگیردار، بررسی انتشار ایده ها در جوامع، بررسی گسترش رضایتمندی مشتریان از یک محصول خاص از جمله ی این مطالعات به حساب می آیند.
با گسترش استفاده از کامپیوتر ها، شبکه های اینترنتی و مخابراتی همچنین فراهم شدن دسترسی سریع و آسان به این امکانات برای تعدادی زیادی از افراد، توجه محققین حوزه ی انتشار فرآیند ها نیز به این زمینه ها معطوف شد. شبکه های کامپیوتری، مخابراتی و اجتماعی بستر مناسبی برای انتشار فرآیند های گوناگون ایجاد کرده اند. در شبکه های کامپیوتری ویروس های کامپیوتری گسترش پیدا می کنند و تعدادی کامپیوتر آلوده سعی در آلوده کردن سایرین دارند. در شبکه های ارتباطی و مخابراتی می توان با بررسی روابط و تماس های افراد گوناگون ساختار شبکه های تروریستی را تشخیص دهیم. در نهایت با تجزیه و تحلیل انتشار اطلاعات در شبکه های اجتماعی نیز می توان به نتایج سودمند گوناگونی دست یافت. امروزه این تحلیل ها به ما کمک می کند تا بتوانیم محصولات خود را بهتر به فروش برسانیم، ایده های جدید را به سرعت گسترش دهیم، افراد را از خطرات گوناگون با خبر نماییم و غیره.
جهت بهره گیری از این امکانات، می بایست مدل های دقیقی را برای تشخیص نحوه ی انتشار اطلاعات در این شبکه ها ارائه نماییم. در طی سال های اخیر افراد گوناگون مدل های متفاوتی را برای تحلیل هر چه

بهتر این فرآیند معرفی کرده اند. هر یک از این مدل ها نیز پایه و اساس تحقیق های گسترده ای را بنیان نموده اند. به طور مسلم هیچ یک از این مدل ها عاری از عیب نبوده اند و همین امر موجب پیشرفت سریع در این حوزه شده است.

در این فصل، ابتدا توصیفی دقیق از مدل های انتشار اطلاعات در شبکه های اجتماعی ارائه می گردد. سپس به مطالعه ی روش های مرسوم برای حل این مسئله می پردازیم.

## 2-2- تعریف مسئله

الگوریتم های ارائه شده در زمینه ی مدل سازی انتشار اطلاعات در شبکه های اجتماعی علی رغم تمام تفاوت هایی که در هدفشان از مدل سازی دارند، در نهایت به صورت احتمالی سعی در بررسی گسترش پدیده هایی مانند انتشار ویروس ها، ایده ها، بیماری ها و غیره دارند. این مدل ها نحوه ی شروع یک پدیده، گسترش آن و مقبولیت[1] آن را مدل سازی می نمایند. به طور کلی مسئله ی مدل سازی انتشار اطلاعات در شبکه های اجتماعی شامل تعاریف زیر می شود:

یک شبکه ی اجتماعی ساختاری مبتنی بر گراف[2] است که ارتباطات و اتصالات داخلی میان افراد خاص و موجودیت ها را نمایش می دهد. رأس های این گراف نشان دهنده ی اشخاص، حیوانات یا کامپیوتر های شبکه شده هستند و گراف یال های گراف اتصال میان دو رأس است که این دو رأس به طریقی با یکدیگر در ارتباط می باشند.

**تعریف 1-** شبکه ی اجتماعی یک گراف جهت دار[3] یا غیر جهت دار[4] $G=(V,E)$ است که در آن رأس برچسب دار $v\in V$ نشان دهنده ی یک موجودیت در یک سیستم فیزیکی است و یال $(u,v)\in E$ نشان دهنده ی یک ارتباط میان دو موجودیت است. یک شبکه ی اجتماعی پویا[5] گراف چند گانه مانند $G=(V,E)$ است که در آن $E$ نشان دهنده ی بسته از یال ها است و هر یال زمان دار $(u,v)_t\in E$ نشان دهنده ی یک ارتباط مانند $(u,v)$ است که در

---

[1] Adoption
[2] Graph
[3] Directed
[4] Undirected
[5] Dynamic Social Network

زمان $t \in \mathbb{Z}^+$ رخ داده است.
با در نظر گرفتن فرض های یک مدل انتشار ساده، مجموعه ای از رأس های یک شبکه را به صورت فعال در نظر می گیریم. با گذشت زمان در هر لحظه ی زمانی $t$ هر رأس که فعال باشد سعی می کند که همسایگان غیر فعال خود را نیز فعال نماید. پس از گذشت زمان و برقراری شرط مدلی که از آن استفاده کرده ایم الگوریتم خاتمه می یابد و طی آن تعدادی از رأس های غیر فعال به مجموعه ی رأس های فعال افزوده شده اند.
**تعریف 2-** یک مدل انتشار یک ساختار گراف $G = (V, E)$، یک بردار وضعیت $S_v^{(t)}$ برای هر رأس $v \in V$ در زمان $t$ و یک بردار از پارامتر های دلخواه $P$ را به عنوان ورودی دریافت می نماید. بر اساس بردار وضعیت تمامی رأس هایی که با یکدیگر در ارتباط هستند، این مدل یک بردار وضعیت جدید $S_v^{(t+1)}$ را برای هر رأس در زمان $t+1$ ایجاد می نماید. برای یک شبکه ی اجتماعی ایستا[1] (غیر پویا)، ساختار گراف در هر زمان ثابت خواهد بود. برای یک شبکه ی اجتماعی پویا، ساختار گراف به اینصورت تعریف می شود که $G_t = (V, E)_t$ به طوری که $E_t = \{(u, v): (u, v)_t \in E\}$ مجموعه ی تمامی یال ها در زمان $t$ باشد.
پس به طور کلی در یک مدل فرآیند انتشار که انتشار اطلاعات حالت پیچیده و خاصی از آن می شود ما با بردار های وضعیت برای هر رأس و به روز رسانی آن ها سر و کار داریم. به روز رسانی این بردار ها نشان دهنده ی نحوه ی گسترش و مقبولیت اطلاعات در بین رأس های یک شبکه خواهد بود.

## 2-3- روش های اخیر جهت مدل سازی انتشار اطلاعات

همانطور که در قسمت قبل نیز گفته شد، مسئله فرآیند انتشار و به طور خاص انتشار اطلاعات تنها مختص به شبکه های اجتماعی نمی باشد. برای مثال مسئله مدل کردن گسترش بیماری های واگیردار از طریق ارتباط شخص به شخص یکی از مسائل مهم در زمینه همه‌گیر شناسی ریاضیاتی[2] به حساب می آید

---
[1] Static Social Network
[2] Mathematical Epidemiology

[24]. از آنجا که ما اطلاعات شبکه های اجتماعی را در اختیار داریم ایده‌ی کلی گسترش بیماری های واگیر دار را می توان برای مدل کردن حالت های کلی تری از فرآیند انتشار استفاده نمود؛ به طور مثال برای مدل کردن انتشار نوآوری ها[1] [19،16] و همچنین مدل کردن نحوه گسترش کرم های الکترونیکی در سرتاسر شبکه های کامپیوتری [25] . هرچند هدف نهایی از مدل کردن هر یک از این فرآیند های انتشار متفاوت است، اما در نهایت تمامی این ها متکی به یک مدل انتشار هستند که این مدل به صورت کاملا آماری مشخص می کند که هر یک از این فرآیندها به چه صورت آغاز شده و چگونه گسترش می یابند.

مطالعه فرآیند انتشار اطلاعات و همچنین ارائه مدل برای آن در شبکه های اجتماعی تاریخچه ای بسیار طولانی داشته و تاکنون مدل های انتشار متفاوتی برای مدل کردن فرآیند های انتشار اطلاعات ارائه شده اند. این مدل ها در قالب سه دسته ی اصلی با عناوین زیر ارئه شده اند:

1. روش های سنتی (انتشار و فراگیری اطلاعات[2] و نوآوری)
2. روش های مدرن (روش های مبتنی بر بازی و الگوریتم های ژنتیک)

روش های زیر مجموعه ی هر یک از این گروه ها را به صورت مختصر شرح می دهیم.

### 2-3-1- روش های سنتی (انتشار و فراگیری اطلاعات و نوآوری)

غالب روش هایی که در زمینه تحقیق در مورد انتشار اطلاعات مورد بررسی قرار گرفته اند بر اساس نظریه ی انتشار بیماری ها و اطلاعات در شبکه ها استوار هستند. این فرض باعث شده است که تحقیقات زیادی در زمینه ی دانش همه گیرشناسی صورت پذیرد تا بتوان این غالب را برای اطلاعات نیز به کار ببریم. برای مثال می توان به کتاب بِیلی [26-50] که به صورت کاملا جامع به این مسئله پرداخته است رجوع نماییم. مدل های همه گیرشناسی کلاسیک گسترش بیماری بر اساس چرخه ی بیماری ها در یک گروه عمل می کنند: یک فرد

[1] Diffusion of Innovations

[2] Information Propagation and Epidemics

در ابتدا در معرض[1] بیماری است. این فرد پس از تماس با یک فرد آلوده به بیماری دچار می شود و با احتمال خاصی آلوده[2] می شود. در نهایت نیز ممکن بیماری از یک شخص که بازیابی[3] شده است خارج شود. در صورتی که آن شخص خاص پس از بازیابی (یا رفع کردن) دیگر در معرض خطر قرار نگیرد ما مدل را SIR می نامیم. برای مثال می توان به بیماری هایی مانند فلج اطفال اشاره نمود. حالت دیگری که ممکن است رخ بدهد این است که فرد بیمار پس از بازیابی و درمان تنها مدتی مشخص ایمن باشد و پس از آن مجددا در معرض آلودگی قرار گیرد و احتمال بیماری وجود داشته باشد که در آن صورت مدل را SIRS می نامیم. برای مثال می توان به بیماری آنفولانزا در این زمینه اشاره کرد که بیمار پس از معالجه همچنان در معرض خطر خواهد بود.

در ابتدا برای بررسی این مدل ها، هنگامی که یک عضو آلوده را انتخاب کرده و می خواستند تاثیر آن را بررسی نمایند می بایست به صورت تصادفی آن عضو را با یک عضو دیگر ارتباط می دادند. امروزه استفاده از شبکه های اجتماعی این ارتباطات را مشخص کرده است. در مدل های دنیای کوچک[4] ارائه شده توسط واتز و استروگتز [27] و مور و نیومن [28] ما قادر هستیم که تاثیر یک عضو آلوده را بررسی نماییم و احتمال انتقال آلوده از آن را بدست بیاوریم.

انتشار یک قسمت از اطلاعات در شبکه های اجتماعی همچنین می تواند به عنوان انتشار ایده ها و نوآوری ها نیز در نظر گرفته شود. برای مثال آدرس یک وب سایت که دربردارنده ی اطلاعات مفیدی برای ما می باشد می تواند یک قسمت از اطلاعات مفید به حساب بیاید. علاوه بر مباحثی که در زمینه ی گسترش بیماری ها مطرح شد، در زمینه ی جامعه شناسی شبکه های اجتماعی نیز تحقیقات گوناگونی در مورد انتشار اطلاعات و نوآوری ها صورت گرفته است. به نظریه ای که در این زمینه مطرح می شود word of mouth گفته می شود.

تمامی بررسی های انجام شده در این زمینه به دو

---

[1] Susceptible

[2] Infected

[3] Recovered

[4] Small-World

مدل اصلی منتهی می شود. حالت ساده ی این مدل ها را به صورت کامل در ادامه توضیح خواهیم داد؛ هرچند که بسته به کاربرد های گوناگون انواع مختلفی برای این مدل ها ارائه شده است. این دو مدل عبارتند از:

- مدل آستانه ی خطی[1] [19]: هنگامی که تاثیر همسایه های $v$ به میزان لازم رسید این رأس از آن ها پیروی می نماید.
- مدل آبشاری مستقل[2] [29]: هر رأس مانند $v$ تنها یکبار شانس فعال کردن همسایه ی خود مانند $w$ را دارد.

## 2-3-1-1- مدل آستانه‌ی خطی

ابتدا مدل آبشاری مستقل را تعریف می کنیم. در این مدل ، ما به ازای هر یال جهت دار مانند $(u,v)$ یک عدد حقیقی (وزن) مانند $p_{u,v}$ تعریف می کنیم به طوری که $0 < p_{u,v} < 1$ و در اینجا $p_{u,v}$ اشاره دارد به احتمال گسترش[3] از طریق یال $(u,v)$. فرآیند گسترش با توجه به مجموعه فعال اولیه[4] با نام S به طریق زیر بدست خواهد آمد. هنگامی که رأس u در ابتدا و در زمان t به صورت فعال در می آید یک شانس برای فعال کردن رأس v دارد به طوری که بین این دو یک یال قرار داشته باشد و همچنین احتمال فعال شدن رأس $v$ ، $p_{u,v}$ خواهد بود.

در صورتی که رأس $u$ موفق شود رأس $v$ به عنوان یک رأس فعال در زمان $t+1$ در خواهد آمد. در صورتی که رأس $v$ بیش از یک والد فعال داشته باشد، تمامی این رأس ها تنها یک بار و آن هم به طور هم زمان در زمان $t$ تلاش خواهند کرد که رأس $v$ را به صورت فعال در آورند و در صورت عدم موفقیت امکان دوباره ای برای فعال سازی رأس $v$ را نخواهند داشت. این فرایند زمانی متوقف خواهد شد که دیگر هیچ فعال سازی در شبکه امکان پذیر نباشد.

[1] Linear Threshold Model
[2] Independent Cascade Model
[3] Propagation Probability
[4] Initial Active Set

## 2-3-1-2- مدل آبشاری مستقل

در اینجا تعریفی از مدل آستانه ی خطی ارائه می دهیم. در این مدل ، برای هر رأس $v \in V$ یک وزن مانند $b_{u,v}$ از والد آن که $u$ است مشخص می کنیم به طوری که :

$$\sum_{w\ active\ neighbor\ of\ v} b_{v,w} \leq 1 \qquad -2)$$

و

$$\Gamma(v) = \{u \in V;\ (u,v)\ \in E\} \qquad -2)$$

فرآیند انتشار با توجه به قوانین تصادفی زیر و همچنین مجموعه فعال اولیه S پی گیری خواهد شد. در ابتدا ، به ازای هر رأس $v \in V$ یک آستانه ی $\theta_v$ به صورت تصادفی از بازه ی [0,1] انتخاب می شود. در زمان $t$ رأس $v$ تحت تاثیر والدین خود با وزن $b_{u,v}$ قرار می گیرد (u) و اگر مجموع وزن هایی که رأس $v$ از والدین فعال خود دریافت می کند از آستانه ی $\theta_v$ بیشتر باشد :

$$\sum_{w\ active\ neighbor\ of\ v} b_{v,w} \leq \theta_v \qquad -2)$$

در آن صورت رأس $v$ به عنوان یک رأس فعال در زمان $t+1$ در نظر گرفته خواهد شد. در اینجا $\Gamma_t^{(v)}$ به معنای تمامی والدین $v$ است که در زمان $t$ فعال بوده اند. این فرایند در صورتی که دیگر فعال سازی رخ ندهد خاتمه خواهد یافت.
دو مدل مطرح شده در این قسمت پایه و اساس بسیاری از مدل های مشابه شده اند. افراد زیادی حالات گوناگونی از این دو مدل به صورت بهبود یافته ارائه داده اند [30،31،32،33]. هدف از ارائه ی این مدل ها ارتقاء دادن دو مدل ساده ی اولیه و آماده سازی آن ها برای حالات کلی تر است. همچنین تعدادی

فعالیت نیز در زمینه ی بهبود فرضیات اولیه ی این مدل ها صورت گرفته است.
همانطور که مشخص است، هر دوی این مدل ها به یک مجموعه ی اولیه مانند $A_0$ وابسته هستند که این مجموعه نمایانگر مجموعه رأس های فعال اولیه است.
همچنین دو معیار $p_{u,v}$ در الگوریتم اول و $b_{u,v}$ در الگوریتم دوم نیز بسیار موثر هستند. الگوریتم های زیادی نیز در راستای بهبود تشخیص این مقادیر ارائه گردیده است. اما نکته ی مهم در مورد این دو الگوریتم کاربرد آن ها در حل مسائل گوناگون است.
برای مثال به تازگی تحقیقات گوناگونی در زمینه ی انتشار دو فرآیند word of mouth و viral marketing و تاثیر آن ها در موفقیت فروش یک محصول جدید صورت گرفته است [18،34، 35،36].

## 2-3-2- روش های مدرن (روش های مبتنی بر بازی و الگوریتم های ژنتیک)

مدل های ارائه شده در قسمت قبل از پرکاربرد ترین مدل ها در زمینه ی انتشار اطلاعات به حساب می آیند. در ادامه ی روند ارائه ی مدل برای این فرآیند ها، مدل های مدرن تری که با الگوریتم های خاص مطابقت دارند ارائه گردیدند. در ادامه دو مدل مهم در این زمینه را مورد بررسی قرار می دهیم. این دو مدل عبارتند از:

1. مدل بر اساس نظریه ی بازی
2. مدل بر اساس الگوریتم های ژنتیک

دلیل دسته بندی این دو الگوریتم در گروه الگوریتم های مدرن، استفاده از روش های نوین تری نسبت به در نظر گرفتن فاکتورهای ساده ای مانند احتمالات است.

### 2-3-2-1- مدل بر اساس نظریه ی بازی

مدل ارائه شده توسط بالا و گویال [37] از جمله مدل هایی است که برای انتشار اطلاعات در شبکه های اجتماعی ارائه شده است. همانطور که پیشتر نیز اشاره کردیم مدل های گوناگونی در رشته های مختلف برای انتشار فرآیند ها ارائه شده اند و در اینجا نیز این مدل برای حل مسائل موجود در رشته ی اقتصاد معرفی شده است.

مدل ارائه شده در واقع بیشتر روی رشد شبکه ی اجتماعی مورد بررسی بر اساس میزان اطلاعات تمرکز دارد به اینصورت که رأس $u$ در صورتی که بتواند اطلاعات مفیدی از $v$ دریافت نماید با آن یک رابطه[1] برقرار می کند. در اینجا برقراری ارتباط برای رأس $u$ یک میزان سودمندی دارد که در واقع همان اطلاعاتی است که در صورت برقراری ارتباط با $v$ به دست می آورد؛ از طرفی این برقراری ارتباط برای رأس $v$ هزینه نیز خواهد داشت. با بررسی و مقایسه ی این دو مقدار یک رأس تصمیم می گیرد که آیا ارتباطی برقرار نماید یا خیر.
فرض کنید که $N = \{1, \ldots, n\}$ یک مجموعه از عامل ها باشد که دو عامل $i$ و $j$ دو عضو از این مجموعه هستند. هر رأس به اینصورت در نظر گرفته می شود که اطلاعاتی را در اختیار دارد که احتمالا برای سایر عامل ها مورد نیاز است. این رأس می تواند میزان اطلاعات خود را با استفاده از برقراری ارتباط با سایرین افزایش دهد. این برقراری ارتباط هزینه هایی مانند زمان، سعی و تلاش و غیره خواهد داشت. در این مدل عامل ها به صورت همزمان ارتباط با یکدیگر برقرار می نمایند و گاهی ممکن است به دلیل غیر قابل اعتماد بود این ارتباط ها، اطلاعاتی رد و بدل نشود. بازی که در اینجا تعریف می شود در مورد ارتباط های میان عامل ها است و نه اطلاعاتی که بین آن ها تبادل می شود. در این مدل استراتژی برای هر عامل $i \in N$ یک بردار به صورت:

$$g_i = (g_{i,1}, \ldots, g_{i,i-1}, g_{i,i+1}, \ldots, g_{i,n}) \qquad -2)$$

است که :

$$g_{i,j} \in \{0,1\} \qquad -2)$$

به ازای هر:

$$j \in N \backslash \{i\} \qquad -2)$$

## 2-3-2-2- مدل بر اساس الگوریتم های ژنتیک

[1] Link

آخرین مدلی که در این بخش بررسی می نماییم، مدلی است که به تازگی توسط لاهیری و سربیان [20] به عنوان یک مدل پیشرفته برای انتشار اطلاعات در نظر گرفته شده است. تفاوت های اصلی این مدل با سایر مدل هایی که تاکنون از آن ها نام برده ایم در توانایی استفاده از یک بردار وضعیت به جای یک بیت وضعیت است. همچنین تمرکز اصلی این مدل روی انتقال اطلاعات است برخلاف مدل قسمت قبل که تنها روی ارتباطات میان رئوس و گسترش یک شبکه تاکید دارد.

این مدل GADM نام دارد. اساس کار در روش GADM استفاده از الگوریتم های ژنتیک است. در این مدل رئوس از بردارهای وضعیت جهت نمایش میزان اطلاعات خود استفاده می نمایند. این بردار ها رشته های بیتی هستند و در واقع نقش کروموزوم ها در جمعیت الگوریتم ژنتیک را به عهده دارند. عملیات انتقال اطلاعات بین رئوس در یک شبکه ی اجتماعی نیز با استفاده از عملگر ترکیب صورت خواهد پذیرفت.

با توجه به اینکه یکی از روش های پیشنهادی حالت ارتقاء یافته ی این مدل است، توضیحات بیشتر در مورد الگوریتم GADM را در فصل سوم ارائه خواهیم داد.

# فصل سوم

# ارائه راه حل و روش های پیشنهادی

## 3-1- مقدمه

روشی که در اینجا ارائه خواهیم داد در ابتدا سعی در برطرف کردن مشکل اساسی روش های قبل را دارد. همانطور که اشاره شد روش های پیشین امکان مدل سازی واحد های چند گانه ی اطلاعات را ندارند و تنها می توانند وجود و یا عدم وجود اطلاعات خاص در یک رأس را نمایش دهند. ما در این پایان نامه دو نکته ی مهم را در نظر گرفته ایم. اول آنکه امروزه در تمامی شبکه هایی که ما با آن ها سر و کار داریم تنها یک واحد اطلاعاتی مفهوم چندانی ندارد و پیچیدگی زیادی در این شبکه ها وجود دارد. تمامی رأس های یک شبکه حاوی اطلاعات و دانش گوناگونی از موضوعات مختلف هستند. برای مثال افرادی را در یک شبکه ی اجتماعی در نظر بگیرید که با هم در تعامل هستند. هر فرد تلاش خواهد کرد که دانش خود را در زمینه های گوناگون علمی، فرهنگی و غیره افزایش دهد. دومین مورد که می بایست مد نظر قرار گیرد این است که دانش و سطح اطلاعات افراد در هر یک از زمینه های اشاره شده الزاما کامل نیست و هر فرد می تواند تمام اطلاعات و یا بخشی از آن را در اختیار داشته باشد. حال آنکه در تمامی روش های قبلی روی این موضوع تاکید شده است که یک رأس در یک شبکه می تواند شامل اطلاعات خاصی باشد یا خیر و این مورد در این مدل ها در نظر گرفته نشده است.

بر مبنای آنچه به آن اشاره کرده ایم در این قسمت دو روش ارائه خواهیم داد. در روش اول که بر اساس مدل ارائه شده توسط لاهیری و سربیان [20] می باشد، سعی در بهینه سازی مدل آن ها خواهیم کرد. در این روش ما با افزودن عملگر جهش[1] از مجموعه عملگر های الگوریتم های ژنتیک، مدل GADM را بهینه خواهیم کرد. در روش دوم نیز با استفاده از مفاهیم پایه ی مشترک مدلی را بر اساس الگوریتم های بهینه سازی ازدحام ذرات ارائه می دهیم که در این مدل امکان در نظر گرفتن واحد های چند گانه ی اطلاعاتی برای ما فراهم خواهد بود. در نهایت روشی نوین بر اساس تشخیص تشکل ها در شبکه های اجتماعی ارائه خواهیم

---

[1] Mutation Operator

داد که این روش معیار مقایسه ی مدل های ارائه شده خواهد بود.
به طور خلاصه موارد انجام شده در این پایان نامه را می توان به شکل زیر بیان نمود:

- گسترش مدل GADM [20] جهت ارائه ی مدلی کارا تر که بتواند شبکه های اجتماعی پیچیده ی امروزی را مدل سازی نماید.
- ارائه ی مدلی بر اساس الگوریتم بهینه سازی ازدحام ذرات جهت رفع تمامی مشکلات ذکر شده در مورد روش های قبلی
- استفاده از انتشار اطلاعات به عنوان یک روش تشخیص تشکل[1] ها در شبکه های اجتماعی و استفاده از آن به عنوان اساس مقایسه ی مدل های ارائه شده

در ادامه مفاهیم اولیه ی مورد نیاز برای رسیدن به روش های پیشنهادی را بررسی خواهیم کرد. الگوریتم های ژنتیک به عنوان پایه ای برای روش پیشنهادی اول، الگوریتم های بهینه سازی ازدحام ذرات به عنوان اساس روش دوم و همچنین نظریه ی بازی ها به عنوان مکانیزم مورد استفاده در الگوریتم تشخیص تشکل ها. همچنین مدل GADM را نیز با توضیحات بیشتری ارائه خواهیم داد تا با استفاده از آن به مدل ارتقاء یافته ی پیشنهادی برسیم. در نهایت روشی را برای مقایسه ی مدل های ارائه شده بیان می نماییم.

## 2-3-مروری بر نظریه بازی ها

نظریه بازی شاخه‌ای از ریاضیات کاربردی است که در علوم اجتماعی و به ویژه در اقتصاد، زیست‌شناسی، مهندسی، علوم سیاسی، روابط بین‌الملل، علوم کامپیوتر، بازاریابی و فلسفه مورد استفاده قرار گرفته است. نظریه بازی در تلاش است توسط ریاضیات رفتار را در شرایط راهبردی یا بازی، که در آنها موفقیت فرد در انتخاب کردن وابسته به انتخاب دیگران می‌باشد، بدست آورد.
یک بازی شامل مجموعه‌ای از بازیکنان، مجموعه‌ای از حرکت‌ها یا استراتژی ها و نتیجه ی مشخصی برای هر ترکیب از استراتژی ها می‌باشد. پیروزی در هر بازی تنها تابع یاری شانس نیست بلکه اصول و قوانین

[1] Community Detection

ویژه ی خود را دارد و هر بازیکن در طی بازی سعی می‌کند با به کارگیری آن اصول خود را به برد نزدیک کند. رقابت دو کشور برای دستیابی به انرژی هسته‌ای، سازوکار حاکم بر روابط بین دو کشور در حل یک مناقشه ی بین‌المللی، رقابت دو شرکت تجاری در بازار بورس کالا نمونه‌هایی از بازی‌ها هستند.
نظریه ی بازی تلاش می‌کند تا رفتار ریاضی حاکم بر یک موقعیت استراتژیک (تضاد منافع) را مدل‌سازی کند. این موقعیت زمانی پدید می‌آید که موفقیت یک فرد وابسته به استراتژی هایی است که دیگران انتخاب می‌کنند. هدف نهایی این دانش یافتن استراتژی بهینه برای بازیکنان است.

### 3-2-1- تاریخچه

در سال ۱۹۲۱ یک ریاضی‌دان فرانسوی به نام امیل برل[1] برای نخستین بار به مطالعه ی تعدادی از بازی‌های رایج در قمارخانه‌ها پرداخت و تعدادی مقاله در مورد آن‌ها نوشت. او در این مقاله‌ها بر قابل پیش‌بینی بودن نتایج این نوع بازی‌ها به طریق منطقی، تأکید کرده بود.
اگرچه برل نخستین کسی بود که به طور جدی به موضوع بازی‌ها پرداخت، به دلیل آن که تلاش پیگیری برای گسترش و توسعه ی ایده‌های خود انجام نداد، بسیاری از مورخین ایجاد نظریه ی بازی را نه به او، بلکه به یوهان ون نیومن ریاضی‌دان مجارستانی نسبت داده اند.
آن چه نیومن را به توسعه ی نظریه ی بازی ترغیب کرد، توجه ویژه ی او به یک بازی با ورق بود. او دریافته بود که نتیجه ی این بازی صرفا با تئوری احتمالات تعیین نمی‌شود. او شیوه ی بلوف زدن در این بازی را فرمول‌بندی کرد. بلوف زدن در بازی به معنای راه‌کار فریب دادن سایر بازیکنان و پنهان کردن اطلاعات از آن‌هاست.
در سال ۱۹۲۸ او به همراه اسکار مونگسترن که اقتصاددانی اتریشی بود، کتاب تئوری بازی‌ها و رفتار اقتصادی را به رشته ی تحریر در آوردند. اگر چه این کتاب صرفا برای اقتصاددانان نوشته شده بود، کاربردهای آن در روان‌شناسی، جامعه‌شناسی، سیاست، جنگ، بازی‌های تفریحی و بسیاری زمینه‌های

[1] Emile Borel

دیگر به زودی آشکار شد.
نیومن بر اساس راهبردهای موجود در یک بازی ویژه شبیه شطرنج توانست کنش‌های میان دو کشور ایالات متحده و اتحاد جماهیر شوروی را در خلال جنگ سرد، با در نظر گرفتن آن‌ها به عنوان دو بازیکن در یک بازی مجموع صفر مدل‌سازی کند.
از آن پس پیشرفت این دانش با سرعت بیشتری در زمینه‌های مختلف پی گرفته شد و از جمله در دهه ی ۱۹۷۰ به طور چشم‌گیری در زیست‌شناسی برای توضیح پدیده‌های زیستی به کار گرفته شد.
در سال ۱۹۹۴ جان نش[1] به همراه دو نفر دیگر به خاطر مطالعات خلاقانه خود در زمینه ی تئوری بازی برنده ی جایزه ی نوبل اقتصاد شدند. در سال‌های بعد نیز برندگان جایزه ی نوبل اقتصاد عموما از میان نظریه‌پردازان بازی انتخاب شدند.

## 3-2-2- کاربردها

نظریه بازی در مطالعه ی طیف گسترده‌ای از موضوعات کاربرد دارد. از جمله نحوه تعامل تصمیم گیرندگان در محیط رقابتی به شکلی که نتایج تصمیم هر عامل موثر بر نتایج کسب شده سایر عوامل می باشد. در واقع ساختار اصلی نظریه بازی ها در بیشتر تحلیلها شامل ماتریسی چند بعدی است که در هر بعد مجموعه ای از گزینه ها قرار گرفته اند که آرایه های این ماتریس نتایج کسب شده برای عوامل در ازاء ترکیب های مختلف از گزینه های مورد انتظار است. یکی از اصلی ترین شرایط بکارگیری این نظریه در تحلیل محیط های رقابتی و وفاداری عوامل متعامل در رعایت منطق بازی است. در صورتی که این پیش شرط به هر دلیل رعایت نگردد، یا بایستی در انتظار نوزایی ساختار جدید دیگری از منطق تحلیلی بازیگران متعامل بود و یا به دلیل عدم پیش بینی نتایج بازی و یا گزینه های مورد انتظار سیستم تصمیم گیرنده به سراغ سایر روش های تحلیل در یک چنین محیط های تصمیم گیری رفت. هر چه قدر توان پیش بینی گزینه ها و نتایج حاصل از انتخاب آنها بیشتر باشد، عدم قطعیت در این تکنیک کاهش می یابد. نوعی از بازی نیز وجود دارد که به دلیل اینکه امکان برآورد احتمال وقوع نتایج در آنها وجود ندارد به بازی

[1] John Nash

های ابهام شهرت دارند.
این نظریه در ابتدا برای درک مجموعه ی بزرگی از رفتارهای اقتصادی به عنوان مثال نوسانات شاخص سهام در بورس اوراق بهادار و افت و خیز بهای کالاها در بازار مصرفکنندگان ایجاد شد.
تحلیل پدیدههای گوناگون اقتصادی و تجاری نظیر پیروزی در یک مزایده، معامله، داد و ستد، شرکت در یک مناقصه، از دیگر مواردی است که نظریه بازی در آن نقش ایفا میکند.
پژوهشها در این زمینه اغلب بر مجموعهای از استراتژی های شناخته شده به عنوان تعادل در بازیها استوار است. این استراتژی ها اصولا از قواعد عقلانی به نتیجه میرسند. مشهورترین تعادلها، تعادل نش است. براساس نظریه ی تعادل نش، اگر فرض کنیم در هر بازی با استراتژی مختلط ، بازیکنان به طریق منطقی و معقول راهبردهای خود را انتخاب کنند و به دنبال حداکثر سود در بازی هستند، دست کم یک استراتژی برای به دست آوردن بهترین نتیجه برای هر بازیکن قابل انتخاب است و چنانچه بازیکن راهکار دیگری به غیر از آن را انتخاب کند، نتیجه ی بهتری به دست نخواهد آورد.
کاربرد نظریه بازیها در شاخههای مختلف علوم مرتبط با اجتماع از جمله سیاست، جامعه شناسی، و حتی روان شناسی در حال گسترش است. در زیست شناسی هم برای درک پدیدههای متعدد، از جمله برای توضیح تکامل و ثبات و نیز برای تحلیل رفتار تنازع بقا و نزاع برای تصاحب قلمرو از نظریه بازی استفاده میشود.
امروزه این نظریه کاربرد فزایندهای در منطق و دانش کامپیوتر دارد. دانشمندان این رشتهها از برخی بازیها برای مدلسازی محاسبات و نیز به عنوان پایهای نظری برای سیستمهای چندعاملی استفاده میکنند. کاربردهای این نظریه تا آن جا پیش رفته است که در توصیف و تحلیل بسیاری از رفتارها در فلسفه و اخلاق ظاهر میشود.

## 3-2-3- انواع بازی

نظریه بازی علی الاصول میتواند روند و نتیجه ی هر نوع بازی از دوز گرفته تا بازی در بازار بورس سهام را توصیف و پیشبینی کند. تعدادی از ویژگیهایی که بازیهای مختلف بر اساس آنها

طبقه‌بندی می‌شوند، در زیر آمده است.

### 3-2-3-1- متقارن - نامتقارن[1]

بازی متقارن  به بازی ای اطلاق می شود که نتیجه و سود حاصل از یک استراتژی تنها به این وابسته است که چه استراتژی های دیگری در بازی پیش گرفته شود؛ و از این که کدام بازیکن این استراتژی را در پیش گرفته است مستقل است. به عبارت دیگر اگر مشخصات بازیکنان بدون تغییر در سود حاصل از به کارگیری راه‌بردها بتواند تغییر کند، این بازی متقارن است. بسیاری از بازی‌هایی که در یک جدول ۲×۲ قابل نمایش هستند، اصولا متقارن اند.
بازی‌های نامتقارن اغلب بازی‌هایی هستند که مجموعه ی راه‌بردهای یکسانی برای بازیکنان در بازی وجود ندارد. البته ممکن است راه‌بردهای یکسانی برای بازیکنان موجود باشد ولی آن بازی نامتقارن باشد.

### 3-2-3-2- مجموع صفر - مجموع غیر صفر[2]

بازی‌های مجموع صفر بازی‌هایی هستند که ارزش بازی در طی بازی ثابت می‌ماند و کاهش یا افزایش پیدا نمی‌کند. در این بازی‌ها، سود یک بازیکن با زیان بازیکن دیگر همراه است. به عبارت ساده‌تر یک بازی مجموع صفر یک بازی برد-باخت مانند دوز است و به ازای هر برنده همواره یک بازنده وجود دارد.  اما در بازی‌های مجموع غیر صفر راهبردهایی موجود است که برای همه ی بازیکنان سودمند است.

### 3-2-3-3- تصادفی - غیر تصادفی

بازی‌های تصادفی شامل عناصر تصادفی مانند ریختن تاس یا توزیع ورق هستند و بازی‌های غیر تصادفی بازی‌هایی هستند که دارای استراتژی هایی صرفا منطقی هستند. در این مورد می‌توان شطرنج و دوز را مثال زد.

### 3-2-3-4- با آگاهی کامل – بدون آگاهی کامل[3]

---

[1] Symmetric - Asymmetric

[2] Zero Sum – Nonzero Sum

[3] Perfect Knowledge – Non-perfect Knowledge

بازی‌های با آگاهی کامل، بازی‌هایی هستند که تمام بازیکنان می‌توانند در هر لحظه تمام ترکیب بازی را در مقابل خود مشاهده کنند، مانند شطرنج. از سوی دیگر در بازی‌های بدون آگاهی کامل ظاهر و ترکیب کل بازی برای بازیکنان پوشیده است، مانند بازی‌هایی که با ورق انجام می‌شود.

### 3-2-4- عناصر یک بازی

در نظریه بازی ها، هر بازی از یک تعداد عناصر تشکیل شده است که برای حل کردن مسائل تعریف کردن همه ی این عناصر از اهمیت ویژه ای برخوردار است. بنابراین برای تعریف فضای بازی، مشخص کردن عناصر زیر لازم و کافی است:

#### 3-2-4-1- بازیکن ها[1]

هر بازی از یک سری بازیکن تشکیل شده است. این بازیکن ها بر سر تعدادی منابع با هم به رقابت می پردازند و برای به دست آوردن سود بیشتر استراتژی هایی را در نظر می گیرد. بنابراین اولین مرحله جهت تعریف یک بازی، تعریف کردن بازیکن های آن بازی بر اساس مسئله ی مورد نیاز است. بدیهی است که تعداد بازیکن ها بسته به هر مسئله ای ممکن است فرق کند.

#### 3-2-4-2- استراتژی[2]

در نظریه بازی‌ها استراتژی یا راهبرد یک بازیکن در یک بازی یک مجموعه کامل از اعمالی است که در هر موقعیت انجام می‌دهد. استراتژی به طور کامل رفتار بازیکن را بیان می‌کند. استراتژی یک بازیکن بیان کننده اعمالی است که بازیکن در هر مرحله از بازی، برای هر مجموعه از اعمالی که بازیکن قبل از این مرحله انجام داده، انتخاب می‌کند.
یک نمایه استراتژی[3] (گاهی آن را ترکیب استراتژی نیز می‌نامند) مجموعه‌ای از استراتژی‌ها برای هر

---

[1] Players

[2] Strategy

[3] Strategy Profile

بازیکن است که به طور کامل همه اعمال در یک بازی را بیان می‌کند. یک نمایه استراتژی باید شامل یک و فقط یک استراتژی برای هر بازیکن باشد.
مفهوم استراتژی گاهی به غلط با حرکت اشتباه گرفته می‌شود. یک حرکت عملی است که توسط یک بازیکن در نقطه‌ای از بازی انتخاب می‌شود (مثلا در شطرنج حرکت فیل سفید از نقطه ی ۲a به نقطه ی ۳b). در حالی که یک استراتژی یک الگوریتم کامل برای انجام بازی است که به بازیکن می‌گوید در هر موقعیت ممکن در طول بازی چه کار کند.
مجموعه استراتژی یک بازیکن تعیین می‌کند که برای این بازیکن، بازی کردن کدام استراتژی‌ها ممکن است. اگر برای یک بازیکن تعدادی استراتژی گسسته وجود داشته باشد، مجموعه استراتژی این بازیکن متناهی است. به عنوان نمونه در بازی سنگ، کاغذ، قیچی، هر بازیکن مجموعه استراتژی متناهی {سنگ، کاغذ، قیچی} را دارد.
در غیر این صورت یک مجموعه استراتژی نامتناهی است. به عنوان مثال، در یک مزایده که میزان افزایش قیمت طبق یک قانون است، استراتژی‌ها گسسته هستند و مجموعه استراتژی نامتناهی است {۱۰ هزار تومان، ۲۰ هزار تومان، ۳۰ هزار تومان و...}. همچنین، بازی بریدن کیک دارای استراتژی‌های کراندار و پیوسته در مجموعه استراتژی‌ها است {بریدن هر جا بین ۰٪ تا ۱۰۰٪ از کیک}.
در یک بازی پویا مجموعه استراتژی شامل قوانین ممکن است که یک بازیکن می‌تواند به یک ربات یا یک عامل نرم افزاری بدهد تا بفهمد که چطور بازی کند. برای مثال در بازی اولتیماتوم، مجموعه استراتژی برای بازیکن دوم شامل همه قوانین ممکن است که برای آن‌ها پیشنهاد می‌شود فرد بپذیرد یا رد کند. در یک بازی بیزی[1] مجموعه استراتژی شبیه بازی پویا است. مجموعه استراتژی شامل قوانینی است که بیان می‌کند برای هر اطلاعات خصوصی ممکن چه عملی انجام شود.
در نظریه بازی‌های کاربردی، تعریف مجموعه‌های استراتژی بخش مهمی از هنر به صورت همزمان معنادار و قابل حل کردن یک بازی است. نظریه پرداز بازی می‌تواند از دانش سراسر مسئله به منظور محدود کردن فضای استراتژی و ساده‌تر کردن راه حل استفاده کند.

[1] Bayesian Game

به عنوان مثال، اگر بخواهیم در مورد بازی اولتیماتوم دقیق صحبت کنیم، یک بازیکن می‌تواند استراتژی‌هایی مثل رد کردن پیشنهاد (۱ تومان، ۲ تومان، ...، ۱۹ تومان) و پذیرفتن پیشنهاد (۰ تومان، ۱ تومان، ...، ۲۰ تومان) را داشته باشد. به حساب آوردن همه ی این استراتژی‌ها، فضای استراتژی بسیار بزرگی را به وجود می‌آورد و مسئله را دشوار می‌کند. یک نظریه‌پرداز بازی در عوض می‌تواند مجموعه استراتژی‌ها را به این صورت بسازد: {رد کردن هر پیشنهاد کمتر یا مساوی x و پذیرفتن هر پیشنهاد بزرگتر از x، برای x در (۰ تومان، ۱ تومان، ...، ۲۰ تومان)}.

یک استراتژی خالص[1] تعریف کاملی از این که یک بازیکن چگونه بازی خواهد کرد ارائه می‌دهد. این استراتژی حرکتی را که یک بازیکن برای هر موقعیتی که با آن روبه‌رو خواهد شد باید انجام دهد، تعریف می‌کند. مجموعه استراتژی یک بازیکن مجموعه‌ای است از استراتژی‌های خاصی که برای آن بازیکن ممکن است.

یک استراتژی مختلط[2] انتصاب یک احتمال به هر استراتژی خالص است. این استراتژی به یک بازیکن اجازه می‌دهد به صورت تصادفی یک استراتژی خالص را برگزیند. چون احتمال‌ها پیوسته هستند استراتژی‌های مختلط نامتناهی برای یک بازیکن وجود دارد، حتی اگر مجموعه استراتژی‌های آن متناهی باشد. البته می‌توان یک استراتژی خالص را نوع خاصی از استراتژی مختلط دانست که در آن یک استراتژی خالص خاص با احتمال ۱ و بقیه استراتژی‌ها با احتمال ۰ انتخاب می‌شوند. یک استراتژی کاملا مختلط، استراتژی مختلطی است که در آن بازیکن یک احتمال اکیداً مثبت به هر استراتژی خالص می‌دهد.

### 3-2-5- تعادل

در تئوری بازی ها، تعادل نش[3] (به نام جان فوربز نش، که آن را پیشنهاد کرد) راه حلی از تئوری بازی است که شامل دو یا چند بازیکن، که در آن فرض بر آگاهی هر بازیکن به استراتژی تعادل بازیکنان دیگر است و بدون هیچ بازیکنی که فقط برای کسب سود خودش

---

[1] Pure Strategy

[2] Mixed Strategy

[3] Nash Equilibrium (NE)

با تغییر استراتژی یک جانبه عمل کند. اگر هر بازیکنی استراتژی را انتخاب کند هیچ بازیکنی نمی تواند با تغییر استراتژی خود در حالی که نفع بازیکن دیگر را بدون تغییر نگه داشته باشد عمل کند، سپس مجموعه انتخاب های استراتژی فعلی و بهره مندی مربوطه، تعادل نش را تشکیل می دهد. به بیان ساده، امی و فیل در تعادل نش است اگر امی در حال انجام بهترین تصمیم گیری که او می تواند با توجه به تصمیم گیری فیل داشته باشد و همچنین فیل بهترین تصمیمی که می تواند با توجه به تصمیم گیری امی داشته باشد . به همین ترتیب یک گروه از بازیکنان در تعادل نش است اگر هر یک در حال انجام بهترین تصمیم گیری باشند که آنها می توانند، با توجه به تصمیمات دیگران داشته باشند. با این حال، تعادلی که نش است لزوما به معنای بهترین بهره وری کل برای همه بازیکنان مربوطه نمی باشد، در بسیاری از موارد ممکن است تمام بازیکنان بهره وری خود را بهبود بخشند در صورتی که چگونه بتوانند به توافق بر روی استراتژی های مختلف از تعادل نش برسند. (به عنوان نمونه، شرکت های تجاری رقابتی به منظور افزایش سود آنها تشکیل کارتل میدهد).

مفهوم تعادل نش برای تجزیه و تحلیل نتایج اثر متقابل استراتژیک چندین تصمیم گیرنده استفاده شده است. به عبارت دیگر، این راهی برای پیش بینی اینکه اگر چند نفر یا چندین موسسه که در تصمیم گیری های همزمان هستند و اگر پیامد های آن وابسته به تصمیم های دیگران است چه نتایجی را خواهد داشت . نگرش ساده و ایده اساسی جان نش این است که اگر ما تصمیم های تصمیم گیرندگان مختلف را به صورت جداگانه تحلیل کنیم در نتیجه نمی توانیم نتیجه انتخاب های آنان را پیش بینی کنیم. در عوض، ما باید بپرسیم آنچه که هر کدام از بازیکنان انجام میدهد، با در نظر گرفتن تصمیم گیری های دیگران است. تعادل نش برای تجزیه و تحلیل شرایط خصمانه شبیه جنگ و مسابقات مسلحانه به کار گرفته شده است (معمای زندانی). همچنین نحوه درگیری ممکن است از طریق اثر متقابل تکرار شده تعدیل داده شود (این به جای آن). همچنین این تئوری برای مطالعه آنچه که اندازه افراد با ترجیحات متفاوت می توانند همکاری کنند استفاده شده است (Battle of the sexes). و اینکه آیا آنها خطراتی را برای دستیابی به نتایج مشارکتی خواهد گرفت (Stag hunt). این تئوری برای

مطالعه تصمیم گیری در مورد استانداردهای فنی گرفته شده است، و همچنین رخداد اجرایی بانک و بحران های ارزی (بازی هماهنگ). برنامه های کاربردی دیگر شامل جریان ترافیک (Wardrop's principle).، نحوه سازماندهی مزایده ها (تئوری مزایده)، نتیجه اقدامات وارد شده توسط احزاب مختلف در فرایند آموزشی و حتی ضربات پنالتی در فوتبال (سکه های مطابق[1])

یک نسخه از مفهوم تعادل نش برای اولین بار توسط آنتونی آگوستین کورنو در نظریه خود در انحصار چند جانبه مورد استفاده قرار گرفت (1838).در تئوری کورنو آمده که بنگاه ها چگونه محصول بیشتر برای تولید حداکثر کردن سود خود انتخاب میکنند. با این حال، بهترین محصول برای یک شرکت بستگی به محصولات دیگران دارد. تعادل کورنو هنگامی که هر بنگاه سود هر محصول خود را با توجه به محصول های شرکت های دیگر حداکثر می کند رخ میدهد. مفهوم نوین تئوری بازی از تعادل نش بر حسب استراتژی های ترکیبی تعریف شده است. جایی که بازیکنان یک توزیع احتمال از اقدامات امکان پذیر را انتخاب می کنند. مفهوم استراتژی ترکیبی تعادل نش توسط جان فون نویمان و اسکار مورگسترن در کتاب سال 1944 خود با عنوان نظریه بازی ها و رفتار اقتصادی معرفی شده است. با این حال، تجزیه و تحلیل آنها برای حالت خاصی از بازی با مجموع صفر محدود بود. آنها نشان دادند که استراتژی ترکیبی تعادل نش برای هر بازی مجموع صفر با یک مجموعه متناهی از اقدامات وجود خواهد داشت.
نقش جان فوربز نش در مقاله خود در سال 1951 با عنوان بازی های غیر مشارکتی به تعریف یک استراتژی مختلط تعادل نش برای هر گونه بازی با یک مجموعه متناهی از اقدامات پرداخت و ثابت کرد که حداقل یک (استراتژی مختلط) تعادل نش باید در چنین بازی وجود داشته باشد. از آنجایی که توسعه مفهوم تعادل نش را نظریه پردازان تئوری بازی کشف کرده اند در شرایط معینی پیش بینی های گمراه کننده ای ایجاد می کند. بنابراین راه حل بسیاری از مفاهیم مرتبط پیشنهاد شده (همچنین اصلاحات تعادل نش) طراحی شده که برای غلبه بر نقص در درک مفهوم نش می باشد. یک موضوع خیلی مهم این است که برخی از تعادل های نش ممکن است مبتنی برآن خطراتی باشد که معتبر نیست.

[1] Matchin Pennies

بنابراین، در سال 1965 توسط راینهارد سلتن پیشنهاد شده که تعادل کامل با بازی فرعی به عنوان یک بهبود که به تعادل در تهدیدهای غیر معتبر بستگی دارد را حذف می کند. ضمیمه های دیگر از مفهوم تعادل نش به این پرداخته که اگر یک بازی تکرار شده است و یا چه چیزی اتفاقی می افتد اگر یک بازی در فقدان اطلاعات کامل بازی شده باشد . با این حال، پس از آن اصلاحات و ضمیمه ها از مفهوم تعادل نش سهم دیدگاه اصلی مفهوم نش استوار به همه ی مفاهیم تعادل تجزیه و تحلیل گزینه هایی خواهد شد که هر بازیکن به اعتبار تصمیم سازی های دیگران می گیرد.

## 3-2-5-1- تعریف غیر رسمی

به بیان غیر رسمی، به مجموعه ای از استراتژی ها در صورتی که هیچ بازیکنی توسط راهبردی یک جانبه ی در حال تغییر خود نمی تواند بهتر عمل کند تعادل نش می باشد. برای مشاهده در مورد این مفهوم، تصور کنید که هر بازیکن بیان کرده که استراتژی های دیگران را می گیرد. فرض کنید که سپس هر بازیکن از خودش می پرسد : دانستن استراتژی های بازیکنان دیگر، و بحث کردن در رابطه با استراتژی های بازیکنان دیگر بعنوان زیر بنا تعیین می شوند، آیا با تغییر استراتژی می توانم بهره مند بشوم؟ اگر پاسخ هر بازیکن "بله " باشد، پس آن مجموعه استراتژی ها تعادل نش نمی باشد. اما اگر هر بازیکن هر تغییری را ترجیح نمی دهد (یا بی تفاوت است بین تغییر و یا نه) پس مجموعه ای از استراتژی های تعادل نش می باشد. به این ترتیب، هر استراتژی در تعادل نش بهترین پاسخ به تمام استراتژی های دیگر در آن تعادل است. تعادل نش گاهی اوقات ممکن است غیر عقلایی در دیدگاه سوم شخص ظاهر می شود. دلیل این است که ممکن است که حالتی از تعادل نش که بهینه پارتو نیست رخ دهد. تعادل نش همچنین می تواند عواقب غیر عقلایی در بازی های پی در پی به خاطر بازیکنان، ممکن است تهدیدی برای یکدیگر با حرکت غیر منطقی داشته باشند. برای مثال در تئوری بازی ها تعادل کامل نش با بازی فرعی می تواند معنی دار تر به عنوان ابزار تجزیه و تحلیل باشد.

## 3-2-5-2- تعریف رسمی

مجموعه$(S,F)$را به عنوان بازی با n بازیکن در نظر بگیرید، که در آن $S_i$ مجموعه استراتژی ها برای بازیکن i است، $S = S_1 \times S_2 \cdots \times S_n$مجموعه ای از فضای استراتژی آن است و $F(x)$ تابع بهره مندی آن است $x_{-i}$ را به عنوان فضای استراتژی همه ی بازیکنان به جز بازیکن $i$ در نظر بگیرید .هر بازیکن به ازای هر i عضو مجموعه اعداد صحیح، استراتژی $x_i$ را انتخاب کند، نمایه استراتژی آن به صورت $x = x_1, \cdots, x_n$ و تابع بهره مندی آن به صورت $F(x_i)$ نتیجه داده می شود. قابل ذکر است که تابع بهره مندی به نمایه استراتژی انتخابی وابسته است. به عنوان مثال، در استراتژی انتخاب شده توسط بازیکن i و همچنین استراتژی های انتخاب شده توسط تمام بازیکنان دیگر. نمایه استراتژی $x^* \in S$ یک تعادل نش است اگر هیچ انحراف یک سویی در استراتژی توسط هر بازیکن واحد با یکی دیگر از بازیکنان سودآور نمی باشد. یعنی

$$\forall i, x_i \in s_i , x_i \neq x_i^* : f_i(x_i^*, x_{-i}^*) \geq f_i(x_i, x_{-i}^*) \qquad (1\text{-}3)$$

یک بازی می تواند یا استراتژی محض و یا تعادل نش ترکیبی باشد (در تعریف اخیر استراتژی محض آن است که به صورت تصادفی با فراوانی ثابت انتخاب شده است). نش نشان داد در صورتی که به ما اجازه استراتژی ترکیب شده بدهند، سپس هر بازی با تعداد محدودی از بازیکنان که در آن هر بازیکن می تواند به صورت غیر محدود از میان بسیاری از استراتژی های کامل که حداقل یک تعادل نش می باشد انتخاب کند. وقتی نابرابری اکید در بالا نگه می دارد برای تمام بازیکنان و تمام استراتژی های جایگزین امکان پذیر است، سپس تعادل طبقه بندی شده به عنوان یک تعادل دقیق نش می باشد. اگر در عوض، برای برخی از بازیکنان، برابری دقیقی بین x و بعضی از استراتژی های مجموعه S وجود دارد. سپس تعادل به عنوان یک تعادل طبقه بندی شده ضعیفی از نش می باشد.

## 3-3-بررسی فرآیند انتشار در شبکه های اجتماعی

یک شبکه ی اجتماعی ساختاری مبتنی بر گراف[1] است که ارتباطات و اتصالات داخلی میان افراد خاص و موجودیت ها را نمایش می دهد. رأس های این گراف نشان دهنده ی اشخاص، حیوانات یا کامپیوتر های شبکه شده هستند و گراف یال های گراف اتصال میان دو رأس است که این دو رأس به طریقی با یکدیگر در ارتباط می باشند. مثال های مدرن تری نیز برای شبکه های اجتماعی وجود دارند. برای مثال شبکه های اجتماعی برخط[2] که در آن رأس ها پروفایل[3] های شخصی افراد هستند و ارتباطات دوستی[4] میان این افراد یال های شبکه را تشکیل می دهند. یا شبکه های ارتباطی[5] که در آن ها رأس ها تشکیل شده اند از آدرس های پست الکترونیکی[6] یا شماره های تلفن و یال ها ارسال پست الکترونیکی و یا تماس تلفنی بین آن ها می باشند.

**تعریف 1-** شبکه ی اجتماعی یک گراف جهت دار[7] یا غیر جهت دار[8] $G = (V, E)$ است که در آن رأس برچسب دار $v \in V$ نشان دهنده ی یک موجودیت در یک سیستم فیزیکی است و یال $(u, v) \in E$ نشان دهنده ی یک ارتباط میان دو موجودیت است. یک شبکه ی اجتماعی پویا[9] گراف چند گانه مانند $G = (V, E)$ است که در آن $E$ نشان دهنده ی بسته از یال ها است و هر یال زمان دار $(u, v)_t \in E$ نشان دهنده ی یک ارتباط مانند $(u, v)$ است که در زمان $t \in \mathbb{Z}^+$ رخ داده است.

با در نظر گرفتن فرض های یک مدل انتشار ساده، مجموعه ای از رأس های یک شبکه را به صورت فعال در نظر می گیریم. با گذشت زمان در هر لحظه ی زمانی $t$ هر رأس که فعال باشد سعی می کند که همسایگان غیر فعال خود را نیز فعال نماید. پس از گذشت زمان و برقراری شرط مدلی که از آن استفاده کرده ایم الگوریتم خاتمه می یابد و طی آن تعدادی از رأس های غیر فعال به مجموعه ی رأس های فعال افزوده

---

[1] Graph
[2] Online Social Network
[3] Online Profile
[4] Friendship
[5] Communication Network
[6] E-mail
[7] Directed
[8] Undirected
[9] Dynamic Social Network

شده اند.
در رابطه با این ساختار سوالات زیر مطرح می شوند.
**وسعت**[1]**:** سوال اولی که در این زمینه مطرح می شود این است که با فعال کردن مجموعه ی خاصی از رأس ها، چه تعداد رأس پس از یک بازه ی زمانی خاص در وضعیت فعال قرار خواهند گرفت.
**هدف گیری**[2]**:** کدام یک از رأس های مجموعه به عنوان رأس های فعال ابتدایی انتخاب شوند تا در نهایت تعداد رئوس فعال بیشینه باشد.
**مسدود کردن**[3]**:** چه تعداد از رأس ها ایمن سازی شوند تا در نهایت تعداد رئوس فعال (در اینجا فعال بودن به معنای آلوده بودن است) کمینه شود.
هدف گیری و مسدود کردن هر دو ارتباط مستقیمی با مسئله ی وسعت در یک مدل دارند و از طرفی وسعت نیز کاملا به مدل ارائه شده بستگی دارد. در ادامه تعریف یک مدل انتشار را ارائه می دهیم که پایه و اساس مدل انتشار اطلاعات در شبکه های اجتماعی به حساب می آید.
**تعریف 2-** یک مدل انتشار یک ساختار گراف $G=(V,E)$، یک بردار وضعیت $S_v^{(t)}$ برای هر رأس $v\in V$ در زمان $t$ و یک بردار از پارامتر های دلخواه $P$ را به عنوان ورودی دریافت می نماید. بر اساس بردار وضعیت تمامی رأس هایی که با یکدیگر در ارتباط هستند، این مدل یک بردار وضعیت جدید $S_v^{(t+1)}$ را برای هر رأس در زمان $t+1$ ایجاد می نماید. برای یک شبکه ی اجتماعی ایستا[4] (غیر پویا)، ساختار گراف در هر زمان ثابت خواهد بود. برای یک شبکه ی اجتماعی پویا، ساختار گراف به اینصورت تعریف می شود که $G_t=(V,E)_t$ به طوری که $E_t=\{(u,v):(u,v)_t\in E\}$ مجموعه ی تمامی یال ها در زمان $t$ باشد.
پس به طور کلی در یک مدل فرآیند انتشار که انتشار اطلاعات حالت پیچیده و خاصی از آن می شود ما با بردار های وضعیت برای هر رأس و به روز رسانی آن ها سر و کار داریم. برای مثال در مدل آبشاری مستقل بردار وضعیت برای هر یک از رأس ها یک بیت است که نشان می دهد که وضعیت آن رأس در زمان $t$ به صورت فعال می باشد و یا غیر فعال. هر رأس فعال با

[1] Extent
[2] Targeting
[3] Blocking
[4] Static Social Network

احتمال $p$ شانس فعال کردن همسایه های غیر فعال خود را خواهد داشت و تنها یک بار می تواند این عمل را انجام دهد [19]. این حالت برای رأس هایی رخ خواهد داد که در یک شبکه ی اجتماعی ایستا حضور دارند. در مورد شبکه های اجتماعی پویا این عمل به اینصورت رخ می دهد که پس از هر تماس میان دو رأس ممکن است فعال سازی رخ بدهد [21].

## 3-4-مروری بر الگوریتم های بهینه سازی ازدحام ذرات

الگوریتم بهینه سازی ازدحام ذرات یک الگوریتم جستجوی اجتماعی است که از روی رفتار اجتماعی دسته های پرندگان مدل شده است. در ابتدا این الگوریتم به منظور کشف الگوهای حاکم بر پرواز همزمان پرندگان و تغییر ناگهانی مسیر آنها و تغییر شکل بهینه ی دسته به کار گرفته شد . در الگوریتم بهینه سازی ازدحام ذرات ، ذرات[1] در فضای جستجو جاری می شوند. تغییر مکان ذرات در فضای جستجو تحت تأثیر تجربه و دانش خودشان و همسایگانشان است. بنابراین موقعیت دیگر توده[2] ذرات روی چگونگی جستجوی یک ذره اثر می گذارد . نتیجه ی مدل سازی این رفتار اجتماعی فرایند جستجویی است که ذرات به سمت نواحی موفق میل می کنند. ذرات از یکدیگر می آموزند و بر مبنای دانش بدست آمده به سمت بهترین همسایگان خود می روند اساس کار الگوریتم بهینه سازی ازدحام ذرات بر این اصل استوار است که در هر لحظه هر ذره مکان خود را در فضای جستجو با توجه به بهترین مکانی که تاکنون در آن قرار گرفته است و بهترین مکانی که در کل همسایگی اش وجود دارد، تنظیم می کند.

### 3-4-1- هوش جمعی[3]

هوش جمعی خاصیتی است سیستماتیک که در این سیستم، عامل ها به طور محلی با هم همکاری می نمایند و رفتار جمعی تمام عامل ها باعث همگرایی در نقطه اي نزدیک به جواب بهینه سراسری میشود نقطه قوت

[1] -Particle

[2] -Swarm

[3] -Swarm Intelligent

این الگوریتم عدم نیاز به یک کنترل سراسري میباشد. هر ذره( عامل) در این الگوریتم ها خود مختاری نسبی دارد که می تواند در سراسر فضای جواب ها حرکت کند و می بایست با سایر ذرات (عامل ها) همکاري داشته باشد . دو الگوریتم مشهور هوش جمعی، بهینه سازي جرگه مورچگان و بهینه سازي توده ذرات می باشند. از هر دو این الگوریتم ها می توان براي تعلیم شبکه هاي عصبی بهره برد.

### 3-4-2- شبکه ی عصبی[1]

شبکه هاي عصبی مصنوعی داراي ویژگی هاي فراوانی از جمله انطباق پذیري، قابلیت یادگیري و تعمیم می باشد .در حوزه تطابق الگو ها، شبکه هاي عصبی مصنوعی قادرند که الگو هاي جدید را بر اساس تعالیم قبلی خود به کلاس هاي مرتبط طبقه بندي نمایند .

استفاده از ایده جدید هوش جمعی در ترکیب با شبکه هاي عصبی مصنوعی می باشد تا راهکاري براي غلبه بر چالش موجود در شبکه هاي عصبی باشد.

### 3-4-3- الگوریتم بهینه سازی ازدحام ذرات

در سال 1995 الگوریتم بهینه سازی ازدحام ذرات براي اولین بار توسط اِبرهارت و کِندی به عنوان یک روش جستجوی غیر قطعی براي بهینه سازي تابعی مطرح گشت. این الگوریتم از حرکت دسته جمعی پرندگانی که به دنبال غذا می باشند الهام گرفته شده است.
گروهي از پرندگان در فضايي به صورت تصادفي دنبال غذا مي گردند. تنها یک تکه غذا در فضا ي مورد بحث وجود دارد. هیچ یک از پرندگان محل غذا را نمي دانند. یکي از بهترین استراتژی ها مي تواند دنبال کردن پرنده ا ي باشد که کمتر ین فاصله را تا غذا داشته باشد. این استراتژي در واقع جانمایه الگوریتم است. هر راه حل که به آن يك ذره گفته میشود، در الگوریتم بهینه سازی ازدحام ذرات معادل يك پرنده در الگوریتم حرکت جمعي پرندگان میباشد. هر ذره يك مقدار شایستگي دارد که توسط يك تابع شایستگي محاسبه میشود. هر چه ذره در فضا ي جستجو

---
[1] -Neural Network

به هدف - غذا در مدل حرکت پرندگان نزدیکتر باشد، شایستگي بیشتری دارد. همچنین هر ذره داراي یك سرعت است که هدایت حرکت ذره را بر عهده دارد. هر ذره با دنبال کردن ذرات بهینه در حالت فعلی، به حرکت خود در فضای مساله ادامه میدهد. به این شکل است که گروه ي از ذرات الگوریتم بهینه سازی ازدحام ذرات آغاز کار به صورت تصادفي به وجود مي آیند و با به روز کردن نسل ها سعي در یافتن راه حل بهینه مي نمایند. در هر گام، هر ذره با استفاده از دو بهترین مقدار به روز میشود. اولین مورد، بهترین موقعیتي است که تاکنون ذره موفق به رسیدن به آن شده است. موقعیت مذکور شناخته و نگهدار ي میشود. بهترین موقعیتی که ذره تاکنون داشته است را با pbest نشان می دهیم. همچنین بهترین موقعیتی که تاکنون توسط جمعیت ذرات به دست آمده است را با نام gbest نگهداری می کنیم.

پس از یافتن مقادیر مورد نظر، سرعت و مکان ذره با استفاده از معادلات زیر به روز رسانی می شوند. در ابتدا می بایست سرعت ذرات به روز رسانی شود:

$$V\,[i+1] = V[i] + \big(C_1 * R_1 * (pbest[i] - position[i])\big) + \big(C_2 * R_2 * (gbest[i] - position[i])\big) \quad \text{-3)}$$

پس از آن نوبت به روز رسانی مکان ذره می رسد:

$$position[i+1] = position[i] + V[i+1] \quad \text{-3)}$$

سمت راست معادله (1-3) از سه قسمت تشکیل شده است که قسمت اول، سرعت فعلی ذره است و قسمت هاي دوم و سوم تغییر سرعت ذره و چرخش آن به سمت بهترین تجربه شخصي و بهترین تجربه گروه را به عهده دارند. اگر قسمت اول را در این معادله در نظر نگیریم، آنگاه سرعت ذرات تنها با توجه به موقعیت فعلي و بهترین تجربه ذره و بهترین تجربه جمع تعیین میشود. به این ترتیب، بهترین ذره جمع، در جاي خود ثابت میماند و سایرین به سمت آن ذره حرکت مي کنند. در واقع حرکت دسته جمعي ذرات بدون قسمت اول معادله (1-3)، پروسه اي خواهد بود که طي آن فضاي جستجو به تدریج کوچك میشود و جستجویي محلي حول بهترین ذره شکل میگیرد. در مقابل اگر فقط قسمت اول معادله (3-

1) را در نظر بگیریم، ذرات راه عادي خود را میروند تا به دیواره محدوده برسند و به نوعي جستجويي سراسري را انجام مي دهند.
شبه کد مربوط به الگوریتم بهینه سازی ازدحام ذرات در زیر آمده است.

**Algorithm 1. PSO**

```
For each Particle
Initialize Particle
End For
Do
For each Particle
Calculate fitness value of the particle fp
\*Updating particle's best fitness value so far*\
If fp is better than pBest
Set current value as the new pBest
End If
\*Updating population's best fitness value so far*\
Set gBest to the best fitness value of all particles
For each Particle
Calculate particle velocity according to 3-1
Update particle position according to 3-2
End For While maximum iterations OR minimum criteria is not attained
```

**الگوریتم 3-1- الگوریتم بهینه سازی ازدحام ذرات**

## 5-3-مروری بر الگوریتم های ژنتیک

الگوریتم ژنتیک (Genetic Algorithm - GA) تکنیک جستجویی در علم رایانه برای یافتن راه حل تقریبی برای بهینه سازی و مسائل جستجو است. الگوریتم ژنتیک نوع خاصی از الگوریتم های تکامل است که از تکنیک های زیست شناسی فرگشتی مانند وراثت و جهش استفاده می کنند.
در واقع الگوریتم های ژنتیک از اصول انتخاب طبیعی داروین برای یافتن فرمول بهینه جهت پیش بینی یا تطبیق الگو استفاده می کنند. الگوریتم های ژنتیک اغلب گزینه خوبی برای تکنیک های پیش‌بینی بر مبنای تصادف هستند. مختصراً گفته می‌شود که الگوریتم ژنتیک (یا GA) یک تکنیک برنامه نویسی است که از تکامل ژنتیکی به عنوان یک الگوی حل مسئله استفاده

می کند. مسأله ای که باید حل شود ورودی است و راه حل ها طبق یک الگو کد گذاری می شوند که تابع fitness نام دارد هر راه حل کاندید را ارزیابی می کند که اکثر آنها به صورت تصادفی انتخاب می شوند. به طور کلی این الگوریتم ها از بخش های زیر تشکیل می شوند: تابع برازش، نمایش، انتخاب، تغییر.

## 3-5-1- مکانیزم الگوریتم ژنتیک

الگوريتم ژنتيك به عنوان يك الگوريتم محاسباتيِ بهينه سازي با در نظر گرفتن مجموعه اي از نقاط فضاي جواب در هر تكرار محاسباتي به نحو مؤثری نواحي مختلف فضاي جواب را جستجو مي كند. در مكانيزم جستجو گرچه مقدار تابع هدف تمام فضاي جواب محاسبه نمي شود ولي مقدار محاسبه شده تابع هدف براي هر نقطه، در متوسطگيري آماري تابع هدف براي هر نقطه، در متوسطگيري آماري تابع هدف در كليه زير فضاهايي كه آن نقطه به آنها وابسته بوده دخالت داده مي‌شود و اين زير فضاها به طور موازي از نظر تابع هدف متوسطگيري آماري مي‌شوند. اين مكانيزم را توازي ضمني[1] مي‌گويند. اين روند باعث مي‌شود كه جستجوي فضا به نواحي از آن كه متوسط آماري تابع هدف در آنها زياد بوده و امكان وجود نقطه بهينه مطلق در آنها بيشتر است سوق پيدا كند. چون در اين روش برخلاف روش‌هاي تك‌مسيري فضاي جواب به طور همه جانبه جستجو مي‌شود، امكان كمتري براي همگرايي به يك نقطه بهينه محلي وجود خواهد داشت.
امتياز ديگر اين الگوريتم آن است كه هيچ محدوديتي براي تابع بهينه شونده، مثل مشتق پذيري يا پيوستگي لازم ندارد و در روند جستجو خود تنها به تعيين مقدار تابع هدف در نقاط مختلف نياز دارد و هيچ اطلاعاتِ كمكي ديگري، مثل مشتق تابع را استفاده نمي‌كند. لذا مي‌توان در مسائل مختلف اعم از خطي، پيوسته يا گسسته استفاده مي‌شود و به سهولت با مسائل مختلف قابل تطبيق است.
در هر تكرار هر يك از رشته‌هاي موجود در جمعيت رشته‌ها، رمزگشايي شده و مقدار تابع هدف براي آن به دست می‌آيد. بر اساس مقادير به دست آمده تابع هدف در جمعيت رشته‌ها، به هر رشته يك عدد برازندگي نسبت داده مي‌شود. اين عدد برازندگي احتمال انتخاب

[1] Implicit Parallelism

را براي هر رشته تعيين خواهد كرد. بر اساس اين احتمال انتخاب، مجموعه‌اي از رشته‌ها انتخاب شده و با اعمال عملكردهاي ژنتيكي روي آنها رشته‌هاي جديد جايگزين رشته‌هايي از جمعيت اوليه مي‌شوند تا تعداد جمعيت رشته‌ها در تكرارهاي محاسباتي مختلف ثابت باشد.
مكانيزم‌هاي تصادفي كه روي انتخاب و حذف رشته‌ها عمل مي‌كنند به گونه‌اي هستند كه رشته‌هايي كه عدد برازندگي بيشتري دارند، احتمال بيشتري براي تركيب و توليد رشته‌هاي جديد داشته و در مرحله جايگزيني نسبت به ديگر رشته‌ها مقاوم‌تر هستند. بدين لحاظ جمعيت دنباله‌ها در يك رقابت بر اساس تابع هدف در طي نسل هاي مختلف، كامل شده و متوسط مقدار تابع هدف در جمعيت رشته‌ها افزايش مي‌يابد. بطور كلي در اين الگوريتم ضمن آنكه در هر تكرار محاسباتي، توسط عملگرهاي ژنتيكي نقاطي جديد از فضاي جواب مورد جستجو قرار مي‌گيرند توسط مكانيزم انتخاب، روند جستجوي نواحي از فضا را كه متوسط آماري تابع هدف در آنها بيشتر است، كنكاش مي‌كند. بر اساس سيكل اجرايي فوق، در هر تكرار محاسباتي، توسط عملگرهاي ژنتيكي نقاط جديدي از فضاي جواب مورد جستجو قرار مي‌گيرند توسط مكانيزم انتخاب، روند جستجو نواحي از فضا را كه توسط آماري تابع هدف در آنها بيشتر است، كنكاش مي‌كند. كه بر اين اساس، در هر تكرار محاسباتي، سه عملگر اصلي روي رشته‌ها عمل مي‌كند؛ اين سه عملگر عبارتند از: دو عملگر ژنتيكي و عملكرد انتخابی تصادفی.
گلد برگ[1] الگوريتم ژنتيكي جان هولند را با عنوان الگوريتم ژنتيك ساده[2] معرفي مي‌كند؛ الگوريتم ژنتيك را از الگوريتم ژنتيك طبيعي اقتباس كردند.
در الگوريتم ژنتيك، مجموعه اي از متغيرهاي طراحي را توسط رشته‌هايي با طول ثابت[3] يا متغير[4] كدگذاري مي‌كنند كه در سيستم‌هاي بيولوژيكي آنها را كرروموزوم يا فرد[5] مي‌نامند. هر رشته يا كروموزوم يك نقطه پاسخ در فضاي جستجو را نشان مي‌دهد. به ساختمان رشته‌ها يعني مجموعه‌اي از پارامترها كه

[1] Gold Berg
[2] Simple Genetic Algorithm - SGA
[3] Fixed Length
[4] variable
[5] Individual

توسط يك كروموزوم خاص نمايش داده مي‌شود ژنوتيپ[1] و به مقدار رمزگشايي آن فنوتيپ[2] مي‌گويند. الگوريتم‌هاي وراثتي فرآيندهاي تكراري هستند، كه هر مرحلهٔ تكراري را نسل و مجموعه‌هايي از پاسخ‌ها در هر نسل را جمعيت ناميده‌اند.

الگوريتم‌هاي ژنتيك، جستجوي اصلي را در فضاي پاسخ به اجرا مي‌گذارند. اين الگوريتم‌ها با توليد نسل[3] آغاز مي‌شوند كه وظيفه ايجاد مجموعه نقاط جستجوي اوليه به نام «جمعيت اوليه»[4] را بر عهده دارند و به طور انتخابي يا تصادفي تعيين مي‌شوند. از آنجايي كه الگوريتم‌هاي ژنتيك براي هدايت عمليات جستجو به طرف نقطه بهينه از روش‌هاي آماري استفاده مي‌كنند، در فرآيندي كه به انتخاب طبيعي وابسته است، جمعيت موجود به تناسب برازندگي افراد آن براي نسل بعد انتخاب مي‌شود. سپس عملگرهاي ژنتيكي شامل انتخاب[5] ، پيوند(تركيب)، جهش و ديگر عملگرهاي احتمالي اِعمال شده و جمعيت جديد به وجود مي‌آيد. پس از آن جمعيت جديدي جايگزين جمعيت پيشين مي‌شود و اين چرخه ادامه مي‌يابد.

معمولاً جمعيت جديد برازندگي بيشتري دارد اين بدان معناست كه از نسلي به نسل ديگر جمعيت بهبود مي‌آيد. هنگامي جستجو نتيجه بخش خواهد بود كه به حداكثر نسل ممكن رسيده باشيم يا همگرايي حاصل شده باشد و يا معيارهاي توقف برآورده شده باشد.

## 3-5-2- عملگر های الگوریتم ژنتیک

به طور خلاصه الگوریتم ژنتیک از عملگر های زیر تشکیل شده است:

### 3-5-2-1- کد گذاری[6]

اين مرحله شايد مشكلترين مرحله حل مسأله به روش الگوريتم باشد. الگوريتم ژنتيك به جاي اينكه بر روي پارامترها يا متغيرهاي مسأله كار كند، با شكل

[1] Genotype
[2] Phenotype
[3] Seeding
[4] Initial Population
[5] Selection
[6] Encoding

كد شده آنها سروكار دارد. يكي از روشهاي كد كردن، كد كردن دودويي مي باشد كه در آن هدف تبديل جواب مسأله به رشته‌اي از اعداد باينري (در مبناي 2) است.

### 3-5-2-2- ارزیابی[1]

تابع برازندگي را از اِعمال تبديل مناسب بر روي تابع هدف يعني تابعي كه قرار است بهينه شود به دست مي‌آورند. اين تابع هر رشته را با يك مقدار عددي ارزيابي مي‌كند كه كيفيت آن را مشخص مي‌نمايد. هر چه كيفيت رشته جواب بالاتر باشد مقدار برازندگي جواب بيشتر است و احتمال مشاركت براي توليد نسل بعدي نيز افزايش خواهد يافت.

### 3-5-2-3- ترکیب[2]

مهمترين عملگر در الگوريتم ژنتيك، عملگر تركيب است. تركيب فرآيندي است كه در آن نسل قديمي كروموزوم‌ها با يكديگر مخلوط و تركيب مي‌شوند تا نسل تازه‌اي از كروموزوم‌ها بوجود بيايد.
جفت‌هايي كه در قسمت انتخاب به عنوان والد در نظر گرفته شدند در اين قسمت ژن‌هايشان را با هم مبادله مي‌كنند و اعضاي جديد بوجود مي‌آورند. تركيب در الگوريتم ژنتيك باعث از بين رفتن پراكندگي يا تنوع ژنتيكي جمعيت مي‌شود زيرا اجازه مي‌دهد ژن‌هاي خوب يكديگر را بيابند.

### 3-5-2-4- جهش[3]

جهش نيز عملگر ديگري هست كه جواب‌هاي ممكن ديگري را متولد مي‌كند. در الگوريتم ژنتيك بعد از اينكه يك عضو در جمعيت جديد بوجود آمد هر ژن آن با احتمال جهش، جهش مي‌يابد. در جهش ممكن است ژني از مجموعه ژن‌هاي جمعيت حذف شود يا ژني كه تا به حال در جمعيت وجود نداشته است به آن اضافه شود. جهش يك ژن به معناي تغيير آن ژن است و وابسته به نوع كدگذاري روش‌هاي متفاوت جهش استفاده مي‌شود.

### 3-5-2-5- رمز گشایی[1]

[1] Evaluation
[2] Crossover
[3] Mutation

رمزگشایی، عکس عمل رمزگذاری است. در این مرحله بعد از اینکه الگوریتم بهترین جواب را براي مسأله ارائه کرد لازم است عکس عمل رمزگذاري روي جوابها یا همان عمل رمزگشایی اعمال شود تا بتوانیم نسخه واقعي جواب را به وضوح در دست داشته باشیم.

### 3-5-3- شبه کد الگوریتم ژنتیک

شبه کد مربوط به الگوریتم ژنتیک را در زیر مشاهده می نمایید.

**Algorithm 2. GA**

1 **Choose** initial population
2 **Repeat**
3 Evaluate the individual fit nesses of a certain proportion of the population
4 Select pairs of best-ranking individuals to reproduce
5 Apply crossover operator
6 Apply mutation operator
7 **Until** terminating condition
9 **End**

الگوریتم 3-2- الگوریتم ساده ی ژنتیک

## 3-6-مدل انتشار بر اساس الگوریتم های ژنتیک (GADM) [20]

در این قسمت به طور تفصیلی توضیح می دهیم که چگونه می توان با استفاده از یک الگوریتم ژنتیک ساده [22] که از شیوه ی نمایش باینری و عملگر ترکیب ساده تک نقطه ای[2] استفاده می نماید برای مدل سازی انتشار اطلاعات در شبکه های اجتماعی ایستا و پویا استفاده نمود.
همانطور که پیش از این اشاره کرده بودیم، یک شبکه ی اجتماعی پویا تشکیل شده از مجموعه ای از رأس های $V = \{v_1, \ldots, v_n\}$ که در $T$ زمان گسسته با یکدیگر ارتباط برقرار می نمایند. ابتدا لازم است که ارتباطی میان رأس های شبکه ی اجتماعی و کروموزوم

---

[1] Decoding
[2] One-Point Crossover

های جمعیت الگوریتم ژنتیک برقرار نماییم. بردارهای وضعیت هر یک از رئوس را که با $S_v$ نمایش داده می شوند به صورت یک کروموزوم با طول $\beta$ در جمعیت الگوریتم ژنتیک در نظر می گیریم. هر یک از اعضای این بردارهای وضعیت را می توان با عدد صفر مقدار دهی اولیه کرد و یا با استفاده از یک تابع تصادفی به آن ها مقداری را اختصاص داد. همچنین یک تابع برازندگی مانند $f(x)$ در نظر می گیریم که به هر یک از بردارهای وضعیت یک مقدار برازندگی اختصاص می دهد. ساختار و شیوه ی کاری این تابع را در ادامه توضیح خواهیم داد. برای هر یال $(u,v)$ الگوریتم ژنتیک را روی رئوس معادل اعمال می نماییم:

1. در زمان $t$، برای یال $(u,v)$ دو بردار وضعیت به صورت $S_v^{(t)}$ و $S_u^{(t)}$ تعریف می شوند. بردار ها برای زمان $t+1$ نیز به صورت $S_v^{(t+1)} = S_v^{(t)}$ و $S_u^{(t+1)} = S_u^{(t)}$ مقدار دهی اولیه می شوند.
2. نقطه ی ترکیب[1] با نام $c$ به صورت تصادفی در بازه ی تعریف $[1,\beta]$ می شود. بردارهای وضعیت جدید با جابجایی دنباله های دو رشته ی $S_v^{(t)}$ و $S_u^{(t)}$ به دست می آیند. دنباله ی این دو رشته یعنی تمامی خانه های شامل و بعد از محل ترکیب $c$. این دو رشته ی جدید را با نام های $y_1$ و $y_2$ نمایش می دهیم.
3. میزان برازندگی این دو رشته ی جدید با استفاده از تابع برازندگی $f(x)$ به دست می آید. اگر هر یک از این دو رشته، میزان برازندگی بیشتری از والدین خود داشته باشند برای نسل بعدی با والد خود جایگزین می شوند.

$$S_v^{(t+1)} = \underset{x=\{S_v^t, y_1, y_2\}}{argmax}\ f(x) \qquad -3)$$

$$S_u^{(t+1)} = \underset{x=\{S_u^t, y_1, y_2\}}{argmax}\ f(x) \qquad -3)$$

در صورتی که میزان برازندگی رشته ی جدید با والد یکسان باشد، والد در جای خود باقی خواهد ماند.
الگوریتمی که ارائه شده است با الگوریتم ژنتیک ساده تفاوت چندانی ندارد. تنها نکته ی مهم در

[1] Crossover Point

مورد این الگوریتم این است که در این الگوریتم از خاصیت ذاتی شبکه های اجتماعی به عنوان عملگر انتخاب استفاده شده است. در واقع دو رأس که با یکدیگر در ارتباط هستند برای انجام عمل ترکیب انتخاب می شوند. فرآیند انتشار در این مدل با انتخاب شدن رأس ها و رخ دادن ترکیب و همچنین به روز رسانی آن ها با استفاده از تابع برازندگی رخ می دهد. به طور مشخص تنها نکته ی مبهم در مورد این الگوریتم این است که از چه تابعی به عنوان تابع برازندگی $f(x)$ استفاده نماییم.

توابع چند بعدی هلند (HDF) به عنوان راه حلی مناسب برای این مسئله انتخاب می شوند. این توابع ترکیبی خواص ویژه ای دارند که به ما برای رسیدن به هدف مورد نظر کمک می نمایند [23]. HDF ها از تعدادی الگو[1] تشکیل شده اند. این الگو ها در واقع رشته های هستند که در بخش هایی از آن ها که به عنوان نقطه ی شروع در نظر گرفته می شوند اعداد 0 و 1 قرار دارند. الگو هایی که تنها شامل یک دنباله از این اعداد شوند در مرتبه ی 1 قرار دارن و ترکیب این الگو های مرتبه ی 1، الگو های مرتبه ی بالاتر را تشکیل می دهند. به این الگو ها اعداد مثبت و منفی به صورت تصادفی اختصاص داده می شوند که نشان دهنده ی ارزش هر یک از آن ها می باشند. HDF یک رشته را به عنوان ورودی دریافت کرده و میزان امتیاز برازندگی آن را محاسبه می نماید. این عدد حاصل جمع مقادیر الگو هایی است که در آن رشته وجود دارند. HDF می تواند جهت تولید توابع با پیچیدگی بالاتر نیز استفاده شوند که در اینجا همین تابع ساده پاسخگوی نیاز ما خواهد بود. مثال زیر نحوه ی عملکرد یک تابع HDF به صورت $f(x) \to \mathbb{R}$ را نشان می دهد. این تابع رشته هایی با طول 10 را به عنوان ورودی دریافت می کند. علامت ("*") نشان دهنده ی حالت غیر مهم[2] است که با 0 و 1 همخوانی پیدا می کند.

---

[2] Schema

[1] Don't care

| | |
|---|---|
| ⋆01⋆⋆⋆⋆⋆⋆⋆ | *score 2, order 1 schema* |
| ⋆⋆⋆⋆⋆110⋆⋆ | *score 2, order 1 schema* |
| ⋆⋆⋆⋆⋆⋆⋆⋆10 | *score 3, order 1 schema* |
| ⋆01⋆⋆110⋆⋆ | *score -4, order 2 schema* |
| ⋆01⋆⋆⋆⋆⋆10 | *score 4, order 2 schema* |

با توجه به تابع تعریف شده در بالا، مقدار امتیاز رشته های زیر به اینصورت به دست می آید:

$$1010000011 \quad score: 2$$
$$0010011000 \quad score: 2 + 2 + (-4) = 0$$
$$1010011010 \quad score: 2 + 2 + 3 + (-4) + 4 = 7$$

یک تمثیل مناسب برای این مدل که از الگوریتم ژنتیک و توابع هلند استفاده می نماید انتشار اطلاعات است. در این مدل هر یک از الگو های بدست آمده از تابع می تواند به عنوان یک واحد اطلاعات در نظر گرفته شود. با توجه به مقدار دهی اولیه بردارهای وضعیت، هر یک از رئوس شبکه میزان خاصی از اطلاعات را حمل می نماید. هر زمان که رأس ها با یکدیگر ارتباط برقرار می نمایند به صورت تصادفی اطلاعاتی را به یکدیگر انتقال می دهند که با استفاده از عملگر ترکیب پیاده سازی شده است. این عملیات می تواند باعث افزایش سطح اطلاعات و یا ثابت ماندن آن برای هر رأس شود. همانطور که پیش تر ذکر شد ادعای ارائه دهندگان این روش بر این است که می تواند از انتقال واحد های چندگانه ی اطلاعاتی پشتیبانی نمایند حال آن که در قسمت های بعد نشان خواهیم داد که این اصل برقرار نخواهد ماند.

## 3-7- الگوریتم های پیشنهادی

در ادامه دو الگوریتم پیشنهادی را مورد بررسی قرار خواهیم داد. در الگوریتم اول که به نوعی حالت ارتقاء یافته ی GADM است ما از یک عملگر جهش برای تقویت مدل پیشنهادی لاهیری و سربیان استفاده خواهیم کرد. در قسمت بعد و در معرفی الگوریتم دوم به بررسی نحوه ی استفاده از الگوریتم های بهینه سازی ازدحام ذرات برای مدل سازی انتشار اطلاعات در شبکه های اجتماعی ایستا و

پویا خواهیم پرداخت.

### 3-7-1- الگوریتم اول: مدل بر اساس الگوریتم های ژنتیک بهبود یافته (EGADM)

یکی از مشکلات مدل ارائه شده توسط لاهیری و سربیان [20] این امر است که امروزه و در شبکه های اجتماعی پیچیده ای که ما با آن ها سر و کار داریم دانش تنها از طریق ارتباط با سایر افراد کسب نمی شود. هر شخص ممکن است خود به تنهایی از طریق ابزارهایی که در اختیار دارد به کسب دانش از منابع گوناگون بپردازد. برای پیاده سازی این امر در یک مدل انتشار اطلاعات در شبکه های اجتماعی می بایست این پارامتر را نیز در نظر بگیریم.

به همین جهت و برای ارتقاء مدل GADM بر آن شدیم که با استفاده از عملگر جهش در الگوریتم های ژنتیک عامل دیگری را نیز در فرآیند انتشار دخیل کنیم. با استفاده از این عملگر نشان خواهیم داد که افراد علاوه بر استفاده از عملگر ترکیب، امکان دارد که بتوانند اطلاعات خود را افزایش دهند. مانند قسمت قبل برای دو رأس $u$ و $v$ بردارهای وضعیت را به صورت $S_u^{(t+1)}$ و $S_v^{(t+1)}$ در نظر می گیریم. پس از اجرای قسمت اول الگوریتم و رخ دادن عملگر ترکیب، مراحل زیر صورت می گیرند:

1. بردارهای و نشان دهنده ی وضعیت فعلی رئوس $u$ و $v$ هستند.
2. به ازای هر بیت از بردارهای $S_v^{(t+1)}$ و $S_u^{(t+1)}$ با احتمال $P_m$ عملیات جهش صورت می گیرد. جهش در این مرحله به اینصورت رخ خواهد داد که ما با تولید یک عدد تصادفی، در صورتی که این عدد بیشتر یا مساوی با $P_m$ باشد، بیت های $0$ را تبدیل به 1 می نماییم و بالعکس. بردارهای حاصل از این مرحله را با $w_1$ و $w_2$ نمایش می دهیم.
3. میزان برازندگی را برای هر یک از بردارهای جدید $S_v^{(t+1)}$ و $S_u^{(t+1)}$ محاسبه می نماییم و در صورتی که این میزان از برازندگی والد بیشتر باشد، بردارهای جدید به بردارهای وضعیت والد جایگزین می شوند. مانند عملگر ترکیب، در صورت به وجود آمدن حالت تساوی از بردارهای والدین استفاده

خواهیم کرد. جایگزینی بردارهای جدید با والدین از طریق فرمول های زیر صورت خواهد گرفت:

$$S_v^{(t+1)} = \underset{x=\{S_v^{(t+1)},w_1,w_2\}}{argmax} \; f(x) \qquad \text{-3)}$$

$$S_u^{(t+1)} = \underset{x=\{S_u^{(t+1)},w_1,w_2\}}{argmax} \; f(x) \qquad \text{-3)}$$

شبه کد مربوط به الگوریتم مدل بر اساس الگوریتم های ژنتیک ارتقاء یافته (EGADM) در زیر نشان داده شده است.

| **Algorithm 3. EGADM** | |
|---|---|
| **1** | **Input:** Initialize state vectors of nodes $u$ and $v$ to $S_v$ and $S_u$. |
| **2** | **Output:** New state vector for nodes $u$ and $v$. |
| **3** | **Repeat** |
| **4** | Set $S_v^{(t+1)} = S_v^{(t)}$ and $S_u^{(t+1)} = S_u^{(t)}$. |
| **5** | Select a random crossover point c between $[1, \beta]$. |
| **6** | Create $y_1$ and $y_2$ by swapping the tails of $S_v^{(t)}$ and $S_u^{(t)}$ where the tail is defined as all positions including and after index c. |
| **7** | Update state vectors: According to 3-4 and 3-5 |
| **8** | Do mutation on each bit of $S_v^{(t+1)}$ and $S_u^{(t+1)}$ based on mutation probability $P_m$. Create $w_1$ and $w_2$. |
| **9** | Update state vectors: According to 3-6 and 3-7 |
| **10** | **Until** (All interactions are checked) |

**الگوریتم 3-3- الگوریتم مدل EGADM**

## 3-7-2- الگوریتم دوم: مدل بر اساس الگوریتم های بهینه سازی ازدحام ذرات (PSODM) [49]

همانطور که پیش تر نیز اشاره کردیم، روش GADM و به تبع آن روش EGADM، که در قسمت قبل به آن اشاره شد، دارای یک مشکل اساسی هستند و تنها ارتقاء GADM نمی تواند به تنهایی مدل مناسبی برای انتشار اطلاعات باشد. مشکل اصلی در این روش ها عدم توانایی در مدل سازی واحد های چندگانه ی اطلاعاتی است. دلیل بروز این مشکل در این دو روش استفاده از عملگر ترکیب و در نظر گرفتن الگوهای

تولید شده توسط HDF به عنوان واحد های اطلاعاتی است و در نتیجه اگر یک کروموزوم چند الگو را شامل شود به عنوان یک کروموزوم که حاوی چند واحد اطلاعاتی است در نظر گرفته می شود. هنگامی که دو رأس از یک شبکه ی اجتماعی با بردارهای وضعیت خاص خود، که همان کروموزوم ها هستند، با یکدیگر ارتباط برقرار نمایند، عملیات ترکیب روی آن ها صورت می گیرد و دنباله های این دو کروموزوم با یکدیگر جا به جا می شوند. طبق الگوریتم در صورتی که رشته های حاصل از این ترکیب میزان برازندگی بیشتری نسبت به والد خود داشته باشند جای آن را در جمعیت خواهند گرفت. هنگام انجام عملیات ترکیب، در صورتی که نقطه ی در نظر گرفته شده جهت انجام این عملیات در محلی قرار گیرد که دنباله ی به وجود آمده اطلاعات کمتری نسبت به دنباله ی جدید داشته باشد، این دنباله به کلی با دنباله ی جدید جایگزین خواهد شد حال آنکه ممکن است دنباله ی قدیمی تناقضی با دنباله ی جدید نداشته باشد. به طور مثال فرض کنید که تابع HDF وجود دارد که رشته هایی با طول 10 را به عنوان ورودی می پذیرد. اگر این تابع شامل دو الگوی زیر باشد:

110 ******* score is 6

001 ******* score is 9

همچنین با رخ دادن عملگر ترکیب در نقطه ی پنج رشته ی زیر:

1101000111

رشته ی زیر حاصل شود:

0010000111

رشته ی جدید ممکن است حاوی اطلاعات بیشتری باشد که با شامل شدن الگوی دوم به آن می رسد اما در حین این ترکیب ساختار رشته ی اصلی تغییرات اساس کرده و بیت های چهارم و پنجم آن که تناقضی با رشته ی جدید نداشته اند از بین رفته اند.
برای رفع این مشکل و همچنین ارائه ی مدلی کارا تر نسبت به مدل موجود، ما از الگوریتم های بهینه سازی ازدحام ذرات به عنوان اساس کار خود استفاده

می نمایم. نام الگوریتم جدید برگرفته شد از ابتدای کلمات Particle Swarm Optimization Diffusion Model یا به طور خلاصه PSODM می باشد. هدف اصلی استفاده از ایده ی به روز رسانی نسل ها در این الگوریتم ها است و همانطور که مشخص است ما قصد نداریم که جواب بهینه برای مسئله ی خاصی بیابیم. در ادامه به توضیح این الگوریتم خواهیم پرداخت.

### 3-7-2-1- واحد های اطلاعاتی در PSODM

برای نمایش واحد های چندگانه ی اطلاعاتی در الگوریتم PSODM، همانند GADM، ما از توابع HDF استفاده می نماییم. به اینصورت که با استفاده از HDF رشته هایی تولید خواهیم کرد که این رشته ها نقش الگو های اصلی یا همان واحد های اطلاعاتی را برای ما بازی خواهند کرد. همانند قسمت قبل فرض می کنیم که تابع زیر رشته هایی با طول 10 را تولید می نماید. تفاوت اساسی در این روش این است که ما از HDF به عنوان تابع ارزیابی برازندگی استفاده نخواهیم کرد و در واقع میزان برازندگی بردارهای وضعیت تر رأس با استفاده از توابع مجزا محاسبه خواهد شد.

| | |
|---|---|
| *01******* | *score 2, order 1 schema* |
| *****110** | *score 2, order 1 schema* |
| ********10 | *score 3, order 1 schema* |
| *01**110** | *score -4, order 2 schema* |
| *01*****10 | *score 4, order 2 schema* |

### 3-7-2-2- شیوه ی نمایش[1] بردارهای وضعیت در PSODM

برخلاف روش های قبل، در PSODM ما از شیوه ی نمایش جدیدی استفاده می نماییم. در این روش بردارهای وضعیت رئوس گوناگون تنها با استفاده از دنباله ای از صفر و یک ها نمایش داده نخواهند شد بلکه هر خانه[2] ی بردار وضعیت به صورت عددی در بازه ی [0,1] خواهد بود. به این ترتیب می توان به ویژگی مهمی

[1] Don’t care

[2] Index

که در روش های پیشین امکان استفاده از آن وجود نداشت دست یافت که اطلاعات جزئی[1] نامیده می شود.
فرض کنید که واحد های اطلاعاتی که در قسمت قبل به آن ها اشاره نمودیم نشان دهنده ی زبان های مختلف برنامه نویسی مانند C، C++، JAVA و غیره باشند. می توان این حالت را در نظر گرفت که شخص خاصی هیچ نوع اطلاعاتی در مورد این زبان ندارد، اما در صورتی که فردی اطلاعات و دانشی را در این زمینه داشته باشد حالت دیگری به جز حالت داشتن تمامی دانش در مورد این زبان ها لازم خواهد بود. به طور دقیق تر اینکه ممکن است افراد زیادی وجود داشته باشند که اطلاعات آن ها در مورد زبان های برنامه نویسی به صورت صفر مطلق و عدم دانش یا یک مطلق و دانش کامل نباشد و تنها بخشی از اطلاعات لازم را در اختیار داشته باشند.
بدین ترتیب استفاده از شیوه ی جدید نمایش بردارهای وضعیت این امکان را برای ما فراهم می کند که بتوانیم مسئله ی اطلاعات جزئی را نیز برای مدل سازی انتشار اطلاعات در نظر بگیریم همانطور که در جوامع و شبکه های اجتماعی امروزی شاهد این امر نیز هستیم. برای مثال بردار زیر، نشان دهنده ی بردار وضعیت یک رأس می باشد:

0.5 0 0.6 0.67 1 0 1 0.23 0 0.87

با توجه به توضیحات ارائه شده ما نیازمند تابع برازندگی ویژه ای برای اندازه گیری امتیاز برازندگی هر رأس در مراحل مختلف اجرای الگوریتم خواهیم داشت.

### 3-7-2-3- تابع برازندگی در PSODM

توابع HDF نمی توانند کارائی مناسب جهت محاسبه ی امتیاز رشته های تولید شده توسط الگوریتم PSODM را داشته باشند. این توابع به نحوی طراحی شده اند که بتوانند رشته های باینری را به عنوان ورودی دریافت نموده و با توجه به مقایسه ی آن ها با الگو های مختلف تولید شده امتیاز آن ها را محاسبه

[3] Partial Information

نمایند حال آن که در شیوه ی جدید نمایش اعداد به صورت اعشاری در بازه ی [0,1] قرار می گیرند.
روش های گوناگونی را می توان برای محاسبه ی امتیاز یک بردار وضعیت استفاده نمود. روش استفاده شده در این قسمت که به صورت تجربی بهترین کارائی را در الگوریتم PSODM داشته است مراحل زیر را طی می کند:

1. محاسبه ی اختلاف میان هر عضو از بردار وضعیت یک رأس با عضو معادل آن در هر یک از الگوهای تولید شده توسط HDF. هدف از بدست آورن این اختلاف محاسبه ی فاصله میان حالت مطلوب (عضو موجود در الگو) با حالت کنونی (عضو موجود در رشته ی ورودی) است. پس برای مثال اگر عضو موجود در الگو عدد صفر باشد هرچه عدد موجود در رشته ی ورودی به صفر نزدیک تر باشد حالت مطلوب تری خواهیم داشت و در مجموع رشته ی ورودی امتیاز بیشتری را کسب می نماید و اگر عدد موجود در الگو یک باشد هر چه این عدد به یک نزدیکتر باشد مطلوب تر خواهد بود. همچنین این اختلاف برای حالات غیر مهم ( don't care) محاسبه نمی شود. اگر الگو و بردار وضعیت به صورت زیر باشند:

$$schema{:}*01**\quad score{:}\,4$$
$$state\ vector{:}*xy**$$

اعداد مورد نظر ما $|0-x|$ و y خواهند بود.

2. پس از محاسبه ی تمامی اختلاف ها و بدون در نظر گرفتن حالات صفر (هنگامی که هر دو عضو متناظر برابر باشند) امتیازی که رشته ی ورودی از یک الگوی خاص کسب می نماید را بدست می آوریم. در واقع رشته ی ورودی درصدی از امتیاز الگو را به خود اختصاص خواهد داد. برای این امر از حاصل ضرب اختلاف های بدست آمده استفاده خواهیم کرد. با ضرب این عدد در امتیاز الگوی مورد نظر، امتیازی کسب شده برای رشته ی ورودی از آن الگو بدست خواهد آمد. در مورد مثال قبل:

$$|0-x|\times y=w$$

عدد بدست آمده همانند وزن عمل می کند و با ضرب آن در امتیاز الگو، امتیاز کسب شده برای رشته ی مورد نظر حاصل می شود:

$$state\ vector\ score = w \times schema\ score$$ -3)

این فرآیند به تعداد الگوهای موجود برای هر رشته ی ورودی صورت می گیرد و امتیاز نهایی نیز با توجه به فرمول زیر به دست می آید:

$$total\ score = total\ score + schema\ score$$ -3)

شبه کد مربوط به این تابع در زیر آمده است.

| **Algorithm 4. PSODM Objective Function** | |
|---|---|
| **1** | **Input** State Vector |
| **2** | **Output** Vector Score |
| **3** | **For** each schema |
| **4** | Calculate differences |
| **5** | Calculate Weight |
| **6** | Calculate state vector score according to 3-8 |
| **7** | Calculate total score according to 3-9 |
| **9** | **End For** |

**الگوریتم 3-4- الگوریتم تابع برازندگی مدل PSODM**

### 3-7-2-4- الگوریتم PSODM

پس از معرفی عناصر لازم برای PSODM هم اکنون می توانیم الگوریتم کلی و نحوه ی کار این روش را بیان نماییم. پیش از آن فرضیات زیر می بایست در نظر گرفته شوند:

1. در این الگوریتم هر رأس به عنوان یک ذره در نظر گرفته می شود. تعداد ذرات به تعداد رأس ها خواهد بود.
2. بردار وضعیت هر رأس نشان دهنده ی مکان آن در الگوریتم بهینه سازی ازدحام ذرات است.
3. ذرات به صورت خودخواهانه سعی در افزایش اطلاعات خود دارن، لذا مقدار pBest در زمان $t$ برابر با وضعیت ذره در همان زمان خواهد بود.

4. هر ذره یا رأس تنها با رأس هایی که بین آن ها یال وجود دارد در ارتباط است لذا gBest می بایست از میان همسایه ها انتخاب شود.

با توجه به این پیش فرض ها هم اکنون می توانیم روند کلی الگوریتم را بررسی کنیم. در ابتدا به هر رأس $v$ یک بردار وضعیت اختصاص داده می شود مانند $S_v$ که این بردار با اعداد در بازه ی [0,1] به صورت تصادفی مقدار دهی می شود:

$$S_v[i] = random(0,1) \qquad \text{-3)}$$

پس از مقدار دهی اولیه ی بردار وضعیت رأس ها، مکان هر رأس برابر با بردار وضعیت آن قرار خواهد گرفت:

$$Position_v = S_v \qquad \text{-3)}$$

پس از آن تمامی یال ها با توجه به زمان ورود مورد بررسی قرار می گیرند. با بررسی هر یال که در واقع بررسی یک ارتباط و تبادل اطلاعات خواهد بود، بردار وضعیت یا همان سطح اطلاعات هر رأس با توجه به فرمول زیر به روز رسانی می شود:

$$Velocity_v^{(t+1)}[i] = Velocity_v^{(t)}[i] + \left(C_1 * R_1 * \left(pBest[i] - Position_v^{(t)}[i]\right)\right) + \left(C_2 * R_2 * \left(gBest[i] - Position_v^{(t)}[i]\right)\right) \qquad \text{-3)}$$

همانطور که قبلا ذکر شد pBest برای هر ذره معادل وضعیت فعلی آن خواهد بود لذا قسمت دوم این معادله همواره صفر است و فرمول به صورت زیر در خواهد آمد:

$$Velocity_v^{(t+1)}[i] = Velocity_v^{(t)}[i] + \left(C * R * \left(gBest[i] - Position_v^{(t)}[i]\right)\right) \qquad \text{-3)}$$

پس از اینکه سرعت ذره در وضعیت فعلی با فرمول 3-12 محاسبه گردید، مکان ذره که در واقع همان میزان اطلاعات آن خواهد بود نیز به روز رسانی می شود:

$$Position_v^{(t+1)}[i] = Position_v^{(t)}[i] + Velocity_v^{(t+1)} \qquad \text{-3)}$$

مکان به روز رسانی شده ی یک ذره در واقع بردار وضعیت جدید آن خواهد بود. این بردار به تابع برازندگی ارسال می شود تا با توجه به الگوریتم توضیح داده شده در قسمت قبل امتیاز این بردار به دست آید. در صورتی که این تغییر مکان باعث افزایش سطح اطلاعاتی رأس شود مکان ذره به روز رسانی می شود و در غیر اینصورت مکان ذره ثابت باقی خواهد ماند. اگر بخواهیم فرمول 3-12 را بهتر بررسی نماییم اینگونه می توان توضیح داد که تغییر سرعت در واقع میزان تغییر سطح اطلاعاتی را نشان می دهد. این میزان تغییر حاصل در نظر گرفتن سطح تغییرات قبلی و تاثیر بهترین همسایه خواهد بود. پس در واقع در این الگوریتم ذرات با توجه به پیشینه ی خود و بهترین همسایه های خود اطلاعات کسب می نمایند. این روند در دنیای واقعی نیز رخ می دهد و افراد به طور طبیعی بیشترین تلاش خود را در کسب دانش از بهترین منابع اطلاعاتی و افراد در ارتباط با خود می نمایند. بدین ترتیب بسیاری از ارتباطات بی حاصل با سایر همسایه ها نیز حذف خواهند شد. این نکته را نیز می بایست در نظر داشت که بهترین همسایه نیز ممکن است به مرور زمان تغییر نماید چرا که همسایه های یک رأس نیز به نوبه ی خود در پی افزایش میزان اطلاعات خود می باشند. در نهایت پس از بررسی تمامی ارتباطات انجام شده، سطح دانشی رأس ها تعیین می گردد. نمودار ها و نتایج تجربی در ان زمینه در فصل چهارم ارائه خواهد گردید. شبه کد زیر الگوریتم PSODM را نشان می دهد:

| **Algorithm 5. PSODM** | |
|---|---|
| 1 | **Input:** Initialize state vectors of nodes $u$ and $v$ to $S_v$ and $S_u$. |
| 2 | **Output:** New state vector for nodes $u$ and $v$. |
| 3 | Set $Position_v^{(t)} = S_v^{(t)}$ and $Position_u^{(t)} = S_u^{(t)}$ for all nodes. |
| 4 | **Repeat** |
| 5 | Update $Velocity_v$ and $Velocity_u$ according to 3-13 |
| 6 | Update $Position_v$ and $Position_u$ according to 3-14 |
| 7 | **Until** (All interactions are checked) |

**الگوریتم 3-5- الگوریتم مدل PSODM**

## 3-8-روش پیشنهادی جهت مقایسه ی مدل های انتشار اطلاعات [24و25]

همانطور که در قسمت های پیشین ملاحظه کردید، با توجه به مشکلاتی که در خصوص مدل های مختلفی که تاکنون برای انتشار اطلاعات ارائه شده اند وجود دارد ما دو مدل پیشنهادی ارائه نمودیم. در این قسمت از پایان نامه روش نوینی را ارئه خواهیم کرد که با استفاده از آن می توان مدل های ارائه شده را با یکدیگر مقایسه نمود.
در این روش رأس های شبکه ی اجتماعی به عنوان بازیگر های[1] یک بازی[2] در نظر گرفته می شوند که هر یک سعی در بیشینه کردن یک تابع منفعت[3] دارند. در اینجا تابع منفعت بدست آوردن حداکثر اطلاعات خواهد بود. با در نظر گرفتن این نکته، افراد سعی در تشکیل اتحاد با سایرین خواهند داشت که این امر موجب شکل گیری تشکل ها می شود. با استفاده از این الگوریتم می توان ساختار تشکل ها در شبکه های اجتماعی را نیز بدست آورد. پیش از این الواری و سایرین [24] از نظریه ی بازی ها برای تشخیص تشکل ها استفاده کرده بودند. الگوریتم ارائه شده با استفاده از تابع منفعت خود و پس از اجرا به حالتی می رسد که اثبات می شود که یک حالت تعادلی نش محلی[4] است. با رسیدن به این تعادل ساختار تشکل ها شکل می گیرد و هر رأس با توجه به حالت خود خواهانه به یک تشکل تعلق می گیرد.
ما در این قسمت برای این الگوریتم تابع منفعت جدیدی را ارئه می دهیم. در ابتدای کار و هنگام اجرای مدل انتشار اطلاعات، میزان اطلاعاتی را که میان رأس ها انتقال پیدا می کند ذخیره می نماییم. این تغییرات اطلاعات در آرایه ای به نام ***I*** ذخیره می گردند:

---

[1] Player
[2] Game
[1] Utility Function
[2] Local Nash Equilibrium (NE)

$$I = \frac{I^{(1)}+\cdots+I^{(n)}}{n} = \begin{bmatrix} \frac{I_{11}^{(1)}+\cdots+I_{11}^{(n)}}{n} & \cdots & \frac{I_{1n}^{(1)}+\cdots+I_{1n}^{(n)}}{n} \\ \vdots & \ddots & \vdots \\ \frac{I_{n1}^{(1)}+\cdots+I_{n1}^{(n)}}{n} & \cdots & \frac{I_{nn}^{(1)}+\cdots+I_{nn}^{(n)}}{n} \end{bmatrix} \qquad \text{-3)}$$

این اطلاعات ذخیره شده و با توجه به این آرایه، تابع منفعتی که در حین اجرای بازی به کار برده می شود به صورت زیر تعریف می گردد:

$$U_i(S) = \frac{1}{m}\sum_{j=1, j\neq i}^{n} I_{ij}\delta_{ij} \qquad \text{-3)}$$

در اینجا $S$ نشان دهنده ی استراتژی است که هم اکنون در بازی به کار گرفته شده و $U_i(S)$ نیز میزان منفعت را برای عامل $i$اُم و در صورت اتخاذ استراتژی $S$ نشان می دهد. این استراتژی می تواند پیوستن به یک تشکل (Join)، جدا شدن از آن (Leave) و یا باقی ماندن در وضعیت فعلی (No Operation) باشد. برای یافتن بهترین استراتژی $S_i$ برای رأس $i$ با فرض اینکه سایر رئوس نیز استراتژی خود را انتخاب نموده اند $S_{-i}$، به اینصورت عمل می نماییم:

$$argmax_{s_i'\subseteq[k]} U_i(S_{-i}, s_i') \qquad \text{-3)}$$

رسیدن به نقطه ی تعادلی نش در این بازی در صورتی رخ خواهد داد که با تکرار الگوریتم به حالتی برسیم که تمامی رئوس بهترین استراتژی خود را انتخاب کرده باشند به این شرط که هیچ یک از آن ها تمایلی به تغییر وضعیت کنونی نداشته باشد:

$$\forall i, s_i' \neq s_i, U_i(S_{-i}, s_i') \leq U_i(S_{-i}, s_i) \qquad \text{-3)}$$

نکته ی مهم در مورد این بازی که در نظر گرفته ایم این است که نمی توانیم با این تعداد رأس و این استراتژی ها به حالت تعادلی نش سراسری برسیم و ثابت می شود که این بازی به حالت تعادلی نش محلی می رسد. در این حالت $S$ تشکیل دهنده ی حالت تعادلی نش محلی خواهد بود اگر تمامی رئوس به استراتژی بهینه ی محلی[1] خود رسیده باشند. در اینجا $ls(s_i)$ به

[1] Local Optimal Strategies

فضای محلی استراتژی ها اشاره می نماید:

$$\forall i, s_i' \in ls(s_i), U_i(S_{-i}, s_i') \le U_i(S_{-i}, s_i) \quad \text{-3)}$$

به طوری که مشخص است یک عامل به سمت تشکل هایی تمایل پیدا خواهد کرد که بتوانند در آن ها حداکثر میزان اطلاعات را کسب کند. این امر می تواند به نحوی شبیه سازی دنیای واقعی اطراف ما باشد. افراد در دنیای واقعی به سمت جمعیت هایی میل خواهند داشت که بتوانند در آن ها بیشترین سود و منفعت را کسب نمایند. همچنین باید در نظر داشت که استفاده از این کاربرد مدل های انتشار اطلاعات می تواند ما را در تحلیل محیط های اطراف و ارائه ی مدل های دقیقی تری برای فرآیند های پیچیده ی اطلاعاتی محیط اطرافمان یاری دهد. شبه کد مربوط به این الگوریتم که با نام GID ارئه شده است [25] را در زیر مشاهده می نمایید:

| **Algorithm 6. GID** | |
|---|---|
| **1** | **Input:** underlying network graph *G*. |
| **2** | **Output:** *community* as a final division of *G*. |
| **3** | Initialize each node of *G* as a selfish agent. |
| **4** | Initialize *community* as a set of all communities. |
| **5** | **I** = IDM (*G*) //Create Information Matrix based on EGADM or PSODM |
| **6** | **Repeat** |
| **7** | Choose a random agent fairly from pool of agents. |
| **8** | Choose the best operation among *join*, *leave* or *no operation*. |
| **9** | **Until** (local Nash equilibria is reached) |

**الگوریتم 3-6- الگوریتم روش GID**

با توجه به اینکه ما در روش های خود از ترکیب Game و الگوریتم های GADM، EGADM و PSODM استفاده نموده ایم در این قسمت سه الگوریتم تشخیص تشکل ها بر اساس ترکیب این ها را به ترتیب GGADM، GEGADM و GPSODM می نامیم.

## 3-9-جمع بندی

در این قسمت از پایان نامه دو روش پیشنهادی جهت

مدل سازی انتشار اطلاعات در شبکه های اجتماعی ارائه گردید. روش اول بر مبنای یکی از جدید ترین مدل های ارائه شده در این زمینه عمل می کند و در واقع به نوعی مدل تعمیم یافته ی آن است و در روش دوم نیز ما با استفاده از الگوریتم های بهینه سازی ازدحام ذرات سعی در مدل سازی انتشار اطلاعات می نمایید.

مدل اول یا همان EGADM که مخفف کلمه ی Extended GADM است، به نحوی طراحی شده تا بتواند ساختار پیچیده تری نسبت به روش GADM را مدل سازی کند. در این روش ما با افزودن عملگر جهش سعی داریم که تلاش افراد در شبکه های اجتماعی را برای ارتقاء سطح دانش خود نمایش دهیم. به اینصورت که اصولا افراد همواره علاوه بر اطلاعات و دانشی که از سایرین دریافت می نمایند خود نیز به کسب اطلاعات و دانش خواهند پرداخت.

در مدل دوم یا PSODM ما روش نوینی را برای مدل سازی انتشار اطلاعات در شبکه های اجتماعی ارائه می دهیم. در این روش پایه و اساس کار الگوریتم های بهینه سازی ازدحام ذرات خواهند بود که به ما کمک می کنند تا بتوانیم عملگر های مناسبی را جهت انتقال اطلاعات میان رأس های یک شبکه به کار گیریم. استفاده از این عملگر ها مشکلات مطرح شده در مورد سایر روش ها را در مورد نبود قابلیت مدل سازی واحد های چندگانه ی اطلاعاتی برطرف می نماید. همچنین این امکان برای ما فراهم می شود تا شیوه ی نمایش پیچیده تری را برای نشان دادن اطلاعات یک رأس به کار بگیریم.

در نهایت برای اولین بار و به صورت کاملا کاربردی از مسئله ی تشخیص تشکل ها در شبکه های اجتماعی برای مقایسه مدل ها استفاده می نماییم. سه روش GGADM، GEGADM و GPSODM حاصل این قسمت هستند. این امر برای ما امکان بهبود مدل ها را نیز فراهم می آورد.

# فصل چهارم

# نتایج تجربی

## 4-1- مقدمه

همانطور که در قسمت های قبل نیز اشاره شد مدل های گوناگونی با توجه به کاربرد های خاص برای انتشار اطلاعات در شبکه های اجتماعی ارائه شده است. نکته ی مهمی که در این مورد باید به آن توجه کرد عدم وجود یک روش مناسب جهت مقایسه ی این مدل ها است. ما در این پایان نامه از مبحث تشکل های اجتماعی برای مقایسه ی مدل های مطرح در این زمینه استفاده می نماییم. در واقع برای اولین به مقایسه ی این مدل ها خواهیم پرداخت. دلیل کارا بودن این روش خاصیت شبکه های اجتماعی است. در شبکه های اجتماعی افراد برای به دست آوردن اطلاعات، اخبار و غیره به عضویت تشکل های گوناگون در می آیند، حال اگر ما بتوانیم مدل کاراتری برای انتشار اطلاعات در این شبکه ها ارائه دهیم در واقع توانسته ایم تشکل های موجود در شبکه های اجتماعی را نیز بهتر تشخیص دهیم و نتایج به دست آمده به نتایج دنیای واقعی نزدیک تر خواهند بود.
در فصل سوم الگوریتم پیشنهادی برای مقایسه ی مدل های اخیر انتشار اطلاعات را معرفی نمودیم و در این فصل برای مقایسه ی مدل های معرفی شده از این الگوریتم استفاده می نماییم. به طور کلی، بررسی کارایی هر الگوریتم پیشنهادی برای تشخیص تشکل ها در شبکه های اجتماعی، با توجه به این که در اکثر موارد، ساختار تشکل واقعی[1] در دسترس نیست، کاری بس دشوار و غیر علمی بنظر می آید. در عمل، شبکه های اجتماعی موجود، دائما در حال تغییر بوده و در نتیجه ساختار ثابت و از پیش تعیین شده ای ندارند. این امر، چالشی بزرگ بر سر راه تحلیل ساختار شبکه های اجتماعی محسوب شده و نیاز برای معرفی معیارهای ارزیابی کاراتر را دوچندان می کند. در واقع، این معیارها باید طوری طراحی شوند که برای ارزیابی روش پیشنهادی، حتی الامکان بدون نیاز به مقایسه با ساختار واقعی عمل کنند. در این بخش، نتایج آزمایشات تجربی برای نشان دادن برتری روش های پیشنهادی با دیگر روش های موجود نشان داده خواهد شد. برای این کار، از مجموعه داده های

[1] Ground truth

مصنوعی *GN*، *LFR* و *Erdös-Réyni* و مجموعه داده های واقعی *Dolphin*، *Zachary Karate Club* و *American College Football* استفاده کرده ایم که در ادامه آن ها را معرفی خواهیم کرد. همچنین برای نمایش توانمندی الگوریتم در تشخیص تشکل ها در شبکه های اجتماعی امروزی که از حجم بالای از اطلاعات برخوردارند نتایج خود روی مجموعه داده های *Flickr* و *BlogCatalog* را نمایش خواهیم داد که این دو مجموعه نیز زیر مجموعه ی داده های واقعی به حساب می آیند. لازم به ذکر است که روش های پیشنهادی ما در *JAVA* و بر روی سیستمی با مشخصات *Processor Intel(R) Corei5 2.53GH* و *4 GB RAM* پیاده سازی و اجرا شده اند.

## 4-2- مجموعه داده ها

در این قسمت به معرفی اجمالی مجموعه داده های استفاده شده و ویژگی های مورد نیاز هر یک از آن ها می پردازیم.

### 4-2-1- مجموعه داده های مصنوعی

همانطور که در مقدمه ذکر شد، با توجه به اینکه ساختار تشکل اغلب شبکه های اجتماعی واقعی در دسترس نیست، نمی توان از این داده ها برای بررسی کارایی الگوریتم پیشنهادی استفاده نمود. به همین دلیل، محققان اغلب به فکر ایجاد مجموعه داده های مصنوعی هستند. به همین منظور، تعدادی از این شبکه های مصنوعی تولید شده و مورد استقبال زیادی قرار گرفته اند. در ادامه تعدادی از مهمترین آن ها را بررسی خواهیم کرد.

#### 4-2-1-1- شبکه های *GN*

این روش برای تولید شبکه های اجتماعی در واقع مشهورترین روش در این زمینه به حساب می آید [38]. از گذشته تاکنون غالب الگوریتم های ارائه شده در زمینه ی تشخیص تشکل ها از این مجموعه جهت نمایش کارایی خود استفاده نموده اند. در این مجموعه ما چهار تشکل داریم و تعداد 128 رأس که در این چهار تشکل حضور دارند. هر یک از رئوس این شبکه درجه ی 16 دارد و در واقع تمامی رئوس هم

درجه هستند. در این شبکه ها از پارامتری با نام $\boldsymbol{\mu}$ برای تنظیم نسبت تعداد یال های متقاطع به تعداد کل یال ها استفاده می نماییم. هر چه مقدار این پارامتر بیشتر باشد تشخیص تشکل ها نیز در آن ها سخت تر خواهد بود.

### 4-2-1-2- شبکه های *LFR*

برای تولید گراف های مصنوعی با ساختار تشکل از پیش معلوم، روشی توسط لانسیچینتی و فورتوناتو ارائه شد [39] که قابلیت تولید گراف هایی با اندازه های دلخواه و همراه با ساختار تشکل آن ها را دارد. پیچیدگی محاسباتی این روش، خطی و بر اساس تعداد یال ها بوده و گراف هایی را به ترتیب قدم های زیر تولید می کند:

1. تعداد تشکل هایی که هر رأس باید عضو آن ها باشد تولید کرده و به هر رأس بر اساس توزیع قانون توانی با نمای $\boldsymbol{\tau_1}$ درجه نسبت می دهد.
2. برای تعداد ثابتی از تشکل ها، اندازه های آن ها را با یک توزیع قانون توانی دیگر با نمای $\boldsymbol{\tau_2}$ نسبت می دهد.
3. گراف دو بخشی بین رأس ها و تشکل ها را بر اساس مدل *configuration* [45] را تشکیل می دهد.
4. برای هر رأس، بر اساس $\boldsymbol{\mu}$، درجه های داخلی[1] هر تشکل و همچنین درجه ی بین تشکل ها[2] را نسبت می دهیم.
5. گراف ها را برای هر تشکل و یال های بین تشکل ها را با توجه به مدل *configuration* می سازیم.

این گراف ها داری پارامترهایی هستند که پیش از تولید باید تنظیم شوند. ما در اینجا برای مقایسه با روش *Game* ( )، پارامترها را به صورت زیر مقدار دهی کرده ایم. مقادیر $\boldsymbol{s_{min}}$ و $\boldsymbol{s_{max}}$ به کمینه و بیشینه ی اندازه های تشکل ها اشاره می کنند که به ترتیب برابر با 10 و 50 و بار دیگر 20 و 100 می باشند. دیگر پارامترها شامل $\boldsymbol{\tau_1 = 2}$ ، $\boldsymbol{\tau_1 = 1}$

[1] Internal degrees
[2] Cross-community

$k_{avg} = 20$ ، $k_{max} = 50$ ، $om = 2$ ، $\mu$ نیز مابین [0و 0.8] ($\mu$ ، نسبت یال هایی است که یکدیگر را قطع می کنند به کل یال های موجود) می باشند.

#### 4-2-1-3- شبکه های تصادفی *Erdös-Réyni*

این شبکه دارای ساختار تشکل معین نبوده و ارتباط معناداری در بین رأس های آن مشاهده نمی شود. طبق ادعای محققانی نظیر *Lancichinetti* هر الگوریتم پیشنهادی برای شناسایی تشکل ها، بعد از اجرا بر روی این مجموعه داده، باید تنها یک تشکل با پوشش کلیه ی رئوس آن را شناسایی کند [39]. در واقع اگر الگوریتمی روابط معنا داری در این شبکه بیابد ار کارایی لازم جهت تشخیص تشکل ها برخوردار نیست.

### 4-2-2- مجموعه داده های واقعی

به علت اینکه همیشه نمی توان بر مبنای مجموعه داده های مصنوعی قضاوت کرد، تعدادی از مجموعه داده های واقعی نیز همواره مورد استفاده قرار می گیرند که در ذیل به برخی از آن ها اشاره خواهد شد. در این میان، تعدادی از آن ها، بسیار بزرگ و پویا بوده و تنها می توان به پاره ای از مشاهدات اکتفا کرد و برخی، بسیار کوچک و ایستا بوده و ساختار تشکل مشخصی دارند و بنابراین برای بررسی روش پیشنهادی مناسب ترند. ناگفته پیداست که همواره چالشی برای تمام روش های پیشنهادی درمورد اجرا بر روی داده های بسیار بزرگ وجود دارد.

#### 4-2-2-1- شبکه ی *Flickr*

این شبکه ی اجتماعی، یک وب سایت اشتراک گذاری محتوا[1] با تمرکز بر روی عکس بوده که کاربران در آن، قادر به ساخت پروفایل و بارگذاری عکس های خود هستند. با توجه به اینکه این مجموعه داده، بسیار بزرگ و همچنین پویا بوده، در اینجا به صورت تصادفی 195 گروه انتخاب کرده ایم[2] [40] که ویژگی های آماری آن را در جدول 4-1 می بینید.

---

[1] Content sharing

[2] www.socialcomputing.asu.edu/pages/datasets

جدول 4-1 ویژگی های آماری مجموعه داده *Flickr*

| تعداد *Category* ها | تعداد رئوس | تعداد یال ها | بیشترین درجه |
|---|---|---|---|
| **195** | **80513** | **5899882** | **5706** |

## 4-2-2-2- شبکه ی *BlogCatalog*

این شبکه در واقع از یک وب سایت اجتماعی به دست آمده است [41]. در این وب سایت هر کاربر می تواند برای خود یک وبلاگ زیر مجموعه ی موضوعات خاصی ایجاد نماید. صاحبان این وبلاگ ها می توانند از metadata جهت بهبود دسترسی افراد به سایت های خود استفاده نمایند. همچنین کاربران می توانند این وبلاگ های یکدیگر را نیز به اشتراک بگذارند. در این مجموعه 39 دسته ی علاقه مندی وجود دارد که هر وبلاگ می تواند شامل چندین دسته بشود.

جدول 4-2 ویژگی های آماری مجموعه داده *BlogCatalog*

| تعداد *Category* ها | تعداد رئوس | تعداد یال ها | بیشترین درجه |
|---|---|---|---|
| **39** | **10312** | **333983** | **3992** |

## 4-2-2-3- شبکه ی *Dolphin*

این شبکه شامل 62 رأس که هر کدام نماینده ی یک دلفین و 159 یال که ارتباط بین آن ها را نشان می دهد، می باشد [42] و به علت اینکه ساختار تشکل این شبکه معلوم است، در اغلب آزمایش های کارایی الگوریتم های تشخیص تشکل، از آن استفاده می شود. طبق مشاهدات انجام شده توسط دانشمندان این شبکه از دو تشکل ساخته شده و در واقع دلفین های مورد بررسی به دلایل نامعلومی در دو دسته قرار می گیرند.

## 4-2-2-4- شبکه ی *Zachary Karate Club*

شبکه ی *Karate* دارای 34 رأس و 78 یال بوده که نشان دهنده ی اعضای باشگاه کاراته ی *Zachary* و بعضی از مسابقاتی است که بین آن ها برگزار شده است. در شکل 4-1 ساختار تشکل واقعی این شبکه نشان داده شده است [43]. در این شبکه افراد به دو دسته

تقسیم می شوند که در وقع نشان دهنده ی افرادی هستند که با رئیس باشگاه ارتباط بهتری دارند و افرادی که با معاون وی در ارتباط هستند.

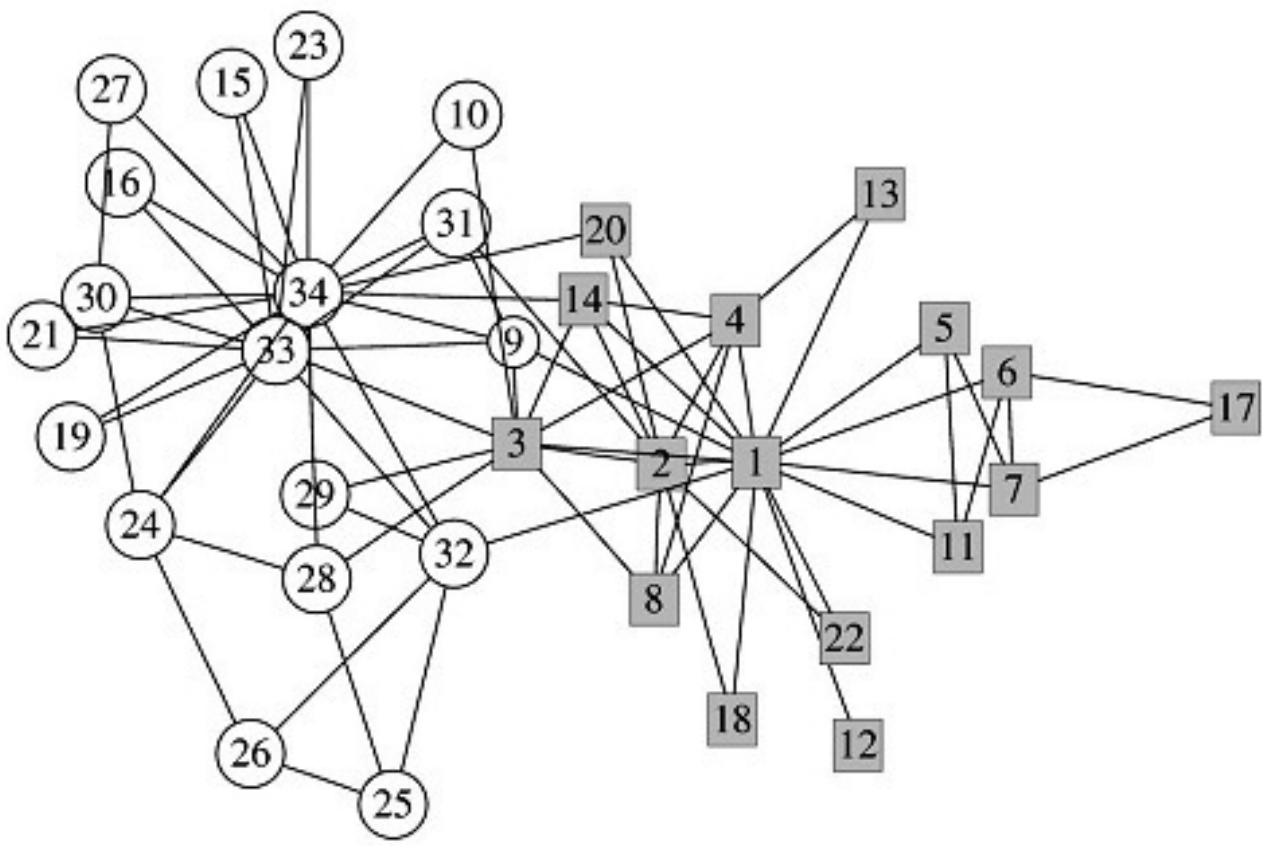

**شکل 4-1- ساختار تشکل واقعی شبکه ی *Karate***

### 4-2-2-5- شبکه ی *American College Football*

این شبکه دارای 115 رأس و 613 یال بوده که هر رأس نماینده ی یک تیم فوتبال و هر یال نشان دهنده ی بازی برگزار شده بین دو تیم می باشد و به طور کلی دارای 12 تشکل می باشد [44]. تشکل های این شبکه نشان دهنده ی کنفرانس ها هستند. در اینجا هر کنفرانس به معنی دسته ای از تیم ها است که در طول یک فصل با یکدیگر بازی می کنند. هرچند گاهی اوقات تعدادی بازی نیز مابین تیم های کنفرانس های مختلف برگزار می گردد.

## 4-3- معیارهای ارزیابی

همانطور که پیشتر نیز بیان شد، برای ارزیابی کارایی یک الگوریتم پیشنهادی برای شناسایی تشکل ها در شبکه های اجتماعی، همواره نیاز به تعریف معیاری است که نشان دهنده ی میزان شباهت ساختار تشکل شناسایی شده توسط الگوریتم و ساختار تشکل واقعی گراف مورد نظر است. برای این کار، معیارهای زیادی ارائه شده است که به طور کلی به سه دسته تقسیم می شوند: معیارهای بر مبنای شمارش زوج[1]، تطابق کلاستر[2] و تئوری اطلاعات[3].

[1] Pair counting

[2] Cluster matching

[3] Information theory

به طور کلی در مبحث شناسایی تشکل ها، بخش ناچیزی از تحقیقات به آزمایش کارایی الگوریتم ها اختصاص داده شده که این امر یک نقصان بزرگ محسوب می شود، چرا که اغلب مقاله ها بعد از پیشنهاد یک الگوریتم برای شناسایی تشکل ها، به علت کمبود معیارهای ارزیابی مطمئن، الگوریتم خود را بر روی مجموعه داده های بسیار کوچکی اجرا می کنند که ساختار تشکل آن ها به راحتی و به صورت چشمی قابل شناسایی باشد. در نتیجه تعداد الگوریتم های پیشنهادی برای شناسایی تشکل ها در شبکه های اجتماعی، به علت عدم مقایسه ی درست، آزادانه در حال زیاد شدن است.

در این پایان نامه، جهت بررسی ارزیابی الگوریتم های پیشنهادی خود، از معروف ترین معیارهای موجود یعنی *Normalized Mutual Information (NMI)*، *Fraction of Correctly Classified Nodes (FNCCN)* و *Modularity (Q)* استفاده کرده ایم.

### 4-3-1- معیار *Normalized Mutual Information*

ما در اینجا از نسخه ی ارتقا یافته ی NMI [45] استفاده کرده ایم که قادر به اجرا بر روی ساختارهای تشکل همپوشان را دارد و از روابط زیر محاسبه می شود.

$$N(X|Y) = 1 - \frac{1}{2}[H(X|Y)_{norm} + H(Y|X)_{norm}] \qquad (1-4)$$

این متغیر در بازه ی [0و1] قرار دارد و زمانی برابر 1 است که دو تشکل $C$ و $C'$ یکسان باشند. برای محاسبه ی این مقدار، به محاسبات زیر نیاز داریم:

$$H(X|Y)_{norm} = \frac{1}{|C'|}\sum_{k} H(X_k|Y)_{norm} \qquad (2-4)$$

که در آن، $\boldsymbol{H(X_k|Y)_{norm}}$ از رابطه زیر بدست می آید:

$$H(X_k|Y)_{norm} = \frac{H(X_k|Y)}{H(X_k)} \qquad (3-4)$$

و داریم:

$$H(X_k|Y)_{norm} = \min_{l \in \{1,2,\dots,|C''|\}} H(X_k|Y_l) \quad (4\text{-}4)$$

$$H(X_k|Y_l) = H(X_k, Y_l) - H(Y_l) \quad (5\text{-}4)$$

در این معادلات، $X_k = (X)_k$ و $H(X)$ و $H(Y)$ به ترتیب آنتروپی متغیر تصادفی $X$ و $Y$ می باشند. همچنین $H(X_k|Y)$ آنتروپی شرطی $X_k$ نسبت به همه ی اجزای تشکیل دهنده ی متغیر $Y$ است. در نهایت $H(X|Y)_{norm}$ آنتروپی شرطی نرمال سازی شده ی $X$ نسبت به $Y$ خواهد بود. برای جزئیات بیشتر به [45] رجوع شود.

## 4-3-2- معیار Fraction of Correctly Classified Nodes

برای محاسبه این معیار از روش موجود در [46] استفاده می کنیم که آن را در اینجا شرح داده ایم. اگرچه هنوز نسخه ای از این معیار ارائه نشده که مفهوم تشکل های همپوشان را پشتیبانی کند، اما معیار فعلی می تواند به عنوان تخمینی مناسب برای نشان دادن برتری یک الگوریتم ارائه شده باشد.
برای محاسبه این معیار به روش زیر عمل می کنیم.
در ابتدا برای یافتن یک تناظر یک به یک با بیشینه ی تعداد اعضای مشترک بین تشکل های تشخیص داده شده و تشکل های واقعی جستجو می کنیم و ایندکس تشکل یافت شده را برابر ایندکس تشکل واقعی متناظر با آن قرار می دهیم و این کار را تا زمانی ادامه می دهیم که تمام تشکل های یافت شده ایندکس دار شوند و یا اینکه ایندکسی برای ادامه ی کار باقی نماند.
در ادامه تعداد اعضای مشترک بین هر تشکل یافت شده ی ایندکس دار با تشکل واقعی متناظر با آن را شمارش کرده ($CCN$) و در نهایت آن ها را با یکدیگر جمع کرده و بر تعداد کل تشکل ها تقسیم می کنیم:

$$FCCN = \frac{\sum_{i \in C} CCN_i}{|C|} \quad (6\text{-}4)$$

## 4-3-3- معیار Modularity

این معیار برای مواقعی مفید است که ساختار واقعی تشکل برای مقایسه موجود نباشد و در نتیجه نتوان

از *NMI* استفاده کرد. اگرچه این معیار دارای مشکلاتی است و زمانی که شبکه ها بسیار تنک باشند غیر قابل اعتماد است [47-48]، اما بهرحال از محبوبیت و معروفیت زیادی برخوردار است. بالاتر بودن این مقدار، نشان دهنده ی نتایج بهتر است و از رابطه زیر بدست می آید:

$$Q = \sum_{s=1}^{k}[\frac{l_s}{l} - (\frac{d_s}{2L})^2] \qquad (7-4)$$

که در آن $k$ تعداد تشکل ها، $l_s$ تعداد یال های موجود در تشکل $s$، $d_s$ مجموع درجات رئوس موجود در تشکل $s$ و $L$ تعداد کل یال های گراف است.

## 4-4- نتایج و تحلیل ها

در این بخش به تحلیل نتایج حاصل از اجرای الگوریتم های ذکر شده در فصل 4 و با توجه به معیار های ارائه شده در بخش قبل فصل جاری می پردازیم.

### 4-4-1- مجموعه داده های مصنوعی

همانگونه که پیشتر نیز مطرح نمودیم، از آنجا که معمولا برای شبکه های اجتماعی ساختار تشکل ها مشخص نیست، ما از داده های مصنوعی جهت نمایش کاربردی بودن الگوریتم های خود استفاده می نماییم. در این بخش نتایج الگوریتم های ارائه شده توسط خود را با تعدادی از الگوریتم های مطرح حوزه ی تشخیص تشکل ها مقایسه می نماییم.

#### 4-4-1-1- شبکه های *GN*

شکل 4-2 نشان دهنده ی نتایج حاصل از الگوریتم های *GGADM*، *GEGADM* و *GPSODM* و همچنین الگوریتم های زیر روی مجموعه داده های *GN* باشد:

- *MMC*
- *LPA*
- *HA*

- *InfoMap*

این چهار الگوریتم تقریبا به صورت مشابه عمل می نمایند و تمامی آنها از الگوریتم های مطرح این حوزه می باشند. مهمترین الگوریتم مورد نظر ما نیز *MMC* می باشد چرا که از یک روش انتشار اطلاعات جهت تشخیص تشکل ها استفاده می نماید.

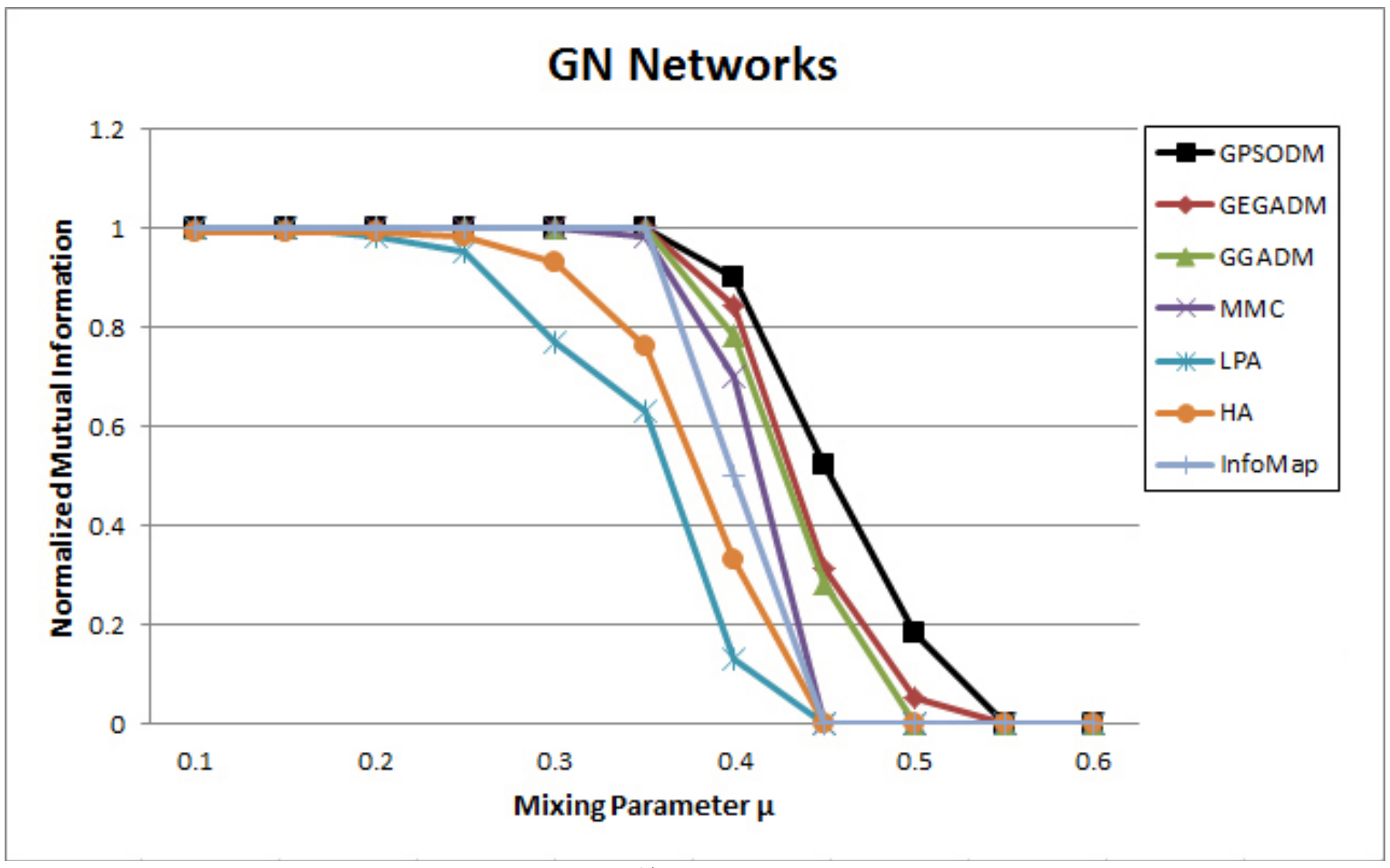

**شکل 4-2- نمودار *NMI* به دست آمده از الگوریتم های مختلف روی شبکه ی *GN***

همانطور که در نمودار می بینیم به ازای افزایش میزان پارامتر **μ** کارائی الگوریتم ها نیز کاهش یافته و *NMI* به سمت صفر میل می کند. تمامی الگوریتم هایی که عمل مقایسه را با آن ها انجام دادیم در نقطه ی به صفر می رسند و تنها سه الگوریتم *GGADM*، *GEGADM* و *GPSODM* برای مقادیر بیشتر به *NMI* بالای صفر می رسند.

## 4-4-1-2- شبکه های LFR

شبکه های *LFR* را پیشتر توضیح دادیم. این قابلیت وجود دارد که بتوانیم شبکه هایی با تعداد رئوس و سایز تشکل های دلخواه را با استفاده از الگوریتم ارائه شده ایجاد نماییم.
به همین منظور و برای مقایسه ی الگوریتم های پیشنهادی از دو شبکه ی *LFR* با تعداد 10000 رأس استفاده می نماییم. همچنین در این دو شبکه سایز تشکل ها را بین 20 و 50 به عنوان تشکل ها با سایز

کوچک و بین 20 و 100 به عنوان تشکل ها با سایز بزرگ در نظر می گیریم. شکل 4-3 نتایج الگوریتم ها را روی این دو شبکه نشان می دهد.

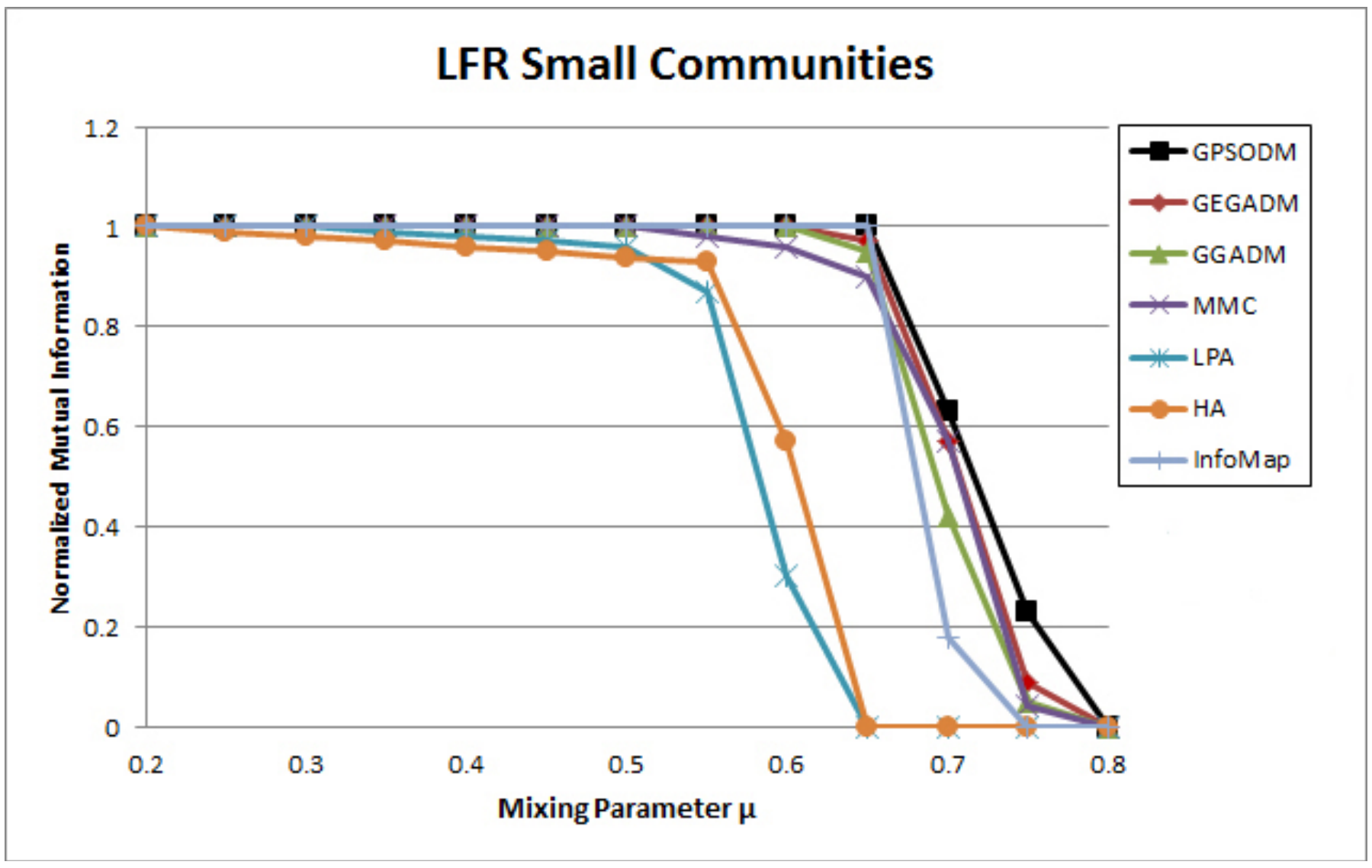

(الف)

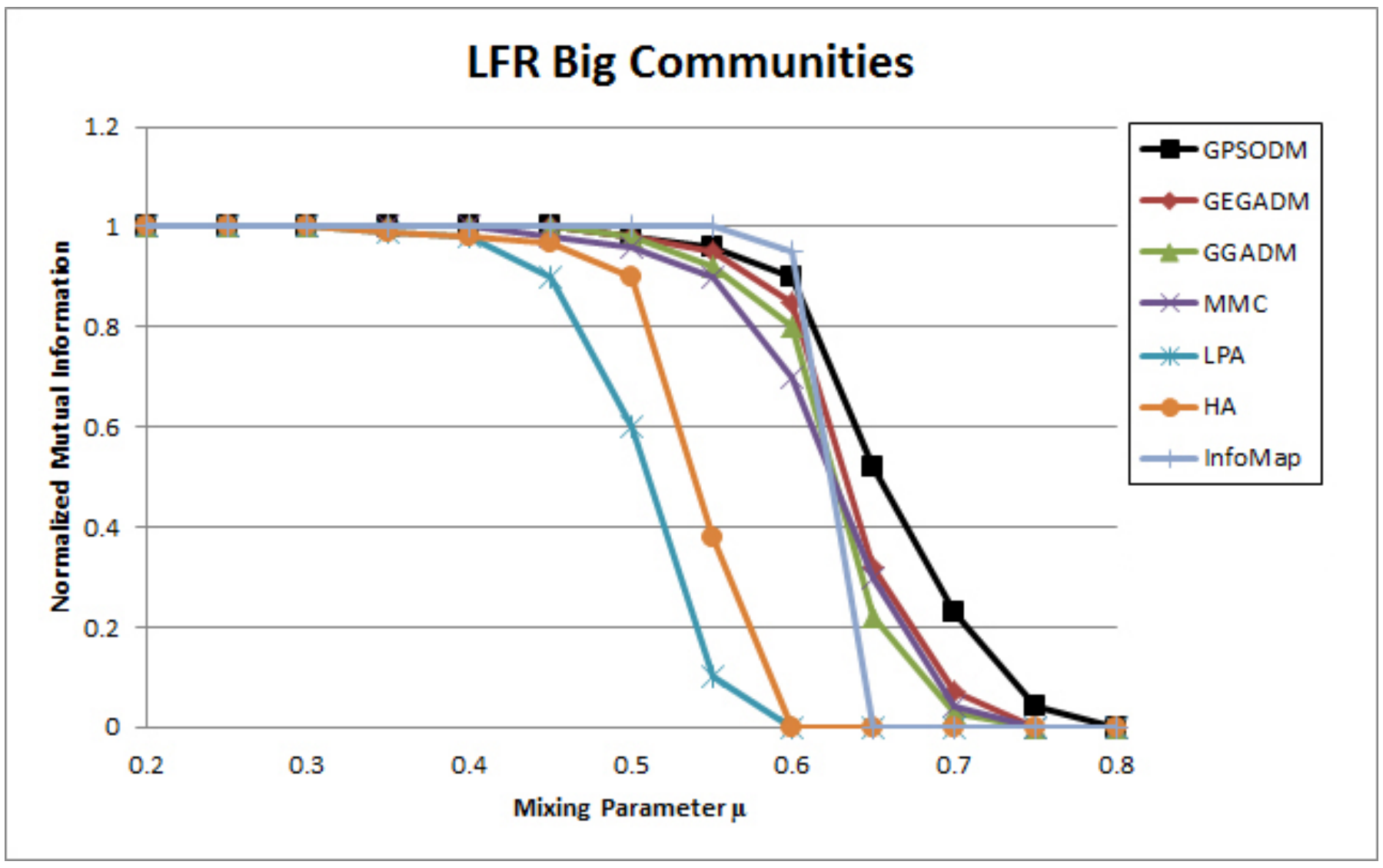

(ب)

**شکل 4-3- نمودار *NMI* به دست آمده از الگوریتم های مختلف روی شبکه ی *LFR***
**(الف) شبکه با تعداد 10000 رأس و سایز تشکل ها بین 20 تا 50. (ب) شبکه با تعداد 10000 رأس و سایز تشکل بین 20 تا 100**

به طور مشخص هر سه روش ارائه شده می توانند نتایج بهتری را نسبت به سایر روش های تشخیص تشکل ها بدست آورند. همچنین مشاهده می شود که بهترین نتیجه توسط الگوریتم *GPSODM* به دست می آید.

همانطور که مشاهده می شود در مورد شبکه های مصنوعی الگوریتم های ارائه شده نتایج مناسبی را کسب می نمایند و همان طور که انتظار می رفت الگوریتم *GPSODM* که بر اساس مدل جدید پایه گزاری شده است در این میان بهترین نتیجه را بدست آورده است لذا تا این مرحله می توان نتیجه گرفت که *PSODM* توانایی های بیشتری نسبت به سایر مدل های ارائه شده دارد.

### 4-4-1-3- شبکه های تصادفی *Erdös-Réyni*

در قسمت اول راجع به این شبکه ها توضیحات لازم را ارئه دادیم. نتایج این قسمت قابل نمایش دادن روی نمودار نیستند چرا که تنها کافی است که الگوریتم های ارائه شده بتوانند هیچ تشکلی در این شبکه ها نیابند. هر سه الگوریتم ارائه شده توسط ما نیز تمامی رئوس این شبکه را به عنوان یک تشکل واحد در نظر گرفته و در واقع ارتباط معنا داری میان رئوس نمی یابند.

### 4-4-2- مجموعه داده های واقعی

در این قسمت نتایج حاصل از تمامی سه روش ارائه شده روی مجموعه داده های واقعی را بررسی می نمایيم. ابتدا به ارائه ی نتایج به دست آمده روی دو مجموعه داده ی *Flickr* و *BlogCatalog* می پردازیم. همانطور که پیشتر نیز اشاره شد این دو شبکه از تعداد زیادی رأس تشکیل شده اند و به همین دلیل برخی از روش های تشخیص تشکل را نمی توان برای آن ها به کار برد. همچنین ساختار مشخصی برای این شبکه ها وجود ندارد و ارائه ی نتایج تنها برای اثبات قابلیت اجرای روش های ارائه شده روی آن ها می باشد.

**جدول 4-3 زمان اجرا و تعداد تشکل های تشخیص داده شده بر روی *Flickr***

| روش | زمان (ثانیه) | تعداد تشکل ها |
|---|---|---|
| **GPSODM** | 42306 | 134095 |
| **GEGADM** | 24654 | 132765 |
| **GGADM** | 22430 | 124347 |
| **سایر روش ها** | غیر قابل اجرا | غیر قابل اجرا |

همانطور که مشخص است با توجه به پیچیدگی روش *GPSODM* زمان اجرای این روش نیز نسبت به سایرین بیشتر است. همچنین این امر در مورد *GPSODM* نیز صادق است چرا که این روش نیز یک عملگر جهش را دارد که می تواند سربار زمانی ایجاد کند.

**جدول 4-4 زمان اجرا و تعداد تشکل های تشخیص داده شده بر روی BlogCatalog**

| روش | زمان (ثانیه) | تعداد تشکل ها |
|---|---|---|
| **GPSODM** | 8834 | 15340 |
| **GEGADM** | 6765 | 15645 |
| **GGADM** | 5430 | 14763 |
| **سایر روش ها** | غیر قابل اجرا | غیر قابل اجرا |

در مورد مجموعه داده های *BlogCatalog* نیز موارد مطرح شده در قسمت قبل صادق است و زمان اجرای الگوریتم ها به دلیل تفاوت در پیچیدگی با یکدیگر متفاوت است.

نتایج آورده شده در ادامه، مربوط به اجرای الگوریتم های مذکور بر روی داده های واقعی کوچک و البته معروف می باشد که تماما به برتری روش پیشنهادی اشاره می کنند. جداول 4-5 تا 4-7 به ترتیب نمایش دهنده ی نتایج میانگین در قالب معیار های *NMI*، *FCCN* و *Modularity* می باشند.

**جدول 4-5 Modularity میانگین بر روی مجموعه داده های واقعی**

| شبکه | InfoMap | LPA | MMC | HA | GGADM | GEGADM | GPSODM |
|---|---|---|---|---|---|---|---|
| **Dolphin** | 0.514 | 0.450 | 0.526 | 0.449 | 0.538 | 0.544 | **0.580** |
| **Karate** | 0.354 | 0.362 | 0.371 | 0.300 | 0.373 | 0.381 | **0.412** |
| **Football** | 0.575 | 0.597 | 0.595 | 0.566 | **0.598** | 0.560 | 0.583 |

**جدول 4-6 NMI میانگین بر روی مجموعه داده های واقعی**

| شبکه | InfoMap | LPA | MMC | HA | GGADM | GEGADM | GPSODM |
|---|---|---|---|---|---|---|---|
| **Dolphin** | 0.695 | 0.710 | 0.579 | 0.707 | 0.715 | **0.736** | 0.723 |
| **Karate** | 0.643 | 0.751 | 1.000 | 0.754 | 1.000 | 1.000 | **1.000** |
| **Football** | 0.899 | 0.927 | 0.885 | 0.907 | 0.838 | 0.910 | **1.000** |

**جدول 4-7 FCCN میانگین بر روی مجموعه داده های واقعی**

| شبکه | InfoMap | LPA | MMC | HA | GGADM | GEGADM | GPSODM |
|---|---|---|---|---|---|---|---|
| **Dolphin** | 0.586 | 0.609 | 0.543 | 0.601 | 0.619 | 0.701 | 0.678 |
| **Karate** | 0.768 | 0.896 | 1.000 | 0.889 | 1.000 | 1.000 | **1.000** |
| **Football** | 0.854 | 0.860 | 0.846 | 0.867 | 0.821 | 0.877 | **1.000** |

در ادامه نتایج بهترین اجرای هر یک از الگوریتم های ارائه شده روی مجموعه داده های Dolphin، Karate و Football Club را در شکل ها مشاهده می نماییم.

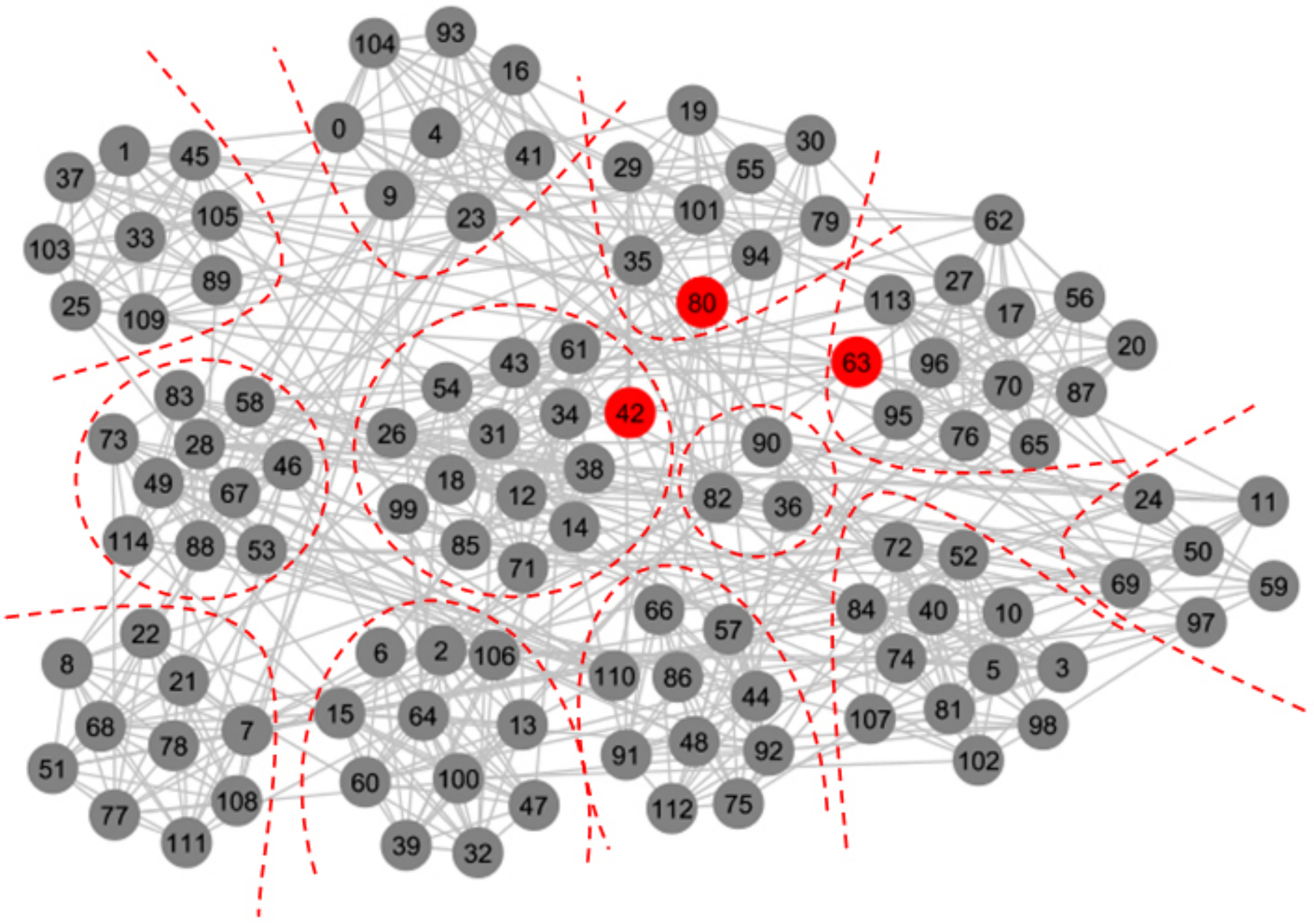

(الف)

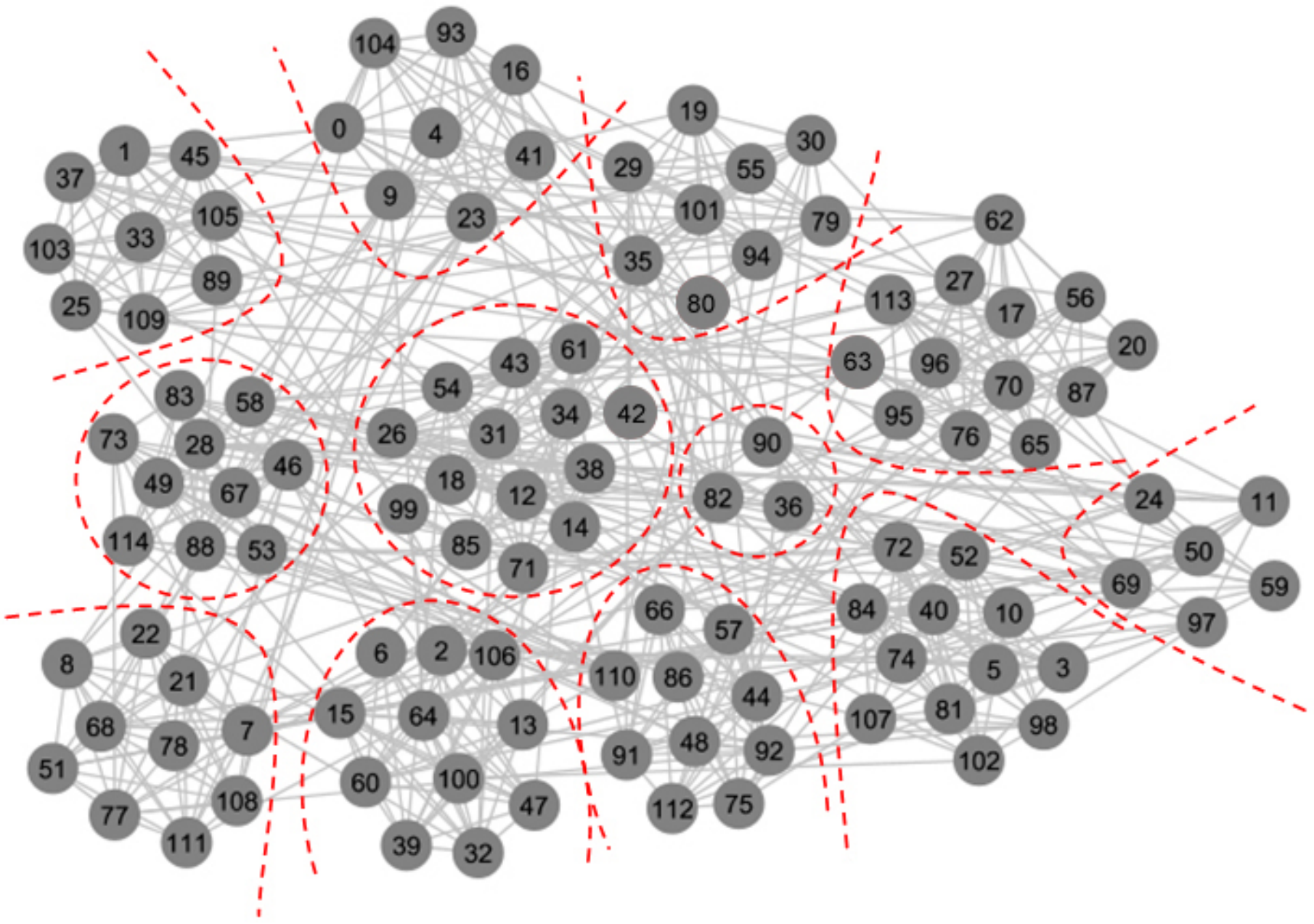

(ب)

شکل 4-4- تشکل های تشخیص داده شده در مجموعه داده ی *Football* (الف) نتیجه حاصل از الگوریتم های *GGADM* و*GEGADM* (ب) نتیجه حاصل از الگوریتم *GPSODM*

همانطور که مشاهده می شود دو الگوریتم *GGADM* و *GEGADM* نمی توانند 3 رأس را به درستی در تشکل های خود قرار دهند اما این مشکل در مورد الگوریتم *GPSODM* وجود ندارد.

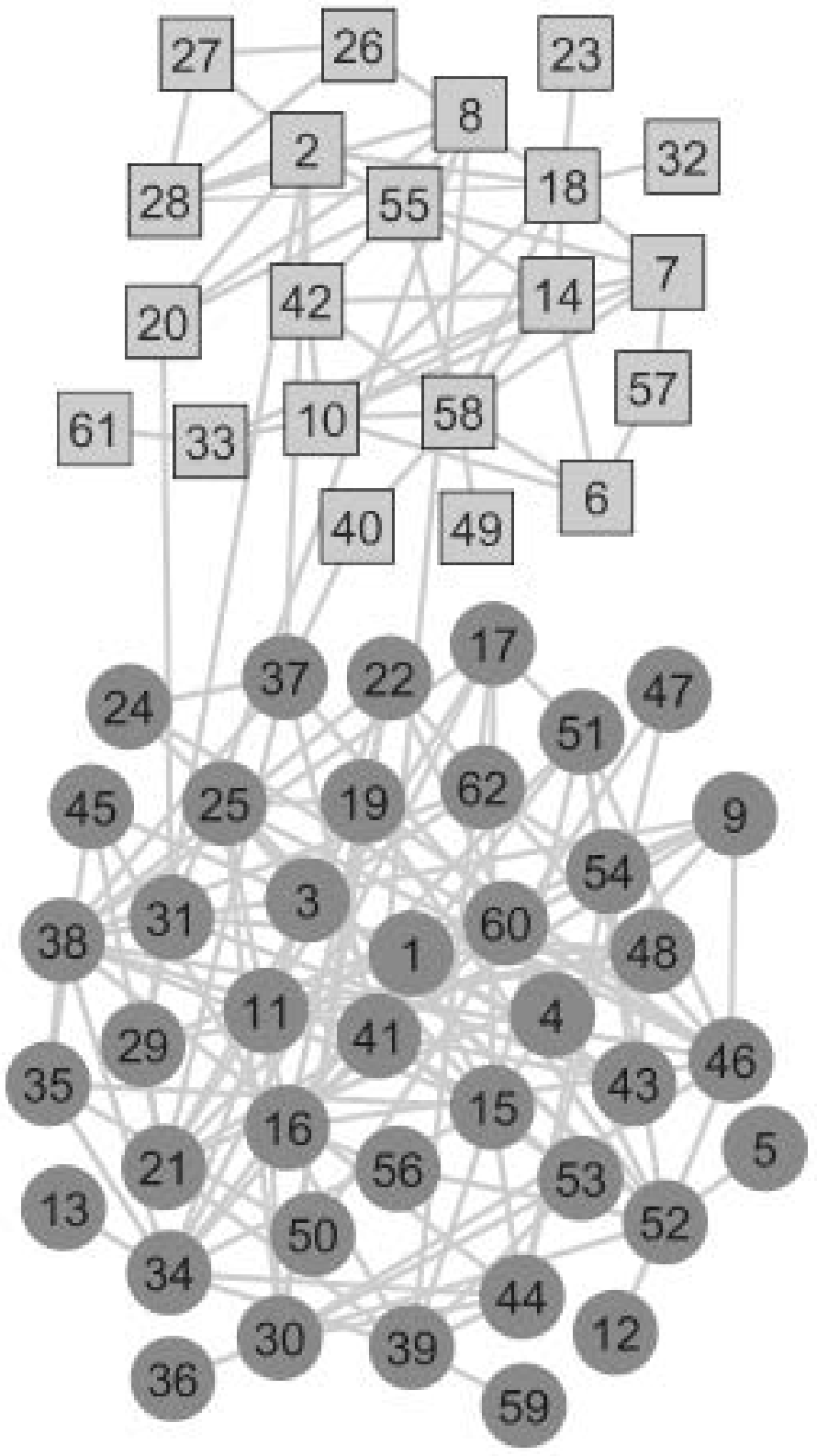

**شکل 4-5- تشکل های تشخیص داده شده در مجموعه داده ی *Dolphin***

شکل 4-5 بهترین نتیجه ی بدست آمده از الگوریتم های *GGADM*، *GEGADM* و *GPSODM* روی مجموعه داده ی *Dolphin* را نشان می دهد. ساختار تشکل از پیش معلوم این گراف ها که پیشتر در بخش معرفی مجموعه داده ها بررسی شد، دلالت بر برتری روش ما با ماکسیمم *NMI* ممکن دارد که البته به دلیل ماهیت تصادفی بودن روش های *GGADM*، *GEGADM* و *GPSODM* در پاره ای از اوقات به بهترین نتایج نمی رسیم و نشان دادن میانگین نتایج در جدول های 4-5 تا 4-7 نیز به همین علت بود.

## 4-4-3- تحلیل پیچیدگی زمانی

در این بخش، در ابتدا در مورد زمان اجرای روش های *GGADM*، *GEGADM*و *GPSODM* توضیح داده و سپس زمان

اجرا و تعداد عملیات انجام شده ی این روش ها را با یکدیگر مقایسه می نماییم.

### 4-4-3-1- پیچیدگی زمانی

پیچیدگی زمانی این سه الگوریتم برابر $O\,(k_1 \times m + k_2.C_1.\bar{d})$ می باشد. همانطور که پیشتر نیز بیان شد، در این روش ها، در ابتدا میزان اطلاعات رد و بدل شده میان رئوس شبکه را محاسبه می نماییم که این مرحله از الگوریتم به صورت آفلاین محاسبه می گردد و پیچیدگی زمانی آن برابر است با $O\,(k_1 \times m)$. سپس وارد بخش اصلی الگوریتم، یا همان بازی مورد نظر، می شویم. در این بخش و در هر دور یک عامل به صورت تصادفی انتخاب شده و این عامل بر اساس تابع سودمندی خود، یکی از اعمال موجود را انتخاب می کند. در نتیجه هر کدام از این روش ها به طور مشابه دارای پیچیدگی زمانی $O(k_2.C_1.\bar{d})$ می باشند که در آن $k_2$ ثابت و $C_1$ میانگین تعداد دفعاتی است که هر عامل انتخاب می شود. همچنین $\bar{d}$ میانگین درجات رئوس (عامل ها) می باشد. همچنین محاسبه ی شباهت ها بین عامل ها به صورت آفلاین حساب می شود و در نتیجه از در نظر گرفتن زمان آن چشم پوشی می کنیم.

شکل 4-6 مقایسه ی بین اعمال انجام شده در سه روش پیشنهادی بر روی گراف هایی با اندازه های مختلف را نشان می دهد. همانطور که انتظار داریم، متوسط اعمال انجام شده در آن هم به مراتب بیشتر است. در نتیجه زمان اجرای بالاتری هم برای آن قابل پیش بینی است که این امر در شکل 4-7 نشان داده شده است. در واقع روش *GPSODM* در کنار نتایج خوبی که دارد دارای مشکل پیچیدگی زمانی بالا نسبت به روش های دیگر می باشد.

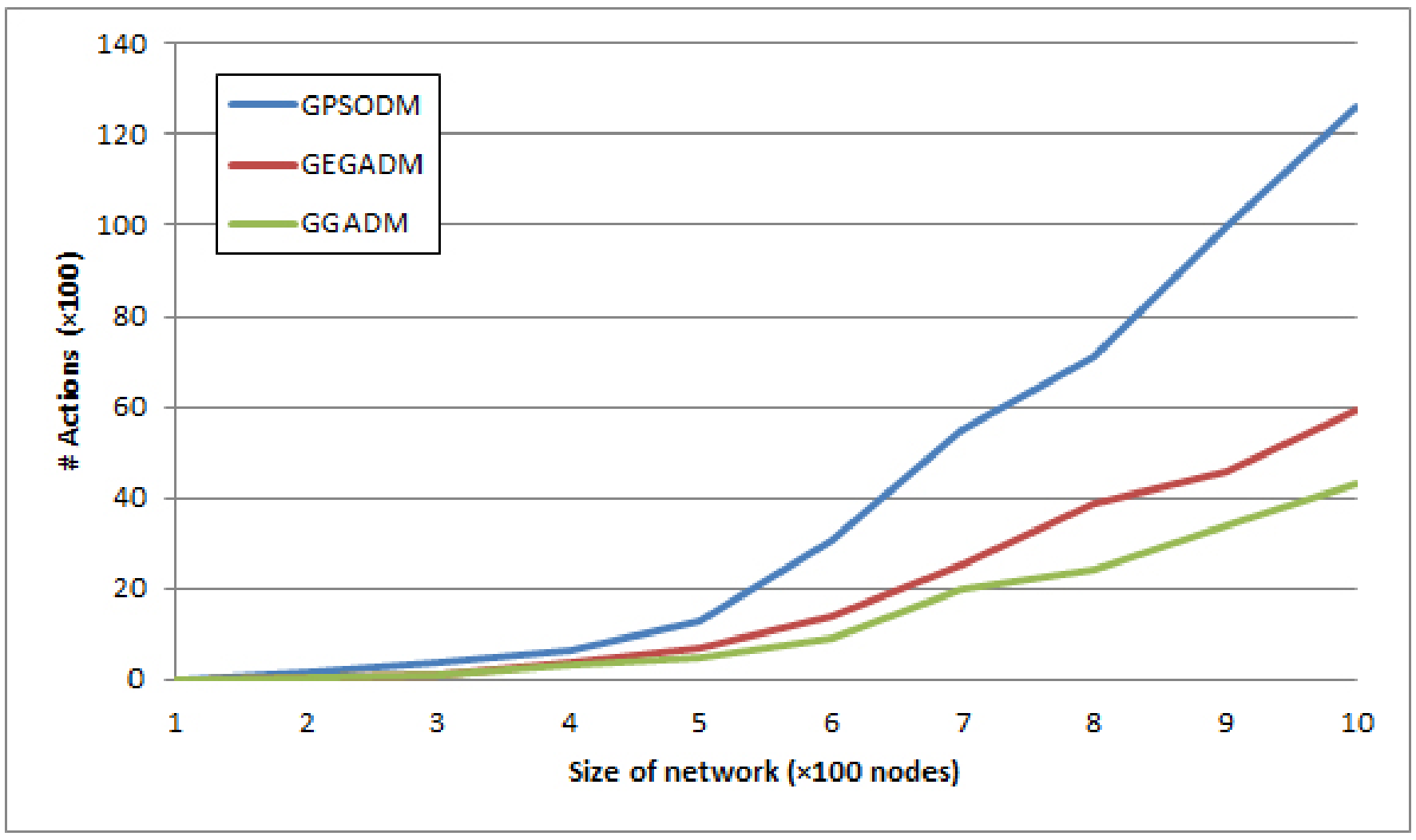

شکل 4-6-مقایسه بین میانگین تعداد عملیات انجام شده بر روی گراف های با اندازه های مختلف توسط روش های GPSODM، GEGADM و GPSODM

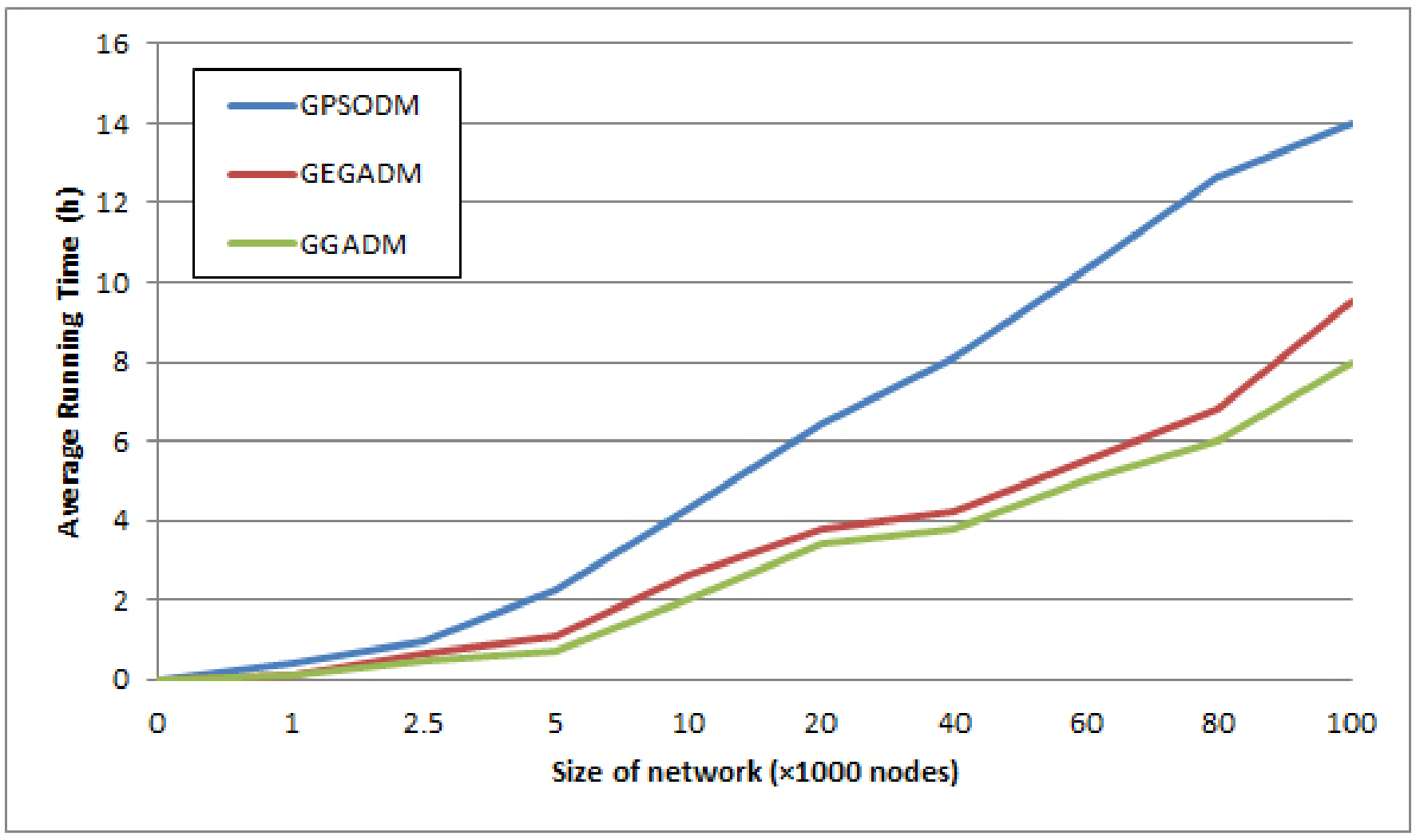

شکل 4-7-مقایسه بین میانگین زمان اجرای روش های GPSODM، GEGADM و GPSODM بر روی گراف های با اندازه های مختلف

# فصل پنجم

# جمع‌بندی

## 5-1- نتیجه گیری

امروزه فراگیری اینترنت و فن آوری های جدید ارتباطی و اطلاعاتی، موجب ظهور فضای مجازی در کنار جهان واقعی شده که این امر، معادلات و الگوهای ارتباطات سنتی، تولید، انتقال و مصرف اطلاعات را به هم زده و موجب تغییر در آن شده است. چنین فضایی که به عنوان واقعیت مجازی یک پارچه، در نظر گرفته می شود، از ویژگی هایی چون بی مکانی، فرا زمان بودن، صنعتی بودن محض، عدم محدودیت به قوانین مدنی متکی بر دولت- ملت ها، از معرفت شناسی تغییر شکل یافته پسا مدرن برخوردار بودن، قابل دسترسی بودن همزمان، روی فضا بودن و برخورداری از فضاهای فرهنگی، اعتقادی، اقتصادی، سیاسی و نیز آزادی از هویت بدنی و جنسی جدید برخوردار است. شبکه های اجتماعی مجازی، امروزه نقش بسیار مهمی در خلق این فضای مجازی دارند. از خلال همین واقعیت های مجازی است که آسیب های روانی و سیاسی بسیار گسترده ای را می توانند برای یک جامعه به وجود آورند.

شبکه های اجتماعی، به مجموعه ای از افراد که به صورت گروهی با یکدیگر ارتباط داشته و مواردی مانند اطلاعات، نیازمندی ها، فعالیت ها و افکار خود را به اشتراک بگذارند، شبکه های اجتماعی گویند. شبکه های اجتماعی را می توان به دو دسته شبکه های مجازی و شبکه های غیرمجازی تقسیم کرد. شبکه های غیرمجازی در واقع شبکه هایی هستند که توسط مجموعه ای از افراد و گروه های به هم پیوسته، در محیط اجتماعی عمل می کنند. شبکه اجتماعی مجازی یا شبکه اجتماعی اینترنتی، وب سایت یا مجموعه ای از وب سایت هایی است که به کاربران امکان می دهد، علاقه مندی ها، افکار و فعالیت های خود را با یکدیگر به اشتراک بگذارند؛ به عبارت دیگر، شبکه های اجتماعی سایت هایی هستند که با استفاده از یک موتور جست و جو گر و افزودن امکاناتی مانند چت، پیام رسانی الکترونیک، انتقال تصویر و صدا و...، امکان ارتباط بیشتر کاربران را در قالب شبکه ای از روابط فردی و گروهی فراهم می آورند.

شبکه های اجتماعی مجازی در جوامع جهانی مورد استقبال بیشتری قرار گرفته است. شبکه اجتماعی مجازی ابزار متنوعی است که به صورت آزاد و مجانی به آن دسترسی می توان داشت. یک فرد می تواند مطالب مورد نظر خود یا یک سری اطلاعات خاص را دریک ثانیه با صدها و حتی هزاران فرد در سراسر جهان به اشتراک بگذارد. این شبکه مزایا و معایب خاص خود را دارد، برای مثال عده ای از محققان معتقدند شبکه های مجازی باعث افزایش معاشرت پذیری می شود و در سوی دیگر نیز افرادی مقابل این تعریف قرار دارند و معتقدند شبکه های اجتماعی فعلی باعث کاهش ارتباط با خانواده می شود.

با گسترش وب اجتماعی، نیاز به تحلیل ساختارها و رفتارهای شبکه های اجتماعی، به عنوان یکی از نیازمندی های اساسی شرکت های تجاری مبدل گشت. تحلیل شبکه های اجتماعی در بسیاری از کاربردها از جمله مدیریت شبکه اجتماعی، تحلیل گرایش بازار، شناسایی افراد تاثیرگذار و حامیان، ارتقاء کارایی سامانه های توصیفگر و.... قابل استفاده است. نیازمندی های تجاری باعث شده است در سال های اخیر در بعد آکادمیک توجه زیادی به تحلیل شبکه های اجتماعی گردد. امروزه این ابزار قدرتمند نه تنها مورد توجه متخصصان فناوری اطلاعات می باشد، بلکه پژوهشگران سایر رشته هایی چون علوم تربیتی، زیست شناسی، علوم ارتباطات، اقتصاد، ....، به عنوان یک تکنیک کلیدی از تحلیل شبکه اجتماعی بهره می برند.

همانطور که پیشتر نیز بیان شد، تحلیل شبکه‌های اجتماعی به عنوان یک تکنیک کلیدی در جامعه شناسی، انسان شناسی، جغرافیا، روانشناسی اجتماعی، جامعه شناسی زبان، علوم ارتباطات، علوم اطلاعات، مطالعات سازمانی، اقتصاد و زیست شناسی مدرن همانند یک موضوع محبوب در زمینه ی تفکر و مطالعه پدیدار شده است.

جدا از ویژگی های آماری مشترک میان شبکه های اجتماعی ، ویژگی دیگری را نیز می توان در نظر گرفت که به تازگی کانون توجهات را به خود جلب کرده و از آن به عنوان راه حلی برای بسیاری از مشکلات موجود در شبکه های اجتماعی استفاده می شود. این ویژگی فرآیند انتشار در شبکه های اجتماعی نام دارد. انتشار اطلاعات را می توان به عنوان یکی از نمونه های فرآیند انتشار نام برد. انتشار اطلاعات یک تعریف عمومی است که شامل هر چیزی که در یک

شبکه گسترش می یابد می شود. بیشینه کردن گسترش اعتبار مانند یک ایده ی برتر و یا حتی تشخیص سریع یک فاجعه مانند دزدی. همه ی اینها نمونه هایی از انتشار اطلاعات هستند.

در نهایت از فرایند انتشاری که در شبکه های اجتماعی رخ می دهد، از جمله انتشار اطلاعات، جهت ارائه ی مدلی برای دسته ای از پدیده ها در این شبکه ها استفاده می شود. پدیده هایی مانند گسترش ویروس ها و بد افزارها در کامپیوتر ها، گسترش اطلاعات مربوط به کالاها در میان مردم و غیره. ارائه این مدل های آماری می تواند راهنمایی باشد جهت بررسی بیشتر ساختار شبکه های اجتماعی ، نحوه گسترش و انتشار اطلاعات در آنها و همچنین تشخیص تأثیرگذارترین رئوس در این شبکه ها ؛ مبحثی که به تازگی مورد توجه بسیاری از محققین قرار گرفته است.

اطلاعات می تواند باعث تشکیل یک ایده ی عمومی، ایجاد ترس و اضطراب در یک جامعه، پذیرش یک محصول توسط خریداران و غیره شود و به این طریق نقش بسیار اساسی را در سازمان های اجتماعی ایفا می کند. اطلاعات از طریق افراد مختلف در سطح شبکه منتشر می شود. افراد می توانند با استفاده از شیوه های گوناگون (شیوه های مرسومی نوشتاری، زبانی و الکترونیکی) باعث انتشار اطلاعات در شبکه ها شوند. ساختار اطلاعاتی با توجه به گسترش روز افزون استفاده از اینترنت و وب بسیار تغییر کرده و از شکل سنتی خود خارج شده است. تا چند سال پیش اگر فردی نیازمند دریافت اطلاعات خاصی می بود، می بایست هزینه ی ایجاد زیربنای ارتباطی با افراد مختلف را نیز می پرداخت. امروزه با توجه به دسترسی گستره و آسان به اینترنت این محدودیت ها از بین رفته است.

طی چند دهه ی اخیر، دانشمندان تنها به مشاهده و انتشار اطلاعات در شبکه های اجتماعی دقت نمی کنند و هدف بزرگتری را دنبال می نمایند. امروزه هدف اصلی ایجاد، گسترش و بیشینه کردن اطلاعات در یک شبکه است. انتشار اطلاعات یک اصل کلی است و هر چیزی که بتواند در یک شبکه منتشر شود زیر چتر انتشار اطلاعات قرار می گیرد. اهداف بررسی انتشار اطلاعات را می توان به دو دسته ی اصلی تقسیم کرد:

1. بیشینه کردن گسترش تاثیر
2. تشخیص سریع حادثه

جهت تحلیل و بررسی بیشتر این نوع مسائل و مشکلات، ما نیاز به درک عمیق تر از ساختارهای درگیر داریم که همین امر باعث می شود که بحث مدل سازی و پیش بینی انتشار اطلاعات بیشتر مورد توجه قرار گیرند. به طور دقیق تر می توان گفت که ما برای حل مشکلات در این زمینه نیازمند مدل های قوی جهت بررسی انتشار اطلاعات داریم.

مدل های انتشار اطلاعاتی که تاکنون مطرح شده اند پایه و اساسی مبتنی بر کارهای انجام شده در زمینه های گوناگونی از جمله جامعه شناسی، فیزیک، بیماری شناسی و بازاریابی دارند. تمامی این مدل ها فرض را بر این قرار می دهند که رأس های یک شبکه دو حالت دارند: فعال و غیرفعال. رأس های فعال اطلاعات را پخش کرده و به سایر رأس ها که در حالت غیر فعال هستند انتقال می دهند. مدل های گوناگون در این زمینه فرضیات متفاوتی را در مورد نحوه ی انتقال اطلاعات میان رأس ها در شبکه ها را در نظر می گیرند و همین امر است که موجب تمایز بین آن ها می شود.

به تازگی نیز مدل های گوناگونی طی سال های اخیر در این زمینه ارائه شده است. مشکلاتی که روش های مطرح شده تاکنون داشته اند به طور خلاصه شامل موارد زیر می شود:

5. نیاز به بهینه سازی تعدادی پارامتر در روش های اولیه
6. عدم توانایی در نمایش مسائل پیچیده ی دنیای امروز
7. عدم توانایی در نمایش واحد های چندگانه ی اطلاعاتی
8. عدم توانایی در تطابق با شبکه های بزرگ و پیچیده ی کنونی

اهمیت ارائه ی مدلی برای انتشار اطلاعات در شبکه های اجتماعی با توجه به تمامی مطالب ذکر شده در بالا و موارد و مشکلات موجود در اکثر مدل های متداول برای انتشار اطلاعات، ما را بر آن داشت تا در این پایان نامه، روشی بر مبنای الگوریتم بهینه سازی ازدحام ذرات جهت مدل سازی انتشار اطلاعات در شبکه های اجتماعی ارائه دهیم به طوریکه بسیاری از مشکلات ذکر شده را برطرف سازد. بررسی ها نشان می دهند که روش پیشنهادی، روشی کارا بوده و نتایج بهتری در مقایسه با سایر روش های موجود، ارائه می دهد.

در این پایان نامه دو روش پیشنهادی جهت مدل سازی انتشار اطلاعات در شبکه های اجتماعی ارائه گردید. روش اول بر مبنای یکی از جدید ترین مدل های ارائه شده در این زمینه عمل می کند و در واقع به نوعی مدل تعمیم یافته ی آن است و در روش دوم نیز ما با استفاده از الگوریتم های بهینه سازی ازدحام ذرات سعی در مدل سازی انتشار اطلاعات می نماییم.

مدل اول یا EGADM که همان مخفف کلمه ی Extended GADM است، به نحوی طراحی شده تا بتواند ساختار پیچیده تری نسبت به روش GADM را مدل سازی کند. در این روش ما با افزودن عملگر جهش سعی داریم که تلاش افراد در شبکه های اجتماعی را برای ارتقاء سطح دانش خود نمایش دهیم. به اینصورت که اصولا افراد همواره علاوه بر اطلاعات و دانشی که از سایرین دریافت می نمایند خود نیز به کسب اطلاعات و دانش خواهند پرداخت.

در مدل دوم یا PSODM ما روش نوینی را برای مدل سازی انتشار اطلاعات در شبکه های اجتماعی ارائه می دهیم. در این روش پایه و اساس کار الگوریتم های بهینه سازی ازدحام ذرات خواهند بود که به ما کمک می کنند تا بتوانیم عملگر های مناسبی را جهت انتقال اطلاعات میان رأس های یک شبکه به کار گیریم. استفاده از این عملگر ها مشکلات مطرح شده در مورد سایر روش ها را در مورد نبود قابلیت مدل سازی واحد های چندگانه ی اطلاعاتی برطرف می نماید. همچنین این امکان برای ما فراهم می شود تا شیوه ی نمایش پیچیده تری را برای نشان دادن اطلاعات یک رأس به کار بگیریم.

در نهایت برای اولین بار و به صورت کاملا کاربردی از مسئله ی تشخیص تشکل ها در شبکه های اجتماعی برای مقایسه مدل ها استفاده می نماییم. سه روش GGADM، GEGADM و GPSODM حاصل این قسمت هستند. این امر برای ما امکان بهبود مدل ها را نیز فراهم می آورد.

## 5-2- پیشنهادات برای کار های آتی

از جمله پیشنهادات برای کار های آتی که مرتبط با این پایان نامه است می توان به ارائه و معرفی الگوریتمی جهت بررسی مدل ها روی شبکه هایی با ساختار غیر مشخص اشاره کرد. شبکه هایی که در آن ها تمامی یال ها و ارتباط ها مشخص نباشند و نتوان

ساختار دقیق آن ها را بدست آورد. همچنین می توان از روش های جدیدی جهت ارزیابی برازندگی رشته های ورودی استفاده نماییم. از طرفی منظور از انتشار اطلاعات در این پایان نامه، سطح دانش از یک موضوع خاص است که می توان از اعتماد نیز به عنوان اطلاعات استفاده نماییم.

برای اجرا بر روی گراف هایی با اندازه های بسیار بزرگ، می توان روش ارائه شده را بر روی *Hadoop* ، چارچوب ارائه شده توسط *Apache™* ، اجرا کرد.

# منابع و مأخذ

## انگلیسی